\documentclass[showpacs,twocolumn,pra]{revtex4-1}
\usepackage{amsmath}
\usepackage{amssymb}
\usepackage{dsfont}
\usepackage{color}
\makeatletter
\usepackage{graphicx}
\usepackage{epstopdf}
\usepackage{epsfig}
\usepackage{subfigure}
\usepackage{enumerate}
\makeatother

\newcommand{\ket}[1]{|#1\rangle}

\renewcommand{\vec}[1]{\mathbf{#1}}

\newcommand{\partder}[1]{\frac{\partial}{\partial #1}}

\DeclareMathOperator{\Tr}{Tr}

\usepackage{cancel,ifthen}
\usepackage[normalem]{ulem}

\newcommand{\comment}[2][NoInPuT]{\ifthenelse{\equal{#1}{NoInPuT}}{}{{\color{blue}\sout{#1}}}{\color{red} #2}}
\allowdisplaybreaks
\begin{document}
\raggedbottom
\title{A Non-Equilibrium Kinetic Theory for Trapped Binary Condensates}

\author{M. J. Edmonds}
\author{K. L. Lee}
\author{N. P. Proukakis}
\affiliation{Joint Quantum Centre (JQC) Durham-Newcastle, School of Mathematics and Statistics,
Newcastle University, Newcastle upon Tyne NE1 7RU, England, UK}

\date{\today{}}

\begin{abstract}\noindent
We derive a non-equilibrium finite-temperature kinetic theory for a binary mixture of two interacting atomic Bose-Einstein condensates and use it to explore the degree of hydrodynamicity attainable in realistic experimental geometries. Based on the standard separation of timescale argument of kinetic theory, the dynamics of the condensates of the multi-component system are shown to be described by dissipative Gross-Pitaevskii equations, self-consistently coupled to corresponding Quantum Boltzmann equations for the non-condensate atoms: on top of the usual mean field contributions, our scheme identifies a total of 8 distinct collisional processes, whose dynamical interplay is expected to be responsible for the system's equilibration. In order to provide their first characterization, we perform a detailed numerical analysis of the role of trap frequency and geometry on collisional rates for experimentally accessible mixtures of $^{87}$Rb-$^{41}$K and $^{87}$Rb-$^{85}$Rb, discussing the extent to which the system may approach the hydrodynamic regime with regard to some of those processes as a guide for future experimental investigations of ultracold Bose gas mixtures.  
\end{abstract}
\maketitle
\section{Introduction}
The possibility of unprecedented control over experimental parameters in ultracold atom experiments, such as the statistics, interactions and dimensionality of trapped gases \cite{weiner_bagnato_1999,leggett_2001,lewenstein_sanpera_2007,giorgini_pitaevskii_2008}, offers an opportunity to elucidate novel many-body quantum effects. At the level of a single-component Bose gas, the study of the Bose-Einstein condensate has already bifurcated into a plethora of directions. Opportunities now exist to investigate a broad spectrum of problems, including mimicking the behaviour of atoms in solids using sophisticated optical manipulations \cite{bloch_dalibard_2008,bloch_2005,morsch_oberthaler_2006}, as well as applications to quantum information and computation \cite{brennen_caves_99,jaksch_cirac_00,schmiedmayer_folman_02}.
Experimental advances have also led to the controlled generation of multi-component~\cite{hall_matthews_1998,matthews_anderson_1999,maddaloni_modugno_2000,modugno_modugno_2002,simoni_ferlaino_2003,papp_pino_2008,tojo_taguchi_2010,sugawa_yamazaki_2011,lercher_takekoshi_2011,mccarron_cho_11,pasquiou_bayerle_2013,xiong_li_2013,wacker_jorgensen_2015} and spinor~\cite{stamperkurn_ueda_2013,stenger_inouye_1998,lewandowski_mcguirk_2003,schweikhard_coddington_2004,kronjager_becker_2005,sadler_higbie_06,beattie_moulder_2013} condensates, with the dynamical interplay between different components leading to even richer physics, including, for example, phase separation~\cite{ho_shenoy_1996,pu_bigelow_1998,pu_bigelow_1998a} and spin-domain formation~\cite{stenger_inouye_1998,sadler_higbie_06}. Recently, condensates have also been used to simulate gauge theories, which has attracted intense experimental and theoretical focus due to the strong analogies with condensed matter systems \cite{dalibard_gerbier_2011,goldman_juzeliunas_2014}. 
The behavior of Bose gas mixtures is also related to the study of doubly-superfluid Bose--Fermi mixtures in the BEC regime \cite{ferrier_delehaye_14}, where the Fermi gas forms a molecular condensate.

For the single-component condensates, an understanding of the dynamics of Bose-condensed systems often relies on the Gross-Pitaevskii equation, which naturally encompasses the wave-like behaviour of the weakly interacting gas, valid deep within the ultracold regime. However, to gain insight into the dynamics of the gas over a broader range of temperatures, one must explicitly consider the behaviour of the normal component of the system, which leads to a rich non-equilibrium behaviour. Numerous approaches have been devised to describe the condensate dynamics in the presence of a thermal cloud (see e.g the reviews \cite{proukakis_gardiner_2013,proukakis_jackson_2008,berloff_brachet_14,griffin_nikuni_2009,blakie_bradley_08,brewczyk_gajda_07}), each with its own merits and drawbacks. 
%
Classical field methods \cite{blakie_bradley_08,brewczyk_gajda_07,kagan_svistunov_92a,*kagan_svistunov_92b,davis_morgan_01,davis_morgan_02,berloff_svistunov_02} cumulatively describe the highly-occupied low-lying "classical" modes of the gas, relying on the ergodic relaxation of a non-equilibrium initial state (to a Rayleigh-Jeans distribution); appropriately-sampled quantum noise could also be added in the initial states to mimic quantum fluctuations (an approach referred to as "truncated Wigner") \cite{steel_olsen_98,sinatra_lobo_02}. Explicitly adding a stochastic coupling to a heat bath, representing the set of high-lying modes largely unaffected by the condensate, one can also introduce fluctuating dynamics into the system 
\cite{stoof_bijlsma_00,duine_stoof_2001,gardiner_davis_2003,proukakis_03,rooney_blakie_12}; this is expected to be mostly relevant for studying equilibrium fluctuations 
\cite{andersen_alkhawaja_02,*alkhawaja_andersen_02b,cockburn_negretti_11,cockburn_gallucci_11,gallucci_cockburn_12,cockburn_proukakis_2012,davis_blakie_2012} and quenched dynamics \cite{weiler_neely_08,proukakis_03,proukakis_schmiedmayer_2006,cockburn_nistazakis_2010,damski_zurek_10}. 
While such approaches are suited for describing the critical region, they only describe dynamics up to a (fixed energy/momentum) cutoff \cite{blakie_bradley_08}, and cannot therefore account for any perturbations of the high-lying, thermal, modes. 

Contrary to such approaches, the dynamics of thermal modes can be accurately handled by an alternative perturbative method, following the usual route of kinetic theory, which describes the coupled condensate and thermal cloud dynamics, based on a separation of timescales argument 
\cite{kirkpatrick_dorfman_1983,kirkpatrick_dorfman_1985,kirkpatrick_dorfman_1985a,zaremba_nikuni_1999,walser_williams_1999,walser_cooper_01,wachter_walser_01a,*wachter_walser_01b,proukakis_burnett_1996,proukakis_burnett_1998,proukakis_2001a};
while this method relies on symmetry-breaking \cite{griffin_1996}, and thus fails to account for the critical fluctuation region, it is particularly suited to studying damping of collective modes and macroscopic excitations, which it has done very successfully \cite {williams_griffin_01,jackson_zaremba_01,jackson_zaremba_02c,jackson_zaremba_03,jackson_proukakis_2007,jackson_proukakis_09,allen_zaremba_13}.
%
Despite its inherent limitation in requiring the assumption of a non-zero condensate mean field (which can however be negligibly small), this method (referred to by many as the "Zaremba--Nikuni--Griffin", or ZNG method \cite{zaremba_nikuni_1999}) has nonetheless been found to perform very well even on the issue of condensate number growth following a sudden truncation in the thermal distribution \cite{bijlsma_zaremba_2000} or on surface evaporative cooling \cite{markle_allen_14}, complementing studies based on other approaches which do not themselves require a symmetry-broken condensate mean field potential when initiating the numerical simulations \cite{davis_gardiner_00,stoof_bijlsma_00,gardiner_lee_1998,kohl_davis_2002,proukakis_schmiedmayer_2006,hugbart_retter_07}. 

A somewhat similar kinetic approach, which is explicitly number-conserving, and does not invoke symmetry-breaking \cite{gardiner_morgan_2007,*gardiner_morgan_2007err} has also been successfully implemented for describing system dynamics \cite{billam_gardiner_12,billam_mason_13}.


In the context of multi-component condensates, which have been extensively studied with coupled Gross-Pitaevskii equations (GPEs) \cite{ho_shenoy_1996,pu_bigelow_1998,pu_bigelow_1998a,myatt_burt_1997,Trippenbach2000a,Busch2001a,Ohberg2001a,Coen2001a,Berloff_05,Kasamatsu2006a,Sasaki2011a,pattinson_billam_2013}, or
their dissipative generalisations \cite{Ronen2008a,Achilleos2012a,pattinson_2014}, 
their finite temperature dynamics remains a partly open problem. Approaches considered to date include classical field \cite{pattinson_proukakis_14}, truncated Wigner
\cite{Sabbatini2011a,Sabbatini2012a,Swislocki2013a}, coupled stochastic projected Gross-Pitaevskii equations 
\cite{bradley_blakie_2014,De2014a,Su2013a,liu_pattinson_2014}, or number-conserving approaches \cite{mason_gardiner_2014}. However, the detailed dynamics {\em far from} the critical region are expected to be better described by a model that fully accounts for all condensate and thermal cloud dynamics. This is particularly important since, parallel to the internal relaxation within each system, the two dynamical thermal clouds will also need to equilibrate together, thus creating a rather involved competition of collisional processes, with distinct timescales. While the promising number-conserving method of Ref.~\cite{mason_gardiner_2014} has not yet been advanced to the self-consistent dynamical level, all other methods (classical field, truncated Wigner and stochastic GPEs) feature a cutoff, and thus ignore the coupling of the high-lying thermal modes within and across the two systems; although such an approximation may be adequate for certain non-equilibrium features (e.g. defect formation following a quench \cite{weiler_neely_08}, persistent current decay \cite{rooney_neely_13}), it is nonetheless known to fail, at least formally, in some cases; a typical example of this is the Kohn mode of oscillation set up by a harmonic trap displacement which is not reproduced by such models \cite{bradley_blakie_05}. Variants of the kinetic model described here, whose single-component limit does not suffer from such a problem \cite{griffin_nikuni_2009}, have been put forward in \cite{nikuni_williams_2003,endo_nikuni_2011,edmonds_lee_2015}; as explained in more detail within the present manuscript, the latter work~\cite{edmonds_lee_2015} undertaken by the present authors, was specifically designed in order to introduce the collisional terms not explicitly dealt with in previous kinetic approaches, in a way which facilitates its numerical implementation.

The aim of this work is twofold: (i) firstly, we provide a detailed derivation (Secs.~II--IV) of our previously proposed multi-component kinetic scheme \cite{edmonds_lee_2015}, which includes both condensate and thermal cloud dynamics and all their cross-collisional terms; (ii) moreover, we show how numerical application of our scheme to near-equilibrium situations  (Sec.~V) can be used to map out regimes of near-hydrodynamic behaviour in accessible experimental mixtures, clearly highlighting the extent to which the relevant degree of "hydrodynamicity" with respect to different collisional processes can be controlled.
For completeness, we also briefly describe hydrodynamic multi-component equations (Sec.~VI) and summarize the relevance of our work in the context of existing multi-component treatments (Sec.~VII). 
The derivations presented in the main text are also supplemented by five more technical appendices.

%
\begin{figure*}[t]
    \centering
    \includegraphics[width=1.06\textwidth]{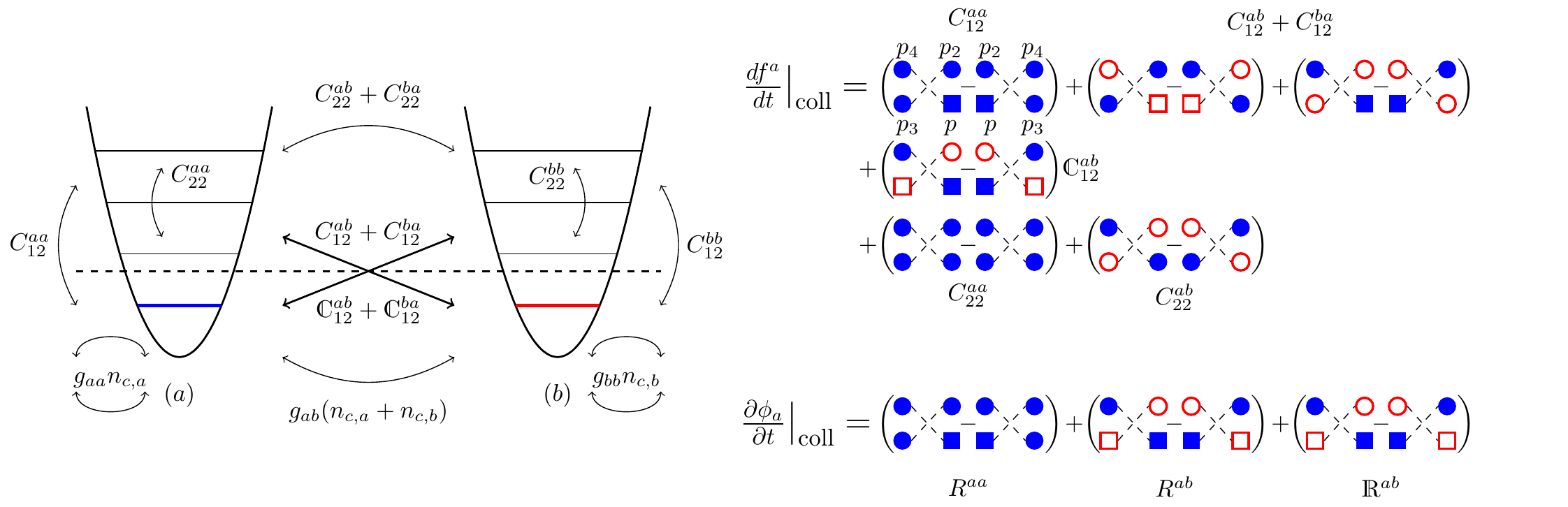}
    \caption{(Color online) Schematic representation of the various scattering processes for the binary system. Left: The binary system  along with all possible transport pathways.
Each component is composed of a condensate (below the dashed line) and a collection of non-condensate modes (above the dashed line), cumulatively comprising the thermal cloud.
Both collisional processes (denoted by $C$ and $\mathds{C}$) and condensate-condensate scattering events that contribute to the mean-field potential, $U_{c}$, seen by each condensate are clearly highlighted. Right: Schematic representation of the coupled equations for the condensates (Eq. \eqref{dschro1}) [bottom] and the thermal clouds (Eq. \eqref{qbe}) [top] are shown for component $a$;
each diagram represents a momentum and energy conserving collision between condensate $a$ ($b$) atoms, shown as closed blue (open red) squares, while thermal $a$ ($b$) atoms are depicted as closed blue (open red) circles. }
    \label{figbig}
\end{figure*}

\section{Coupled Dynamical Equations}
The starting point for our derivation will be the general Hamiltonian describing an interacting bosonic binary system, with the two components labeled $a$ and $b$ respectively. The Hamiltonian describing the binary system is written in second-quantized form as
\begin{equation}
\hat{H}=\int d{\bf r}\bigg\{\sum_{j}\hat{\Psi}^{\dagger}_{j}\bigg[-\frac{\hbar^2}{2m_j}\nabla^2+V_{j}({\bf r})\bigg]\hat{\Psi}_{j}\bigg\}+\hat{H}_{I},\label{ham1}
\end{equation}
and the two-body interactions are given by
\begin{equation}\label{hami}
\hat{H}_{I}{=}{\int}{d{\bf r}}\bigg\{\sum_{j}\frac{g_{jj}}{2}\hat{\Psi}^{\dagger}_{j}\hat{\Psi}^{\dagger}_{j}\hat{\Psi}_{j}\hat{\Psi}_{j}{+}\sum_{k\neq j}g_{kj}\hat{\Psi}^{\dagger}_{j}\hat{\Psi}^{\dagger}_{k}\hat{\Psi}_{k}\hat{\Psi}_{j}\bigg\}
\end{equation}
where $\hat{\Psi}_{j}\equiv\hat{\Psi}_{j}({\bf r})$ is the bosonic annihilation operator for an atom of species-$j$, which obey the usual commutation relationships for bosons 
\begin{align}
&[\hat{\Psi}_j({\bf r}),\hat{\Psi}^{\dagger}_{k}({\bf r}')]=\delta_{kj}\delta({\bf r}-{\bf r}'),\\ &[\hat{\Psi}_{j}({\bf r}),\hat{\Psi}_{k}({\bf r}')]=[\hat{\Psi}^{\dagger}_{j}({\bf r}),\hat{\Psi}^{\dagger}_{k}({\bf r}')]=0. 
\end{align}
The $s$-wave collisions between atoms in different components are encompassed by $g_{kj}=2\pi\hbar^2 a_{kj}/m_{kj}$, where $a_{kj}$ defines the scattering length between atoms in components $j$ and $k$, and $m^{-1}_{kj}=m^{-1}_{j}+m^{-1}_{k}$ defines the reduced mass. 
The underlying single-particle Hamiltonian appearing in Eq. \eqref{ham1} can in general contain external potentials, coherent couplings as well as the effective trapping and kinetic energies of the atoms.
Here $V_{j}({\bf r})$ denotes the trapping potential for atoms of species $j$ and can be of any form. 

In the language of symmetry breaking the condensed and non-condensed degrees of freedom are separated by means of the Beliaev decomposition
\begin{equation}
\hat{\Psi}_{j}({\bf r})=\phi_{j}({\bf r})+\hat{\delta}_{j}({\bf r}).
\label{beliaev}
\end{equation}
The condensate of component $j$ is described by the classical field $\phi_{j}({\bf r})\equiv\langle\hat{\Psi}_{j}({\bf r})\rangle$, while the non-condensate for component $j$ is encapsulated by the fluctuation operator $\hat{\delta}_{j}({\bf r})$, whose symmetry breaking average is taken as zero \footnote{Subtle issues associated with the validity of this statement are addressed in the related number-conserving approaches of Refs.~\cite{morgan_2000,morgan_2004,gardiner_morgan_2007,*gardiner_morgan_2007err}.}, i.e. $\langle\hat{\delta}_{j}^{(\dagger)}\rangle\equiv0$. 
Using the equations of motion for the Bose field operators obtained from the Heisenberg picture and taking averages with respect to a broken-symmetry non-equilibrium ensemble, one obtains the equation of motion for component $j$ (for $j \in\{a,b\}$) of the condensate field $\phi_{j}\equiv\phi_j({\bf r},t)$ in the form \cite{griffin_1996,proukakis_burnett_1996,proukakis_jackson_2008}
\begin{align}\nonumber
i\hbar\frac{\partial\phi_j}{\partial t}=&\bigg[-\frac{\hbar^2}{2m_j}\nabla^2+U^{j}_{c}\bigg]\phi_j+g_{jj}\bigg[\langle\hat{\delta}_{j}\hat{\delta}_{j}\rangle\phi^{*}_{j}+
\langle\hat{\delta}^{\dagger}_{j}\hat{\delta}_{j}\hat{\delta}_{j}\rangle\bigg]\\&+g_{kj}\bigg[\langle
\hat{\delta}^{\dagger}_{k}\hat{\delta}_{j}\rangle\phi_k+\langle\hat{\delta}_{k}\hat{\delta}_{j}
\rangle\phi^{*}_{k}+\langle\hat{\delta}^{\dagger}_{k}\hat{\delta}_{j}\hat{\delta}_{j}\rangle\bigg].\label{eom_exp}
\end{align}
Here we have defined an effective potential for the component $j$ condensate, to encompass, in addition to the trap potential, the mean fields of both condensate and thermal clouds of both components, via
\begin{equation}
U^{j}_{c}({\bf r},t)=V_{j}({\bf r})+g_{jj}(n_{c,j}+2\tilde{n}_{j})+g_{kj}(n_{c,k}+\tilde{n}_{k}),
\end{equation}
where $n_{c,j}=|\phi_j|^2$ is the condensate density for component $j$, 
and $\tilde{n}_{j}=\tilde{n}_{jj}=\langle\hat{\delta}^{\dagger}_{j}\hat{\delta}_{j}\rangle$
is the (diagonal) non-condensate density; we also introduce the off-diagonal non-condensate density $\tilde{n}_{kj}=\langle\hat{\delta}^{\dagger}_{j}\hat{\delta}_{k}\rangle$ valid for $j\neq k$. The total density of component $j$ is defined by
\begin{equation}
n_{j}=n_{c,j}+\tilde{n}_{j} = |\phi_j|^2 + \langle\hat{\delta}^{\dagger}_{j}\hat{\delta}_{j}\rangle \;.
\end{equation}
Equation \eqref{eom_exp} can then be written in the simpler form
\begin{equation}
i\hbar\frac{\partial\phi_j}{\partial t}=\bigg[-\frac{\hbar^2}{2m_j}\nabla^2+U^{j}_{c}-i(R^{jj}+R^{kj}+\mathds{R}^{kj})\bigg]\phi_j,\label{dschro1}
\end{equation}
where the important source terms $R^{jj}$, $R^{kj}$ and $\mathds{R}^{kj}$ account for atomic transport between the condensate and non-condensate for the two components of the gas, and are defined in terms of triplet and pair {\em anomalous} averages of the fluctuation operators $\hat{\delta}^{(\dagger)}_{j}$ as:
\begin{itemize}
\item $R^{jj}{=}{-}ig_{jj}\langle\hat{\delta}^{\dagger}_{j}\hat{\delta}_{j}\hat{\delta}_{j}\rangle/\phi_{j}$ describes the intra-component scattering of condensate and non-condensate atoms, as in the usual single-component kinetic equations~\cite{proukakis_burnett_1996,zaremba_nikuni_1999,bijlsma_zaremba_2000,proukakis_jackson_2008};
\item $R^{kj}{=}{-}ig_{kj}\langle\hat{\delta}^{\dagger}_{k}\hat{\delta}_k\hat{\delta}_j\rangle/\phi_j$ describes scattering between different components;
\item $\mathds{R}^{kj}{=}{-}ig_{kj}\langle\hat{\delta}_{k}^{\dagger}\hat{\delta}_{j}\rangle\phi_{k}/\phi_{j}$ differs qualitatively from the first two (see later) and accounts for an important "condensate collisional exchange" process not explicitly included in previous treatments.
\end{itemize}
Within the so called ``Popov approximation'' (see Ref. \cite{griffin_1996}), pair anomalous terms appearing as $\langle\hat{\delta}_{j}\hat{\delta}_{j}\rangle$ (diagonal) and $\langle\hat{\delta}_k\hat{\delta}_j\rangle$ (off-diagonal) in Eq. \eqref{eom_exp} are dropped (this is justified on energy conservation considerations -- see Appendix \ref{app_a} for a discussion of their physical meaning). 

The corresponding dynamics of the non-condensate atoms are encapsulated by coupled quantum Boltzmann equations for each component of the gas. Adopting the notational shorthand $f^{jj}({\bf r},{\bf p},t)\equiv f^{j}$ for the distribution function, the kinetic equation for component $j$ is written as
\begin{align}\nonumber\label{qbe}
&\frac{\partial}{\partial t}f^{j}+\frac{1}{m_j}{\bf p}\cdot\nabla_{\bf r}f^{j}-\nabla_{\bf p}f^{j}\cdot\nabla_{\bf r}U^{j}_{\text{n}}\\&=\bigg(C^{jj}_{12}+C^{kj}_{12}\bigg)+\mathds{C}^{kj}_{12}+\bigg(C^{jj}_{22}+C^{kj}_{22}\bigg).
\end{align}
Equation \eqref{qbe} defines the quantum Boltzmann equation for component $j$ of the binary system, where each of the collision integrals on the right hand side describe qualitatively distinct scattering processes occurring within the multi-component partially condensed bosonic mixture. The non-condensate density associated with component $j$ is given by
\begin{equation}
\tilde{n}_{j}({\bf r},t)=\int\frac{d{\bf p}}{(2\pi\hbar)^3}f^{j}({\bf r},{\bf p},t).\label{eq:tilden}
\end{equation} 
A schematic representation of all arising collisional processes for the binary mixture is shown in
Figure \ref{figbig}. Equations \eqref{dschro1}-\eqref{qbe} represent a closed system of equations, with the three source terms of Eq. \eqref{dschro1} related to the collision integrals in Eq. \eqref{qbe} via the relationships
\begin{subequations}
\begin{align}
R^{jj}({\bf r},t)&=\frac{\hbar}{2n_{c,j}}\int\frac{d{\bf p}}{(2\pi\hbar)^3}C^{jj}_{12},\label{rjj_pair}\\
R^{kj}({\bf r},t)&=\frac{\hbar}{2n_{c,j}}\int\frac{d{\bf p}}{(2\pi\hbar)^3}C^{kj}_{12},\label{rkj_pair}\\
\mathds{R}^{kj}({\bf r},t)&=\frac{\hbar}{2n_{c,j}}\int\frac{d{\bf p}}{(2\pi\hbar)^3}\mathds{C}^{kj}_{12}.\label{rkj_ex}
\end{align}
\end{subequations}
In this work we first detail the derivation of the above equations, which is similar in spirit to the established methodology \cite{zaremba_nikuni_1999,griffin_nikuni_2009,proukakis_jackson_2008} and subsequently use them to analyze the relative importance of the emerging collisional processes and the degree of hydrodynamicity of typical experimental configurations.  

\section{Kinetic Formalism\label{sec:kf}}

In order to correctly account for all of the relevant scattering channels amongst atoms in the binary mixture, a careful microscopic analysis is required. Pioneering work \cite{kirkpatrick_dorfman_1983,kirkpatrick_dorfman_1985,kirkpatrick_dorfman_1985a} demonstrated how quantum kinetic theory could be used to understand the dynamics of the non-condensate. 

\subsection{Separation of Timescales: Identification of Slowly-Varying "Master" Variables}

Trapped atoms within the gas are treated as undergoing motion within the trapping potential, which is occasionally interrupted by the $s$-wave collisions between particles. As such, two important collision timescales emerge: the duration of a single collision event between a pair of particles, which is defined as $\tau_0=a_{kj}/v$, where $v$ is the average velocity of the particles during the collision event, and the time in-between collisions $\tau_c=1/(na^{2}_{kj}v)$ where $n$ denotes the particle density \cite{walser_williams_1999}. As the kinetic and interaction energies of the particles are typically small, the dynamics of the gas are encapsulated by a separation of timescales that satisfies $\tau_0\ll\tau_c$, implying that for weak interactions, we can apply an effective perturbative treatment, which is fully equivalent to the adiabatic elimination of anomalous averages used in Refs.~\cite{proukakis_burnett_1996,proukakis_burnett_1998,proukakis_2001a,davis_ballagh_2001}.

Such an action requires the explicit identification of a few {\em slowly-varying} "master" variables.
This task should not be taken lightly, as it is the key step determining the final form of the equations. By identifying the slowly-varying variables, one effectively characterises the mean field potentials of relevance in the system (or vice versa), thus also fixing the form of the unperturbed Hamiltonian. The latter essentially fixes the "basis" in which the equations are formulated, i.e. whether one deals with bare harmonic oscillator states (as e.g. in \cite{walser_williams_1999,walser_cooper_01,proukakis_2001a}), dressed Hartree-Fock states (as most commonly the case~\cite{zaremba_nikuni_1999}), or even in quasi-particle basis (more challenging, but see also~\cite{imamovic_griffin_1999,imamovic_2001}). Clearly, all above are inter-related, and the importance is to be consistent within a particular treatment. In any finite temperature system, we expect to have non-negligible components of both the condensate and the thermal cloud of the system: this already defines the two slowly-varying quantities as $|\phi_j|^2$ and $\tilde{n}_j$. 

The important question is whether other quantities should also be considered as slowly-varying - in this context we should consider the following quantities appearing explicitly in Eq. \eqref{eom_exp}:
\begin{enumerate}[i]
\item Off-diagonal normal pair averages $\langle \hat{\delta}_j^\dagger \hat{\delta}_k \rangle$ ($j \neq k$);
\item Anomalous pair averages of the form $\langle \hat{\delta}_j \hat{\delta}_k \rangle$ (both for $j=k$ and $j \neq k$);
\item Anomalous triplet averages of the form $\langle \hat{\delta}^{\dagger}_k \hat{\delta}_j \hat{\delta}_j \rangle$.
\end{enumerate}
 Reflecting from our knowledge of the single-component case \cite{zaremba_nikuni_1999}, we note that, as pointed out in \cite{proukakis_burnett_1996}, the main condensate kinetics, i.e. its particle exchange with the thermal cloud should come through the latter term.
Pair anomalous averages could also be included into the treatment through additional self-consistently coupled equations of motion, as done, for example within the context of a bare particle basis formulation in \cite{walser_williams_1999,walser_cooper_01,proukakis_2001a}. Their role is discussed in Appendix \ref{app_a}, which shows why such terms can, to lowest order, be neglected due to violating energy conservation. More generally, their inclusion would describe many-body effects~\cite{proukakis_burnett_1998,rusch_burnett_1999}, which are however not expected to be significant in weakly-interacting atomic condensates. Based on this, we are thus justified in only including such terms in the perturbing Hamiltonian, or even dropping such terms altogether from our formalism (the so-called Popov approximation \cite{griffin_1996}).

This leaves us with the off-diagonal normal pair averages of the form $\langle \hat{\delta}_j^\dagger \hat{\delta}_k \rangle$. In general, these could be thought of as describing coherences between the two physical systems and could be treated on equal footing to condensate and excited state populations~\cite{mason_gardiner_2014,endo_nikuni_2011}. However, in the absence of any external coupling, one would expect such terms to evolve on the more rapid collisional timescale, and thus be suitable candidates for adiabatic elimination. The fact that they give rise to the highly intuitive, but never yet numerically characterized, "condensate exchange collisional process" confirms {\em a posteriori} that such treatment was indeed justified. 


Having made such an explicit identification, we can now proceed with  the perturbative treatment, or adiabatic elimination, of all rapidly-varying {\em off-diagonal normal} and {\em anomalous} averages.

\subsection{Identification of a Perturbing Hamiltonian}

To continue we should now explicitly partition the system Hamiltonian given by  Eq.~\eqref{ham1} as
\begin{equation}
\hat{H}=\hat{H}_{\text{MF}}+(\hat{H}-\hat{H}_{\text{MF}})=\hat{H}_{\text{MF}}+\hat{H}',
\end{equation}
where $\hat{H}'$ defines the perturbation, and $\hat{H}_{\text{MF}}$ is the quadratic mean-field (unperturbed) Hamiltonian containing only the identified slowly varying quantities (condensate mean field and diagonal non-condensate densities). 
To proceed, we consider the usual separation of the full quartic system Hamiltonian into terms identified by a label indicating the number of non-condensate operators appearing in each, i.e. from $H_0$ to $\hat{H}_4$ (see e.g. Refs.~\cite{morgan_2000,proukakis_jackson_2008}). This takes the form
\begin{equation}
\hat{H}=H_{0}+\hat{H}_{1}+\hat{H}_{2}+\hat{H}_{3}+\hat{H}_{4}
\end{equation}
where upon defining $\hat{h}_0=-(\hbar^2/2m_j)\nabla^{2}+V_j({\bf r})$ as the single-particle contribution from Eq. \eqref{ham1}, one obtains
\begin{widetext}
\begin{subequations}
\begin{align}
&H_{0}=\int d{\bf r}\bigg\{\sum_{j}\phi_{j}^{*}\bigg[\hat{h}_{0,j}+\frac{g_{jj}}{2}|\phi_j|^2\bigg]\phi_{j}\label{h0}+\sum_{k\neq j}g_{kj}|\phi_j|^2|\phi_k|^2\bigg\},
\\&\hat{H}_{1}=\int d{\bf r}\bigg\{\sum_{j}\bigg(\phi_{j}^{*}\bigg[\hat{h}_{0,j}+\frac{g_{jj}}{2}|\phi_{j}|^2\bigg]\hat{\delta}_{j}+\text{h.c.}\bigg)\label{h1}+\sum_{k\neq j}g_{kj}\bigg(\phi^{*}_{j}|\phi_{k}|^2\hat{\delta}_{j}+\phi_{k}|\phi_{j}|^2\hat{\delta}^{\dagger}_{k}+\text{h.c.}\bigg)\bigg\},
\\&\hat{H}_{2}=\int d{\bf r}\bigg\{\sum_{j}\bigg(\hat{\delta}^{\dagger}_{j}\bigg[\hat{h}_{0,j}{+}g_{jj}|\phi_{j}|^2\bigg]\hat{\delta}_{j}{+\frac{g_{jj}}{2}}\phi_{j}^{2}\hat{\delta}_{j}^{\dagger}\hat{\delta}_{j}^{\dagger}{+}\text{h.c.}\bigg)\label{h2}{+}\sum_{k\neq j}{g_{kj}}\bigg[|\phi_j|^2\hat{\delta}^{\dagger}_{k}\hat{\delta}_{k}+\phi^{*}_{j}\phi^{*}_{k}\hat{\delta}_{k}\hat{\delta}_{j}{+}\phi^{*}_{j}\phi_{k}
\hat{\delta}^{\dagger}_{k}\hat{\delta}_{j}{+}\text{h.c.}\bigg]\bigg\},
\\&\hat{H}_{3}=\int d{\bf r}\bigg\{\sum_{j}g_{jj}\bigg(\phi^{*}_{j}\hat{\delta}^{\dagger}_{j}\hat{\delta}_{j}\hat{\delta}_{j}+\text{h.c.}\bigg)\label{h3}+\sum_{k\neq j}g_{kj}\bigg(\phi^{*}_{j}\hat{\delta}^{\dagger}_{k}\hat{\delta}_{k}\hat{\delta}_{j}+\phi_{k}^{*}\hat{\delta}^{\dagger}_{j}\hat{\delta}_{j}\hat{\delta}_{k}+\text{h.c.}\bigg)\bigg\},
\\\label{h4}&\hat{H}_{4}=\int d{\bf r}\bigg\{\sum_{j}\frac{g_{jj}}{2}\hat{\delta}^{\dagger}_{j}\hat{\delta}^{\dagger}_{j}\hat{\delta}_{j}\hat{\delta}_{j}+\sum_{k\neq j}g_{kj}\hat{\delta}^{\dagger}_{j}\hat{\delta}^{\dagger}_{k}\hat{\delta}_{k}\hat{\delta}_{j}\bigg\}
\end{align}
\end{subequations}
\end{widetext}
We wish to work with a reduced unperturbed Hamiltonian which is (at most) quadratic, and so we perform conventional (but not exact) mean-field approximations \cite{morgan_2000,proukakis_2001a} to only include the leading part from the beyond-quadratic Hamiltonian into the unperturbed Hamiltonian.
Our perturbative treatment of the multi-component gas is motivated by Wick's theorem \cite{bruus_flensberg_2006}. We apply mean-field approximations to $\hat{H}_{3}$ and $\hat{H}_{4}$ above in order to reduce these terms to quadratic form. These are defined as
\begin{widetext}
\begin{subequations}
\begin{align}\label{wick1}
\hat{\delta}^{\dagger}_{k}\hat{\delta}_{k}\hat{\delta}_{j}&\simeq\langle\hat{\delta}^{\dagger}_{k}
\hat{\delta}_{k}\rangle\hat{\delta}_{j}+\langle\hat{\delta}^{\dagger}_{k}\hat{\delta}_{j}\rangle\hat{\delta}_{k}
+\langle\hat{\delta}_{k}\hat{\delta}_{j}\rangle\hat{\delta}^{\dagger}_{k},\\\nonumber
\hat{\delta}^{\dagger}_{j}\hat{\delta}^{\dagger}_{k}\hat{\delta}_{k}\hat{\delta}_{j}&{\simeq}
\langle\hat{\delta}^{\dagger}_{j}\hat{\delta}_{j}\rangle\hat{\delta}^{\dagger}_{k}\hat{\delta}_{k}
{+}\langle\hat{\delta}^{\dagger}_{k}\hat{\delta}_{k}\rangle\hat{\delta}^{\dagger}_{j}\hat{\delta}_{j}
{+}\langle\hat{\delta}^{\dagger}_{j}\hat{\delta}_{k}\rangle\hat{\delta}^{\dagger}_{k}\hat{\delta}_{j}
{+}\langle\hat{\delta}^{\dagger}_{k}\hat{\delta}_{j}\rangle\hat{\delta}^{\dagger}_{j}\hat{\delta}_{k}
{+}\langle\hat{\delta}^{\dagger}_{k}\hat{\delta}^{\dagger}_{j}\rangle\hat{\delta}_{j}\hat{\delta}_{k}
\\&{+}\langle\hat{\delta}_{k}\hat{\delta}_{j}\rangle\hat{\delta}_{j}^{\dagger}\hat{\delta}^{\dagger}_{k}
\label{wick2}{-}\bigg[\langle\hat{\delta}^{\dagger}_{j}\hat{\delta}_{j}\rangle\langle\hat{\delta}^{\dagger}_{k}\hat{\delta}_{k}\rangle
{+}\langle\hat{\delta}^{\dagger}_{j}\hat{\delta}_{k}\rangle\langle\hat{\delta}^{\dagger}_{k}\hat{\delta}_{j}\rangle
{+}\langle\hat{\delta}^{\dagger}_{j}\hat{\delta}^{\dagger}_{k}\rangle\langle\hat{\delta}_{k}\hat{\delta}_{j}\rangle\bigg],
\end{align}
\end{subequations}
\end{widetext}
valid both for $j{=}k$ and $j{\neq}k$. 
The reason we like to work with an approximate quadratic Hamiltonian, is because this is, at least in principle, diagonalizable by a Bogoliubov transformation to quasiparticle basis. In what follows, we do not however consider the dressing of particles to quasi-particles, but choose to work instead with dressed single-particle modes in the Hartree--Fock limit \cite{zaremba_nikuni_1999}.
Hence, our unperturbed mean-field Hamiltonian defining the energy basis of the system ultimately takes the form \cite{proukakis_2001a,proukakis_jackson_2008}
\begin{equation}\label{ham_mf}
\hat{H}_{\text{MF}}\approx(H_0+\delta H_0)+(\hat{H}_1+\delta\hat{H}_1)+(\hat{H}^{\text{diag}}_{2}+\delta\hat{H}^{\text{diag}}_{2}).
\end{equation}
Here, the shifts ($\propto\delta\hat{H}_{i}$) that appear in each bracket are found by applying the mean-field approximations of Eqs.~\eqref{wick1}-\eqref{wick2} to $\hat{H}_3$ and $\hat{H}_4$. The first term in each of the brackets in Eq. \eqref{ham_mf} describes a contribution from Eq. \eqref{hami} above with the subscript indicating the number of fluctuation operators appearing within the operator $\hat{H}_{i}$ (see Eq. \eqref{h0}-\eqref{h4}), while the second term $\delta\hat{H}_{i}$ arises from mean-field approximations, reducing products of three or more fluctuation operators to quadratic form. 
The `diag' superscript appearing in the final terms refer to diagonal contributions with equal component indices. 

The definition of the mean-field Hamiltonian (Eq. \eqref{ham_mf}) along with Eqs. \eqref{h0}-\eqref{h4} and \eqref{wick1}-\eqref{wick2} then allow us to write down the form of the perturbing Hamiltonian, a detailed account of which is given in Appendix \ref{app_b}. 

\subsection{Perturbative Description of Condensate and Thermal Clouds}

The chosen perturbing Hamiltonian $\hat{H}_{i}'(t)$, (see Appendix \ref{app_b}, Eqs. \eqref{ham_pet1}-\eqref{ham_pet4}) will allow us to construct our multi-component kinetic theory. It is straightforward to check that the definitions of $\hat{H}_{i}'(t)$ along with our choice of mean-field Hamiltonian of Eq. \eqref{ham_mf} recovers the Schr\"odinger equation given by Eq. \eqref{eom_exp},
\begin{equation}
i\hbar\frac{\partial\phi_j}{\partial t}=\langle[\hat{\Psi}_{j},\hat{H}_{\text{MF}}]\rangle+\langle[\hat{\Psi}_{j},\hat{H}'(t)]\rangle.\label{eom2}
\end{equation} 
Indeed, it can be seen that the first term on the right hand side of Eq. \eqref{eom2} generates the condensate potential, mean-field potentials and anomalous pair averages which go into the definition of $U_{c}^{j}$, while the second yields the two triplet terms, in agreement with Eq. \eqref{eom_exp}.

To describe the dynamical evolution of the non-condensed degrees of freedom, we define the multi-component single-particle Wigner operator as \cite{nikuni_williams_2003,endo_nikuni_2011}
\begin{equation}\label{wigner}
\hat{f}^{kj}\equiv\int d{\bf r}'e^{i{\bf p}\cdot{\bf r}'/\hbar}\hat{\delta}^{\dagger}_{j}({\bf r}+{\bf r}'/2,t)\hat{\delta}_{k}({\bf r}-{\bf r}'/2,t),
\end{equation}
where the corresponding phase-space distribution function is defined as $f^{kj}({\bf r},{\bf p},t)\equiv\Tr\tilde{\rho}(t,t_0)\hat{f}^{kj}({\bf r},{\bf p},t_0)$, and $\tilde{\rho}(t,t_0)$ defines the general density matrix of the system, which is related to the initial density matrix $\hat{\rho}(t_0)$ by the unitary transformation $\tilde{\rho}(t,t_0)=\hat{U}(t,t_0)\hat{\rho}(t_0)\hat{U}^{\dagger}(t,t_0)$. The unitary evolution operator satisfies the equation of motion
\begin{align}
  i\hbar\partder{t}\hat{U}(t,t_0)=\hat{H}(t)\hat{U}(t,t_0)
\end{align}
while the density matrix $\tilde{\rho}(t,t_0)$ evolves according to
\begin{equation}\label{dmatrix}
i\hbar\partder{t}\tilde{\rho}(t,t_0)=[\hat{H}(t),\tilde{\rho}(t,t_0)].
\end{equation}
In general the coherences of the non-condensate are non-zero only when an optical or magnetic coupling exists between the states $\ket{a}$ and $\ket{b}$. This particular case was explored in \cite{nikuni_williams_2003,endo_nikuni_2011} for spinor Bose gases. In those works it was assumed that the (matrix valued) non-condensate potential $\underline{\underline{U_{\text{n}}}}({\bf r},t)$ along with the optical coupling strength $\vec{\Omega}_{\text{n}}({\bf r},t)$ vary slowly in space, which leads to a qualitatively different expression for the kinetic equation. Within this approximation the off-diagonal terms in the Wigner operator $\hat{f}^{kj}$ are explicitly computed within the perturbing Hamiltonian, leading to matrix valued kinetic equations describing the non-condensate dynamics. For the two cases of optically coupled condensates with either spin-$\frac{1}{2}$ or spin-$1$ internal degrees of freedom, the relevant kinetic equations are given by Eqs. (52) and (41) in Ref. \cite{nikuni_williams_2003} and \cite{endo_nikuni_2011} respectively. We are however interested in understanding an incoherent binary mixture, hence for $j \neq k$ we set in the final calculations $f^{kj}({\bf r},{\bf p},t)=0$, i.e. no explicit long-lived coherences between off-diagonal normal pair averages. The Wigner operator directly allows us to calculate relevant non-equilibrium expectation values for the multi-component system. The corresponding equation of motion for the diagonal elements of the phase-space distribution function $f^{j}({\bf r},{\bf p},t)$ is written
\begin{align}\nonumber
\partder{t}f^{j}({\bf r},{\bf p},t)=&\frac{1}{i\hbar}\Tr\tilde{\rho}(t,t_0)[\hat{f}^{j}({\bf r},{\bf p},t_0),\hat{H}_{\text{MF}}(t)]\\&+\frac{1}{i\hbar}\Tr\tilde{\rho}(t,t_0)[\hat{f}^{j}({\bf r},{\bf p},t_0),\hat{H}'(t)].\label{eom_wig}
\end{align}
where the first term on the right hand side gives the free streaming terms and the second term generates the individual collisional integrals.

\section{Derivation of collisional integrals}
\subsection{Mathematical Formalism}
In order to calculate a closed set of equations describing the finite temperature dynamics of the bosonic mixture, the collision integrals appearing in the dissipative Schr\"odinger equation~\eqref{dschro1} and the quantum Boltzmann equation~\eqref{qbe} are derived using the perturbation Hamiltonian, $\hat{H}'$, defined by Eqs. \eqref{ham_pet1}-\eqref{ham_pet4} in Appendix \ref{app_b}. This is in turn accomplished by expanding the fluctuation operators in terms of their Fourier components, and calculating the non-equilibrium expectation values of the various products of such operators. The non-equilibrium average of an arbitrary time-dependent operator $\hat{O}(t)$ can be computed using the general density matrix $\tilde{\rho}(t,t_0)$ defined previously, as well as the mean-field evolution operator $\hat{S}(t,t_0)$ that satisfies the equation of motion,
\begin{align}
  i\hbar \partder{t}\hat{S}(t,t_0) = \hat{H}_{\text{MF}}(t)\hat{S}(t,t_0).\label{eq:unperturb_s0}
\end{align}
It can then be shown that the expectation value of the operator $\hat{O}(t)$ can be written as \cite{zaremba_nikuni_1999}
\begin{align}\nonumber
\langle\hat{O}_{t}\rangle=&\Tr\hat{\rho}_{t_0}\bigg\{\hat{S}^{\dagger}_{t,t_0}\hat{O}_{t_0}
\hat{S}_{t,t_0}\\&-\frac{i}{\hbar}\int\limits_{t_0}^{t}dt'\hat{S}^{\dagger}_{t',t_0}[\hat{S}^{\dagger}_{t,t'}\hat{O}_{t_0}\hat{S}_{t,t'},\hat{H}'_{t'}]\hat{S}_{t',t_0}\bigg\},\label{avg}
\end{align}
where $\hat{A}_{t_1,t_2}\equiv\hat{A}(t_1,t_2)$ has been used here and in what follows to abbreviate the time dependence of time evolution operators. We use Eq. \eqref{avg} to compute closed expressions for the source terms appearing in Eq. \eqref{dschro1}, along with the collision integrals appearing in Eq. \eqref{qbe} above. The first term on the right hand side of Eq. \eqref{avg} can be dropped, as it is assumed that for long times any initial correlations present in the system vanish (the Markov approximation). Thus, Eq. \eqref{avg} becomes
\begin{align}
&\langle\hat{O}_{t}\rangle\simeq-\frac{i}{\hbar}\int\limits_{t_0}^{t}dt'\langle\hat{S}^{\dagger}_{t',t_0}[\hat{S}^{\dagger}_{t,t'}\hat{O}_{t_0}\hat{S}_{t,t'},\hat{H}'_{t'}]\hat{S}_{t',t_0}\rangle,\label{avg_approx}
\end{align}
where $\langle\dots\rangle\equiv\Tr\tilde{\rho}_{t_0}(\dots)$ for the right-hand side. 
Since we have identified the condensate and non-condensate fields as slowly varying, we write $n_{c,j}({\bf r}',t')\simeq n_{c,j}({\bf r},t), \tilde{n}_{j}({\bf r}',t')\simeq\tilde{n}_{j}({\bf r},t)$ and $U^{j}_{\text{n}}({\bf r}',t')\simeq U^{j}_{\text{n}}({\bf r},t)$. It is useful to write the condensate wave function for component $j$ in the density-phase representation using the Madelung transformation $\phi_j=\sqrt{n_{c,j}}\exp({i\theta_j})$, in which case $\theta_{j}({\bf r}',t')$ can be expressed as 
\begin{align}
\theta_{j}({\bf r}',t')&\simeq\theta_{j}({\bf r},t)+\frac{\partial\theta_j}{\partial t}(t'-t)+\nabla\theta_j\cdot({\bf r}'-{\bf r}),\\&\simeq\theta_{j}({\bf r},t){-}\frac{\varepsilon_{c}^{j}({\bf r},t)}{\hbar}(t'-t){+}\frac{{\bf p}^{j}_{c}}{\hbar}\cdot({\bf r}'-{\bf r}).\label{theta_approx}
\end{align}
In writing Eq. \eqref{theta_approx} we have used the Euler equation for component $j$ (see Eq. \eqref{hydro_c2} in Sec. \ref{sec:2fhydro}) in order to introduce the local condensate energy,
\begin{equation}
\varepsilon_{c}^{j}({\bf r},t)=\mu_{c}^{j}({\bf r},t)+\frac{1}{2}m{v_{c}^{j}}^{2}. 
\end{equation}
Finally, the Fourier transform of $\hat{H}'(t)$ allows us to derive non-equilibrium expectation values for arbitrary products of operators in the following subsections. 
In order to calculate closed expressions for the collisional integrals appearing in Eq. \eqref{dschro1} and \eqref{qbe}, we must express the higher order correlation functions (those formed from non-equilibrium expectation values of products of fluctuation operators) in terms of the distribution functions $f^{j}({\bf r},{\bf p},t)$. As such, the perturbing Hamiltonian $\hat{H}'(t)$ is used to extract collision integrals to second order in the scattering length $a_{kj}$, while maintaining the effect of interactions in the collective mode energies and chemical potentials to first order in $a_{kj}$, in the spirit of the single-component ZNG approach \cite{nikuni_zaremba_1999}. 

The evaluation of non-equilibrium quantities requires the Fourier expansion of the non-condensate field operators, which for component $j$ is given by
\begin{equation}
\hat{\delta}_{j}({\bf r},t_0)=\frac{1}{\sqrt{V}}\sum_{\bf p}\hat{a}_{j,{\bf p}}e^{i{\bf p}\cdot{\bf r}/\hbar}.\label{delf}
\end{equation}
The expansion defined by Eq. \eqref{delf} allows us to write the Fourier transform of the Wigner operator defined by Eq. \eqref{wigner} above. This is best handled by switching to the center of mass and relative momenta for the two-component system. 
Hence, the general Wigner operator in momentum space for a binary mixture is written as
\begin{align}\nonumber
&\hat{f}^{kj}({\bf r},{\bf p},t_0)=e^{i2\frac{m_k-m_j}{M}{\bf r}\cdot{\bf p}/\hbar}\\&\times\sum_{\bf q}\hat{a}^{\dagger}_{j,{2(m_j/M){\bf p}-{\bf q}/2}}\hat{a}_{k,{2(m_k/M){\bf p}+{\bf q}/2}}e^{i{\bf r}\cdot{\bf q}/\hbar},\label{wig_bin}
\end{align}
and the total mass is $M=m_j+m_k$. Since we are only interested in the (incoherent) processes involving the diagonal elements of the Wigner operator, (see Refs. \cite{nikuni_williams_2003,endo_nikuni_2011} for generalizations that include the off-diagonal contributions to the Wigner operator.) we work in what follows with Eq. \eqref{wig_bin} in the limit $j=k$. Hence
\begin{equation}
\hat{f}^{j}({\bf r},{\bf p},t_0)=\sum_{\bf q}\hat{a}^{\dagger}_{j,{\bf p}-{\bf q}/2}\hat{a}_{j,{\bf p}+{\bf q}/2}e^{i{\bf r}\cdot{\bf q}/\hbar}.\label{wig_mom}
\end{equation}
Equation \eqref{wig_mom} will be used to calculate the collision integrals in the following sections.

\subsection{Source Terms from Anomalous Averages}

\subsubsection{Condensate Growth Terms $R^{jj}$ and $R^{kj}$ \\(from triplet anomalous correlations)}
We begin by considering the triplet contributions to Eq. \eqref{dschro1}, $R^{jj}$ and $R^{kj}$. We will explicitly compute $R^{kj}$, using Eq. \eqref{avg}. The triplet contributions can be decomposed as
\begin{equation}\label{trip_split}
\langle\hat{\delta}^{\dagger}_{k}\hat{\delta}_{k}\hat{\delta}_{j}\rangle=\langle\hat{\delta}^{\dagger}_{k}
\hat{\delta}_{k}\hat{\delta}_{j}\rangle_{\text{(1)}}+\langle\hat{\delta}^{\dagger}_{k}\hat{\delta}_{k}
\hat{\delta}_{j}\rangle_{\text{(3)}},
\end{equation} 
The two terms in Eq. \eqref{trip_split} above require the computation of averages from the perturbation Hamiltonian involving one [Eqs. \eqref{hpf_1j}-\eqref{hpf_1kj} for $j{=}k$] and those for three [Eqs. \eqref{hpf_3j}-\eqref{hpf_3kj} for $j{\neq}k$] fluctuation operators respectively. 
We first compute $\langle\hat{\delta}^{\dagger}_{j}\hat{\delta}_{j}\hat{\delta}_{k}\rangle_{(3)}$, i.e. those contributions arising exclusively from commutations involving the perturbing Hamiltonian $\hat{H}_{3,kj}'(t)$. As such we first calculate
\begin{align}\nonumber
&\langle\hat{\delta}^{\dagger}_{k}\hat{\delta}_{k}\hat{\delta}_{j}\rangle_{\text{(3)}}{=}\\&{-}\frac{i}{\hbar}\int\limits^{t}_{t_0}{dt'}\langle\hat{S}^{\dagger}_{t',t_0}[\hat{S}^{\dagger}_{t,t'}\hat{\delta}^{\dagger}_{k}\hat{\delta}_{k}\hat{\delta}_{j}\hat{S}_{t,t'},\hat{H}_{3,kj}']\hat{S}_{t',t_0}\rangle.\label{trip1}
\end{align}
After using the definition of the Fourier transform of $\hat{H}_{3,kj}(t)$ defined as
\begin{widetext}
\begin{equation}\label{eqn:hf3kj}
\hat{H}_{3,kj}'(t)=\frac{1}{\sqrt{V}}\sum_{k\neq j}\sum_{{\bf p}_2,{\bf p}_3,{\bf p}_4}g_{kj}\sqrt{n_{c,j}}\bigg\{\delta_{{\bf p}^{j}_{c}+{\bf p}_{2},{\bf p}_{3}+{\bf p}_{4}}e^{-i(\theta_j-\varepsilon^{j}_{c}(t'-t)/\hbar-{\bf p}^{j}_{c}\cdot{\bf r}/\hbar)}\hat{a}^{\dagger}_{k,{\bf p}_{2}}\hat{a}_{k,{\bf p}_{3}}\hat{a}_{j,{\bf p}_{4}}+\text{h.c.}\bigg\},
\end{equation}
\end{widetext}
Eq. \eqref{trip1} becomes
\begin{align}\nonumber
&\langle\hat{\delta}^{\dagger}_{k}\hat{\delta}_{k}\hat{\delta}_{j}\rangle_{(3)}{=}{-}\frac{i\pi}{V^2}g_{kj}\phi_{j}\sum_{{\bf p}_2,{\bf p}_3,{\bf p}_4}\delta(\varepsilon^{j}_{c}+\varepsilon^{k}_{p_2}-\varepsilon^{k}_{p_3}-\varepsilon^{j}_{p_4})\\&{\times}{\delta_{{\bf p}^{j}_{c}+{\bf p}_2,{\bf p}_3+{\bf p}_4}}[f^{k}_{2}(f^{k}_{3}+1)(f^{j}_{4}+1){-}(f^{k}_{2}+1)f^{k}_{3}f^{j}_{4}]\label{trip_d},
\end{align}
where the shorthand $f^{k}_{\nu}\equiv f^{k}({\bf r},{\bf p}_{\nu},t)$ has been used in the above and what follows, $\varepsilon^{j}_{c}=\mu^{j}_{c}+\frac{1}{2}m_{j}v_{c,j}^{2}$ and $\varepsilon_{p}^{j}=p^{2}/2m_{j}+U_{n}^{j}$ define the non-equilibrium condensate and thermal energy for atoms in component $j$ respectively. By writing Eq. \eqref{trip_d} we let $t_0\rightarrow\infty$ in order to evaluate the integral over $t'$ in Eq. \eqref{trip1} (See Eq. \eqref{eqn:cpv} and discussion in Appendix \ref{appendix:avgs} for an explanation of this important step). Calculation of $\langle\hat{\delta}^{\dagger}_{k}\hat{\delta}_{k}\hat{\delta}_{j}\rangle_{(1)}$ allows us to write the first term in Eq.~\eqref{trip1}. Then by taking the continuum limit, we obtain the source terms
\begin{widetext}
\begin{align}
&R^{kj}{=}\frac{g_{kj}^{2}}{2(2\pi)^5\hbar^6}\int d{\bf p}_{2}\int d{\bf p}_{3}\int d{\bf p}_{4}\ \delta({\bf p}^{j}_{c}{+}{\bf p}_{2}{-}{\bf p}_{3}{-}{\bf p}_{4})\delta(\varepsilon^{j}_{c}{+}\varepsilon^{k}_{p_2}{-}\varepsilon^{k}_{p_3}{-}\varepsilon^{j}_{p_4})\bigg[f^{k}_{2}(f^{k}_{3}{+}1)(f^{j}_{4}{+}1){-}(f^{k}_{2}{+}1)f^{k}_{3}f^{j}_{4}\bigg]\label{trip_anom},\\
&R^{jj}=\frac{g_{jj}^{2}}{(2\pi)^5\hbar^6}\int d{\bf p}_{2}\int d{\bf p}_{3}\int d{\bf p}_{4}\ \delta({\bf p}^{j}_{c}{+}{\bf p}_{2}{-}{\bf p}_{3}{-}{\bf p}_{4})\delta(\varepsilon^{j}_{c}{+}\varepsilon^{j}_{p_2}{-}\varepsilon^{j}_{p_3}{-}\varepsilon^{j}_{p_4})\bigg[f^{j}_{2}(f^{j}_{3}{+}1)(f^{j}_{4}{+}1){-}(f^{j}_{2}{+}1)f^{j}_{3}f^{j}_{4}\bigg].\label{trip_anomjj}
\end{align}
\end{widetext}
Here, Eq. \eqref{trip_anomjj} has been obtained by repeating the steps following Eq. \eqref{trip_split} for $R^{kj}$.

\subsubsection{Condensate Exchange Terms $\mathds{R}^{kj}$ \\(from off-diagonal normal pair averages)}

The final dissipative source term $\mathds{R}^{kj}$ appearing in Eq. \eqref{dschro1} is comprised of a normal average of an off-diagonal pair of fluctuation operators, $\langle\hat{\delta}^{\dagger}_{k}\hat{\delta}_{j}\rangle$. 
Hence we calculate
\begin{equation}
\langle\hat{\delta}^{\dagger}_{k}\hat{\delta}_{j}\rangle{=}{-}\frac{i}{\hbar}\int\limits^{t}_{t_0} dt'\langle\hat{S}^{\dagger}_{t',t_0}[\hat{S}^{\dagger}_{t,t'}\hat{\delta}^{\dagger}_{k}\hat{\delta}_{j}\hat{S}_{t,t'},\hat{H}_{2,kj}']\hat{S}_{t',t_0}\rangle\label{pair1}.
\end{equation}
Computation of Eq. \eqref{pair1} requires the Fourier transform of $\hat{H}_{2,kj}'(t)$, the relevant contribution being
\begin{widetext}
\begin{align}\label{eqn:hf2kj}
\hat{H}_{2,kj}'(t)=&\sum_{k\neq j}\sum_{{\bf p}_1,{\bf p}_2}g_{kj}\sqrt{n_{c,j}n_{c,k}}\bigg\{\delta_{{\bf p}_1+{\bf p}^{j}_{c},{\bf p}_2+{\bf p}^{k}_{c}}e^{-i((\theta_j-\theta_k)-(\varepsilon^{j}_{c}-\varepsilon^{k}_{c})(t'-t)/\hbar-({\bf p}^{c}_{j}-{\bf p}^{k}_{c})\cdot{\bf r}/\hbar)}\hat{a}^{\dagger}_{k,{\bf p}_1}\hat{a}_{j,{\bf p}_2}+\text{h.c.}\bigg\},
\end{align}
\end{widetext}
Using Eqs.\eqref{pair1} and \eqref{eqn:hf2kj} yields the expression
\begin{align}\nonumber
&\langle\hat{\delta}^{\dagger}_{k}\hat{\delta}_{j}\rangle=-\frac{i\pi}{V}g_{kj}\phi_{j}\phi^{*}_{k}\sum_{{\bf p}_1,{\bf p}_2}\delta(\varepsilon^{k}_{c}+\varepsilon^{j}_{p_1}-\varepsilon^{j}_{c}-\varepsilon^{k}_{p_2})\\&\times\delta_{{\bf p}^{k}_{c}+{\bf p}_1,{\bf p}^{j}_{c}+{\bf p}_2}[(f^{j}_{1}+1)f^{k}_{2}-f^{j}_{1}(f^{k}_{2}+1)].\label{pair_d}
\end{align}
By taking the continuum limit of Eq. \eqref{pair_d}, the off-diagonal pair average becomes
\begin{widetext}
\begin{equation}
\mathds{R}^{kj}=\frac{g_{kj}^{2}}{2(2\pi)^2\hbar^3}n_{c,k}\int d{\bf p}_{1}\int d{\bf p}_{2}\ \delta({\bf p}^{k}_{c}{+}{\bf p}_{1}{-}{\bf p}^{j}_{c}{-}{\bf p}_{2})\delta(\varepsilon^{k}_{c}+\varepsilon^{j}_{p_1}-\varepsilon^{j}_{c}-\varepsilon^{k}_{p_2})\bigg[(f^{j}_{1}+1)f^{k}_{2}-f^{j}_{1}(f^{k}_{2}+1)\bigg].\label{pair_anom}
\end{equation}
\end{widetext}

The three expressions derived in this section, Eqs. \eqref{trip_anom}, \eqref{trip_anomjj} and \eqref{pair_anom} are the important source terms that appear in the dissipative Schr\"odinger equation. Equation \eqref{pair_anom} arises due to our explicit separation of slowly and rapidly varying quantities in the system Hamiltonian, and can be understood as a collisional energy exchange process between the two condensates, whereby a condensate and thermal atom in differing components scatter into corresponding thermal and condensed states respectively. 

\subsection{Quantum Boltzmann contributions}
\subsubsection{\label{sec:c12}Collisional $C^{jj}_{12}$ and $C^{kj}_{12}$ terms}
To complete the derivation, we require the collision integrals appearing on the right hand side of Eq. \eqref{qbe}. These are computed using the definition of the multi-component single-particle Wigner operator (Eq. \eqref{wigner}), along with the Fourier transform of $\hat{H}_{3,kj}'(t)$, as defined by Eq. \eqref{eqn:hf3kj}.
The multi-component nature of the problem leads us to partition the ``$C_{12}$'' collision integral into two parts, the first $C^{jj}_{12}$ defines the intra-component scattering of atoms, while the second $C^{kj}_{12}$ gives the inter-component collision rate. We wish to calculate both
\begin{equation}
C^{jj}_{12}=\frac{1}{i\hbar}\Tr\tilde{\rho}(t,t_0)[\hat{f}^{j}({\bf r},{\bf p},t_0),\hat{H}_{3,j}'(t)],\label{c12_jj}
\end{equation}
and
\begin{equation}
C^{kj}_{12}=\frac{1}{i\hbar}\Tr\tilde{\rho}(t,t_0)[\hat{f}^{j}({\bf r},{\bf p},t_0),\hat{H}_{3,kj}'(t)].\label{c12_def}
\end{equation}
As before, let us illustrate the derivation of these two terms by computing the off-diagonal contribution, Eq.~\eqref{c12_def}. This is accomplished by using Eqs.\eqref{eqn:hf3kj} and \eqref{c12_def}, giving  
\begin{widetext}
\begin{align}\nonumber
&C^{kj}_{12}=\frac{g_{kj}}{i\hbar\sqrt{V}}\sum_{\bf q}e^{i{\bf q}\cdot{\bf r}/\hbar}\sum_{{\bf p}_2,{\bf p}_3,{\bf p}_4}\bigg\{\delta_{{\bf p}^{k}_{c}+{\bf p}_2,{\bf p}_3+{\bf p}_4}\phi^{*}_{k}\bigg(\delta_{{\bf p}_2,{\bf p}+{\bf q}/2}\langle\hat{a}^{\dagger}_{j,{\bf p}-{\bf q}/2}\hat{a}_{j,{\bf p}_3}\hat{a}_{k,{\bf p}_4}\rangle-\delta_{{\bf p}_3,{\bf p}-{\bf q}/2}\\&\times\langle\hat{a}^{\dagger}_{j,{\bf p}_2}\hat{a}_{j,{\bf p}+{\bf q}/2}\hat{a}_{k,{\bf p}_4}\rangle\bigg)-\delta_{{\bf p}^{j}_{c}+{\bf p}_2,{\bf p}_3+{\bf p}_4}\phi^{*}_{j}\delta_{{\bf p}_4,{\bf p}-{\bf q}/2}\langle\hat{a}^{\dagger}_{k,{\bf p}_2}\hat{a}_{k,{\bf p}_3}\hat{a}_{j,{\bf p}+{\bf q}/2}\rangle-\text{h.c.}\bigg\}.\label{c12_calc1}
\end{align}
\end{widetext}
Then, by using the definition of the multi-component three-field correlation function, the continuum limit can be obtained as before by replacing the summations with integrations, giving
\begin{widetext}
\begin{align}\nonumber
C^{kj}_{12}=&\frac{g_{kj}^{2}}{(2\pi)^2\hbar^4}n_{c,k}\int d{\bf p}_{2}\int d{\bf p}_{3}\int d{\bf p}_{4}\delta({\bf p}_{c}^{k}+{\bf p}_{2}-{\bf p}_{3}-{\bf p}_{4})\delta(\varepsilon^{k}_{c}+\varepsilon^{j}_{p_2}-\varepsilon^{j}_{p_3}-\varepsilon^{k}_{p_4})\\\nonumber
&\times\bigg[(f^{j}_{2}+1)f^{j}_{3}f^{k}_{4}-f^{j}_{2}(f^{j}_{3}+1)(f^{k}_{4}+1)\bigg]\bigg[\delta({\bf p}-{\bf p}_{2})-\delta({\bf p}-{\bf p}_{3})\bigg]\\\nonumber-&\frac{g_{kj}^{2}}{(2\pi)^2\hbar^4}n_{c,j}\int d{\bf p}_{2}\int d{\bf p}_{3}\int d{\bf p}_{4}\delta({\bf p}_{c}^{j}+{\bf p}_{2}-{\bf p}_{3}-{\bf p}_{4})\delta(\varepsilon^{j}_{c}+\varepsilon^{k}_{p_2}-\varepsilon^{k}_{p_3}-\varepsilon^{j}_{p_4})
\\&\times\bigg[(f^{k}_{2}+1)f^{k}_{3}f^{j}_{4}-f^{k}_{2}(f^{k}_{3}+1)(f^{j}_{4}+1)\bigg]\delta({\bf p}-{\bf p}_4),\label{c12jk}\\\nonumber C^{jj}_{12}=&\frac{2g_{jj}^{2}}{(2\pi)^2\hbar^4}n_{c,j} \int d{\bf p}_{2}\int d{\bf p}_{3}\int d{\bf p}_{4}\delta({\bf p}_{c}^{j}+{\bf p}_{2}-{\bf p}_{3}-{\bf p}_{4})\delta(\varepsilon^{j}_{c}+\varepsilon^{j}_{p_2}-\varepsilon^{j}_{p_3}-\varepsilon^{j}_{p_4})\\&{\times}\bigg[(f^{j}_{2}+1)f^{j}_{3}f^{j}_{4}-f^{j}_{2}(f^{j}_{3}+1)(f^{j}_{4}+1)\bigg]\bigg[\delta({\bf p}-{\bf p}_{2})-\delta({\bf p}-{\bf p}_3)-\delta({\bf p}-{\bf p}_4)\bigg],\label{c12jj}
\end{align}
\end{widetext}
with Eq. \eqref{c12jj} obtained by repeating the same steps for the collisional integral defined by Eq. \eqref{c12_jj}. 
It can be seen that Eq. \eqref{c12jj} is equivalent to the $C_{12}$ collision integral from the single component kinetic theory \cite{zaremba_nikuni_1999,griffin_nikuni_2009}.
\subsubsection{\label{sec:c12e}Exchange collisional term $\mathds{C}^{kj}_{12}$}
To complete our discussion of collisions involving condensate and non-condensate particles, we compute the exchange collisional integral, which is defined by
\begin{align}
\mathds{C}^{kj}_{12}=\frac{1}{i\hbar}\Tr\tilde{\rho}(t,t_0)[\hat{f}^{j}({\bf r},{\bf p},t_0),\hat{H}_{2,kj}'(t)].
\end{align}
Following the methodology discussed above, the Fourier transformed Wigner operator along with Eq. \eqref{eqn:hf2kj} allows us to compute an expression for $\mathds{C}^{kj}_{12}$ given by
\begin{align}\nonumber
\mathds{C}^{kj}_{12}=&\frac{2g_{kj}}{i\hbar}\sum_{{\bf p}_1,{\bf p}_2}\delta_{{\bf p}_{c}^{j}+{\bf p}_{1},{\bf p}^{k}_{c}+{\bf p}_{2}}\bigg(\phi_{j}\phi_{k}^{*}\langle\hat{a}^{\dagger}_{j,{\bf p}_2}\hat{a}_{k,{\bf p}_1}\rangle\\&-\phi_{j}^{*}\phi_{k}\langle\hat{a}^{\dagger}_{k,{\bf p}_1}\hat{a}_{j,{\bf p}_2}\rangle\bigg).\label{cC12kj1}
\end{align}
upon inserting the pair correlation function into Eq.\eqref{cC12kj1} and taking the continuum limit yields the exchange integral $\mathds{C}^{kj}_{12}$, given by
\begin{widetext}
\begin{equation}
\mathds{C}^{kj}_{12}=\frac{2\pi g_{kj}^{2}}{\hbar}\ n_{c,k}\,n_{c,j}\,\int d{\bf p}_{1}\int d{\bf p}_{2}\delta({\bf p}_{c}^{j}+{\bf p}_{1}-{\bf p}_{c}^{k}-{\bf p}_{2})\delta(\varepsilon^{j}_{c}+\varepsilon^{k}_{p_1}-\varepsilon^{k}_{c}-\varepsilon^{j}_{p_2}
)\bigg[f^{k}_{1}(f^{j}_{2}+1)-(f^{k}_{1}+1)f^{j}_{2}\bigg]\delta({\bf p}-{\bf p}_2).
\label{c12jk2}
\end{equation}
\end{widetext}
\subsubsection{\label{sec:c22}Collisional $C^{kj}_{22}$ and $C^{jj}_{22}$ terms}
The final collisional processes described by Eq. \eqref{qbe} are the terms that account for interactions exclusively between non-condensate atoms, $C^{kj}_{22}$ and $C^{jj}_{22}$. The quantities we wish to evaluate are given by
\begin{equation}
C^{jj}_{22}=\frac{1}{i\hbar}\Tr\tilde{\rho}(t,t_0)[\hat{f}^{j}({\bf r},{\bf p},t_0),\hat{H}_{4,j}'(t)],\label{c22_j}
\end{equation}
and
\begin{equation}
C^{kj}_{22}=\frac{1}{i\hbar}\Tr\tilde{\rho}(t,t_0)[\hat{f}^{j}({\bf r},{\bf p},t_0),\hat{H}_{4,kj}'(t)]\label{c22_kj}.
\end{equation}
As with the $C^{jj}_{12}$ and $C_{12}^{kj}$ collision integrals, the commutation of the Wigner operator $\hat{f}^{j}({\bf p},{\bf r},t)$ with $\hat{H}_{4,j}'(t)$ and $\hat{H}_{4,kj}'(t)$ generates the {\it intra} ($C^{jj}_{22}$) and {\it inter} ($C^{kj}_{22}$) collision integrals respectively. We illustrate the derivation by considering the collisional integral $C^{kj}_{22}$. This requires the Fourier transform of $\hat{H}_{4,kj}'(t)$, which is given by
\begin{widetext}
\begin{align}\nonumber
\hat{H}_{4,kj}'(t)=&\frac{1}{V}\sum_{k\neq j}g_{kj}\bigg\{\sum_{{\bf p}_1,{\bf p}_2,{\bf p}_3,{\bf p}_4}\delta_{{\bf p}_1+{\bf p}_2,{\bf p}_3+{\bf p}_4}\hat{a}^{\dagger}_{j,{\bf p}_1}\hat{a}^{\dagger}_{k,{\bf p}_2}\hat{a}_{k,{\bf p}_3}\hat{a}_{j,{\bf p}_4}-\tilde{n}_{jk}\sum_{{\bf p}_1,{\bf p}_2}\delta_{{\bf p}_1,{\bf p}_2}\hat{a}^{\dagger}_{j,{\bf p}_1}\hat{a}_{k,{\bf p}_2}\\&-\tilde{n}_{kj}\sum_{{\bf p}_1,{\bf p}_2}\delta_{{\bf p}_1,{\bf p}_2}\hat{a}^{\dagger}_{k,{\bf p}_1}\hat{a}_{j,{\bf p}_2}\bigg\}.\label{eqn:hf4kj}
\end{align}
\end{widetext}
By inserting Eq.\eqref{eqn:hf4kj} into Eq.\eqref{c22_kj} and taking the continuum limit, the collisional integral $C^{kj}_{22}$ is found to be
\begin{widetext}
\begin{align}
C^{kj}_{22}{=}&{\frac{g_{kj}^2}{(2\pi)^5\hbar^7}}{\int{d{\bf p}_2}}{\int d{\bf p}_3}{\int d{\bf p}_4}\delta({\bf p}{+}{\bf p}_2{-}{\bf p}_3{-}{\bf p}_4){\delta(\varepsilon^{j}_{p}{+}\varepsilon^{k}_{p_2}{-}\varepsilon^{k}_{p_3}{-}\varepsilon^{j}_{p_4})}\bigg[{(f^{j}{+}1)(f^{k}_{2}{+}1)f^{k}_{3}f^{j}_{4}{-}f^{j}f^{k}_{2}(f^{k}_{3}{+}1)(f^{j}_{4}{+}1)}\bigg],\label{c22jk}\\
C^{jj}_{22}{=}&{\frac{2g_{jj}^2}{(2\pi)^5\hbar^7}}{\int d{\bf p}_2}{\int d{\bf p}_3}{\int d{\bf p}_4}\delta({\bf p}{+}{\bf p}_2{-}{\bf p}_3{-}{\bf p}_4)\delta(\varepsilon^{j}_{p}{+}\varepsilon^{j}_{p_2}{-}\varepsilon^{j}_{p_3}{-}\varepsilon^{j}_{p_4})\bigg[(f^{j}{+}1)(f^{j}_{2}{+}1)f^{j}_{3}f^{j}_{4}{-}f^{j}f^{j}_{2}(f^{j}_{3}{+}1)(f^{j}_{4}{+}1)\bigg].
\label{c22jj}
\end{align}
\end{widetext}
Equation \eqref{c22jj} is obtained by repeating the above steps for the collisional integral defined by Eq. \eqref{c22_j}. This formally completes the derivation of all collisional integrals appearing on the right hand side of Eq. \eqref{qbe}. The expressions given by Eqs. \eqref{c12jk}, \eqref{c12jj}, \eqref{c12jk2}, \eqref{c22jk} and \eqref{c22jj} will be used in the subsequent sections to study the equilibrium properties of binary condensates.

\section{Numerical Simulations} 
At finite temperatures, the various collisional processes have a heavy influence on the coupled dynamics between the condensates and the non-condensed atoms, such as the damping of collective modes~\cite{jackson_zaremba_2001,jackson_zaremba_2002a,jackson_zaremba_2002}, the decay of solitons~\cite{jackson_proukakis_2007} and vortices~\cite{allen_zaremba_2013}, as well as the growth of the condensate~\cite{bijlsma_zaremba_2000,markle_allen_14}, as seen in the corresponding single-component kinetic theory. In this section, after obtaining our equilibrium distribution (Sec.~\ref{sec:eq_sol}), we compare the roles of different collisional processes under the variation of isotropic trap frequencies (Sec. \ref{sec:varytrap}) and trap geometries (Sec. \ref{sec:trap_geometry}). This is achieved by calculating the collisional rates (Sec.~\ref{sec:collrate}) and hydrodynamic parameters (Sec.~\ref{sec:hydrodynamic}) for various equilibrium binary systems. We show and explain the scaling relations between the hydrodynamic parameters and isotropic trap frequency in Sec.~\ref{sec:varytrap}. In Sec.~\ref{sec:trap_geometry}, we demonstrate the generic dominance of the exchange collisional process $\mathds{C}_{12}$ across the different trap geometries, even though all collision rates strongly depend on the relevant scattering lengths. Importantly, our results in Sec.~\ref{sec:varytrap} and~\ref{sec:trap_geometry} illustrate different ways to control the hydrodynamicity of the collisional processes, which can be of high interest to experiments. These include 
\begin{enumerate}[(i)]
\item bringing the hydrodynamic parameters of the various processes closer in magnitude by increasing the trap frequency and the temperature,
\item increasing the hydrodynamicity of all processes towards the hydrodynamic regime by changing trap geometry,
\item controlling the hydrodynamicity of the intraspecies and interspecies collisional processes by tuning the relevant scattering lengths through inter- or intra-species Feshbach resonances.
\end{enumerate}
In the final subsection~\ref{subsect:temp_g12}, we briefly explore the validity of the usual high-temperature approximation ($\beta(\epsilon-\mu)\ll1$) in the context  of collisional rates, specifically for the exchange collision $\mathds{C}_{12}$.

Our numerical analysis focuses on experimentally relevant equilibrium $^{87}$Rb-$^{41}$K and $^{87}$Rb-$^{85}$Rb mixtures with a total atom number $N_j = 10^5$ in each component trapped in harmonic potentials
\begin{align}
  V_j(\vec{r}) = \frac{m_j}{2}\left(\omega_\perp^2(x^2+y^2)+\omega_z^2 z^2\right).
\end{align}
These mixtures were chosen as their tunable scattering lengths ($a_{\rm Rb87}=99a_0$, $a_{\rm K41}=60a_0$, $a_{\rm Rb87-K41}=20a_0$ or $163a_0$~\cite{modugno_modugno_2002,thalhammer_barontini_2008}; $a_{\rm Rb87-Rb85}=213a_0$, $a_{\rm Rb85}=900a_0$ or $51a_0$~\cite{papp_pino_2008}) enable the probing of both miscible ($ \Lambda= g_{12}/\sqrt{g_{11}g_{22}} < 1$) and immiscible ($ \Lambda>1$) regimes. In a previous work~\cite{edmonds_lee_2015}, we have presented our numerical results for such systems in an isotropic harmonic trap (frequency $\omega=\omega_\perp=\omega_z=2\pi\times20$Hz). In particular, we have highlighted the dominance of the exchange collisions $\mathds{C}_{12}$ over the other collisional processes within the temperature range $0.3<T/T_c<0.9$. Here, we perform a more detailed analysis that compares rates for different isotropic trap frequencies and different trap geometries, using our results for $\omega=2\pi\times20$Hz as a reference.

\subsection{\label{sec:eq_sol}Equilibrium solutions and condensate fractions}
The equilibrium density distributions at temperature $T$ are numerically obtained by setting the source terms ($R^{jj}, R^{kj},\mathds{R}^{kj}$) and the collision integrals ($C_{12},\mathds{C}_{12},C_{22}$) to zero and self-consistently solving Eqs.~\eqref{dschro1} and \eqref{qbe}. In order to speed up the computation, we adopt the semi-classical approximation~\cite{holzmann_krauth_1999} for the local non-condensate density
\begin{align}
  \tilde{n}_j(\vec{r},t) = \int \frac{d\vec{p}}{(2\pi\hbar)^3} f^j(\vec{p},\vec{r},t) = \frac{1}{\lambda_j^3}g_{3/2}(z_j),
\end{align}
where $\lambda_j=\sqrt{2\pi\hbar^2/(m_j k_BT)}$ is the thermal de Broglie wavelength, $z_j(\vec{r})=\exp\{[\mu^j_c-U^j_{\text{n}}(\vec{r})]/(k_B T)\}$ is the local fugacity and the chemical potential $\mu^j_c$ is obtained from the imaginary-time evolution of the condensate equation~\eqref{dschro1},
\begin{align}
\bigg[-\frac{\hbar^2}{2m_j}\nabla^2+U^{j}_{c}\bigg]\phi_j=\mu^j_c\phi_j.\label{chempot}
\end{align}

\begin{figure*}
  \includegraphics[width=0.7\textwidth]{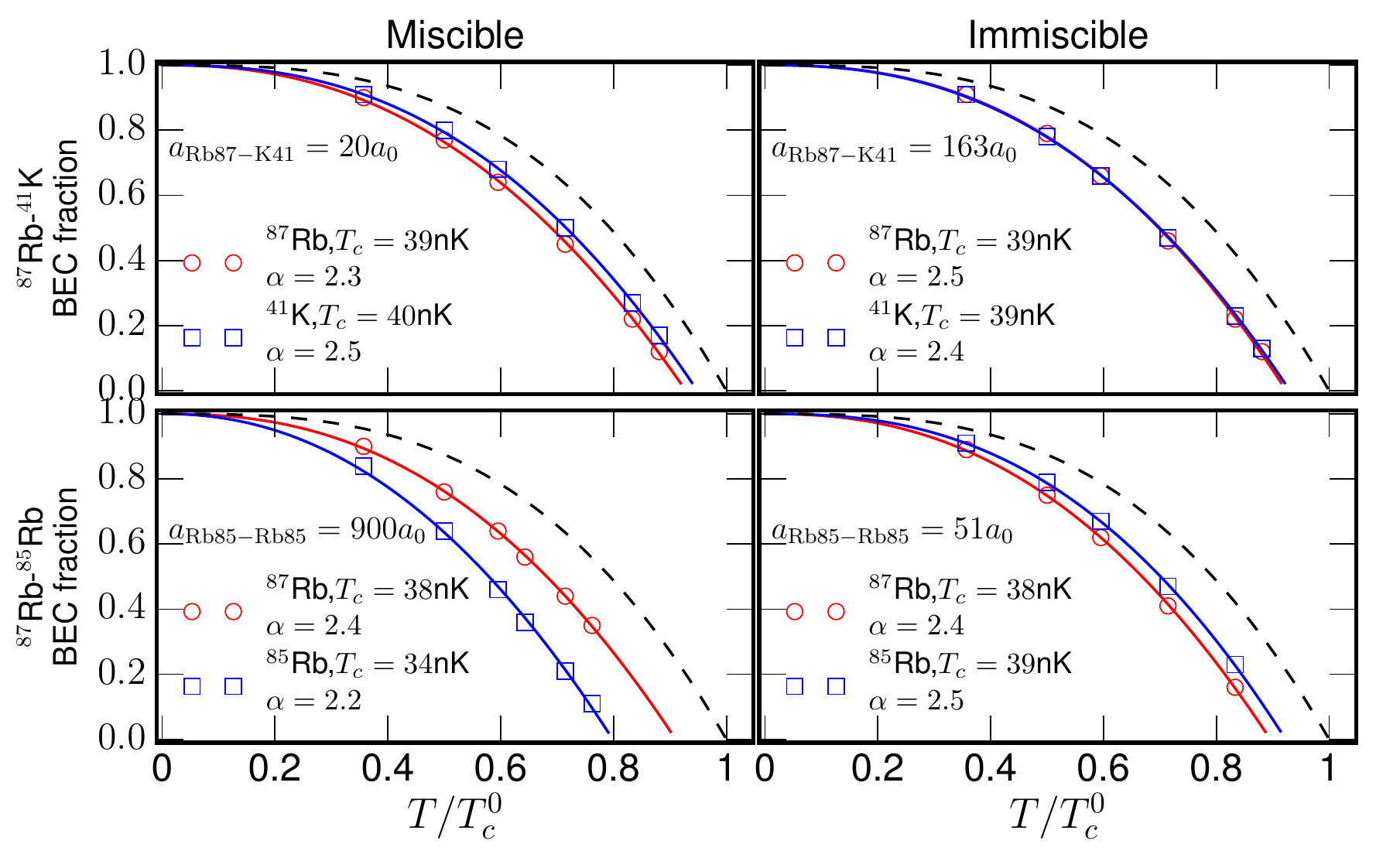}\caption{\label{con_frac}(Color online) Condensate fractions of $^{87}$Rb-$^{41}$K (top) and $^{87}$Rb-$^{85}$Rb (bottom) mixtures at different temperatures in an isotropic harmonic trap (trap frequency $\omega=2\pi\times20$Hz) with scattering lengths $a_{\rm Rb87-Rb87} = 99a_0$ , $a_{\rm K41-K41} = 60a_0$ , $a_{\rm Rb87-K41} = 20a_0$ (miscible) or $163a_0$(immiscible~\cite{modugno_modugno_2002,thalhammer_barontini_2008}, $a_{\rm Rb87-Rb85} = 213a_0$, and $a_{\rm Rb85-Rb85} = 900a_0$ (miscible) or $51a_0$(immiscible)~\cite{papp_pino_2008}; each species has a total of $N = 10^5$ atoms. Dashed lines give the prediction for non-interacting single-component trapped gas, with condensate fraction $N_c/N=1-(T/T^0_c)^3$, with critical temperature $T^0_c=42$nK. The solid lines is our numerical fit using the condensate fraction $N_c/N=1-(T/T_c)^\alpha$, where the extracted $T_c$ (indicated in the legend) are lower than the mean-field single-component $T_c$~\cite{dalfovo_giorgini_1999} by at most $5\%$. }
\end{figure*}

We start our analysis by first considering the condensate fractions of the binary mixture at different temperatures $T$, as shown in Fig.~\ref{con_frac}. While our method is strictly not valid for $T$ close to the critical temperature $T_c$ due to critical fluctuations, we can nevertheless extract $T_c$ by fitting the fractions with $N_c/N=1-(T/T_c)^\alpha$~\cite{hutchinson_zaremba_1997} and compare the extracted $T_c$ to the expected shift in $T_c$ due to finite-size corrections~\cite{grossmann_holthaus_1995,*ketterle_vandruten_1996,*kirsten_toms_1996} and mean-field corrections~\cite{giorgini_pitaevskii_1996}. 

For a single-component Bose gas and using our simulation parameters, $T_c$ decreases by approximately $0.73 N_j^{-1/3}=$2\% due to the finite number of atoms. The mean-field shift, calculated by $-1.3 (a/a_{\rm ho})N_j^{1/6}$, where $a$ is the relevant scattering length and $a_{\rm ho}$ is the relevant harmonic length, further decreases our $T_c$ by 1-2\%. However, for the $T_c$ of $^{85}{\rm Rb}$ in the miscible mixture, the mean-field shift amounts to approximately $17\%$ due to the large scattering length $a_{\rm Rb85-Rb85}=900a_0$. Note that we did not take into account many-body effects beyond mean-field theory~\cite{gruter_ceperley_1997,*baym_blaizot_1999}, which can instead increase the critical temperature. Overall, our extracted $T_c$ are close to the mean-field predictions for a single-component gas~\cite{dalfovo_giorgini_1999}. 

\begin{figure*}[!htbp]
  \includegraphics[width=0.7\textwidth]{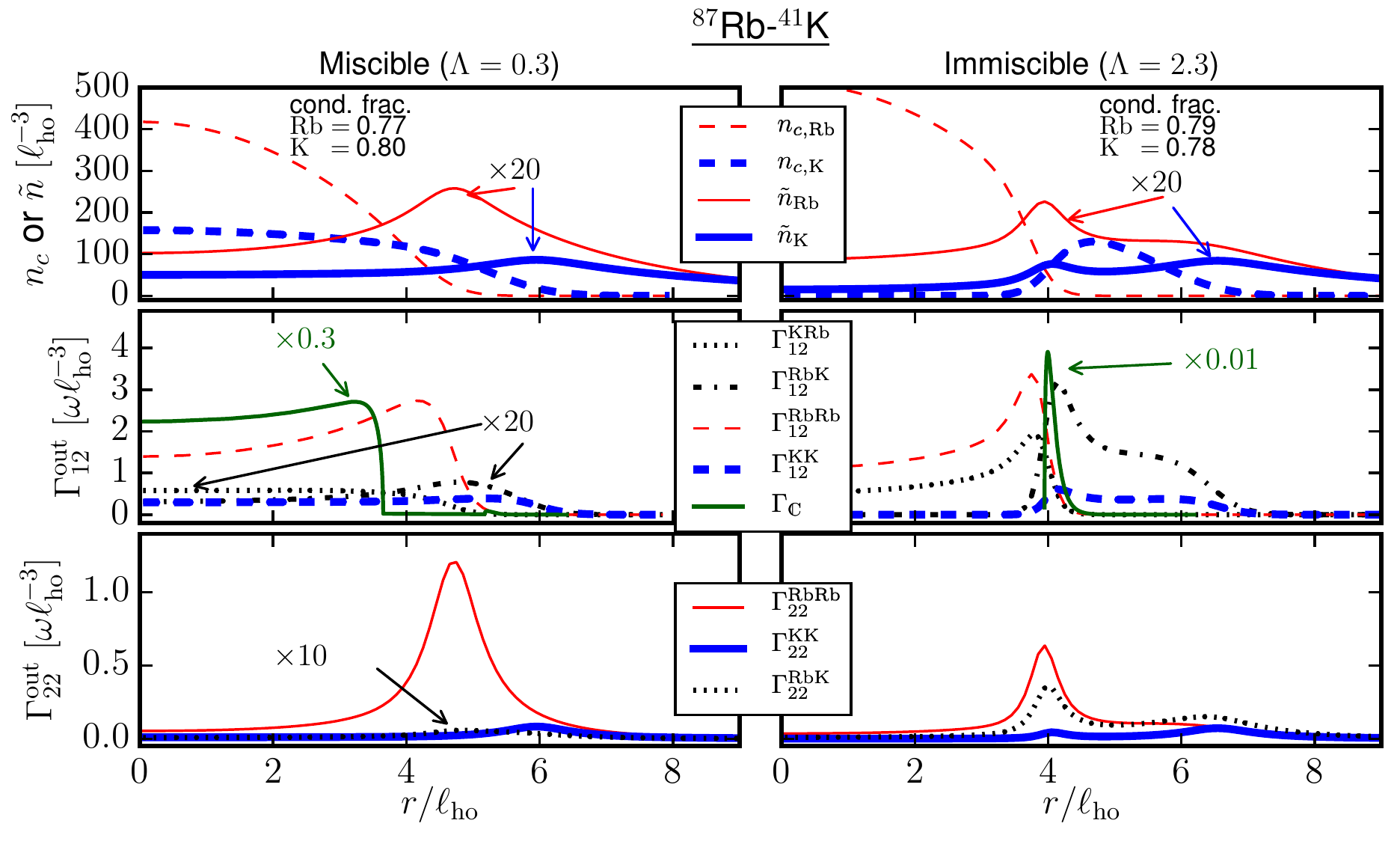}\caption{\label{rbk_T21_density} (Color online) Miscible (left) and immiscible (right) $^{87}$Rb-$^{41}$K mixtures in an isotropic harmonic trap (frequency $\omega=2\pi\times20$Hz) at temperature 21nK. Other simulation parameters (scattering lengths and total numbers of atoms) are the same as Fig.~\ref{con_frac}. The reference harmonic length is $\ell_{\rm ho}=\sqrt{\hbar/(m_{\rm Rb87}\omega)}$. Top: Condensate and thermal densities. Middle: Spatially resolved collision rates between condensate and thermal atoms. Bottom: Spatially resolved collision rates between thermal atoms. }
\end{figure*}

\begin{figure*}[!htbp]
  \includegraphics[width=0.7\textwidth]{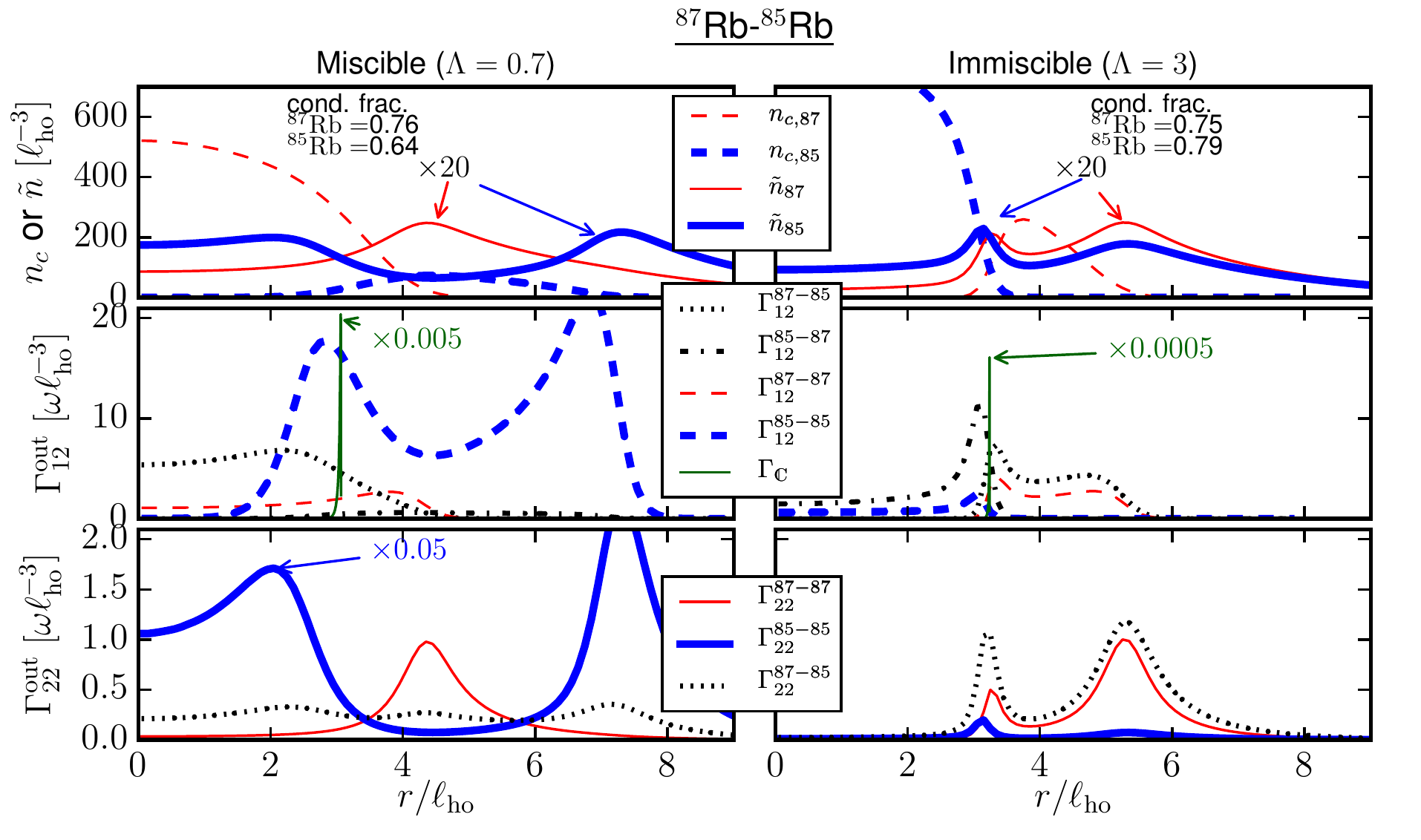}\caption{\label{rbrb_T21_density} (Color online) Same as Fig.~\ref{rbk_T21_density} but computed for $^{87}$Rb-$^{85}$Rb mixtures.}
\end{figure*}

Typical density profiles of binary systems are shown in the top panels of Figs.~\ref{rbk_T21_density} ($^{87}$Rb-$^{41}$K) and \ref{rbrb_T21_density} ($^{87}$Rb-$^{85}$Rb), where the two condensates (dashed lines) mix (left columns) or phase-separate (right columns), but always sit on top of more diffused non-condensate clouds (solid lines). These non-condensate clouds have long tails that extend much further than the condensate clouds, a feature that is also seen in the single-component Bose gas~\cite{hutchinson_zaremba_1997,jackson_zaremba_2002}. However, in contrast to the single-peak structure in a single-component gas, mean-field repulsion from both condensates in a binary mixture leads to a double-peaked thermal structure at the condensate edges, where the effective mean-field potentials of the non-condensed atoms are local minima. The middle and bottom panels give the spatial collisional rates involving collisions between thermal-condensate atoms ($\Gamma_{12}$ and $\Gamma_{\mathds{C}}$) and thermal-thermal atoms ($\Gamma_{22}$), respectively. These spatial rates depend strongly on the condensate and thermal cloud density profiles. In the next subsection, we give more details on the calculation and analysis of these collisional rates.

\subsection{\label{sec:collrate}Collisional rates}
With the equilibrium density profiles, we can proceed to evaluate the local collision integrals~\eqref{c12jk},\eqref{c12jj},\eqref{c12jk2},\eqref{c22jk} and \eqref{c22jj} (where the condensate energy $\varepsilon_c^j=\mu_c^j$ and the condensate momentum $\vec{p}_c^j=0$ at equilibrium). Since these integrals are identically zero at equilibrium, we re-express them in the form 
\begin{align}
  C^{kj}_{12}=C_{12}^{kj,\text{out}}-C_{12}^{kj,\text{in}}
\end{align}
(and analogously for $\mathds{C}_{12}$ and $C_{22}$) to explicitly identify the ``in'' and ``out'' scattering rates. In this way, we can assess the importance of the various collisional processes by comparing either their ``in'' rates or their ``out'' rates. By integrating over momentum space, we obtain the collisional rate
\begin{align}
  \Gamma_{12(22)}^{kj,{\rm out}} = \int \frac{d\vec{p}}{(2\pi\hbar)^3} C_{12(22)}^{kj,\text{out}}\label{eq:gamma_gen_out}
\end{align}
that measures the number of non-condensed atoms that have collided through a particular ``out'' process per unit volume per unit time. The mathematical steps needed to compute Eq.~\eqref{eq:gamma_gen_out} are given in Appendix \ref{appendix:com_transform}. In the following, we only give the final formulae used in our numerical computation.

To calculate the collision rates between non-condensed atoms (for both $k=j$ and $k\neq j$), it is convenient to transform to the center-of-mass frame. We therefore obtain
\begin{align}
\Gamma^{kj,{\rm out}}_{22}=&\int \frac{d\vec{p}_1}{(2\pi\hbar)^3} f_1^{j} \int \frac{d{\bf p}_2}{(2\pi\hbar)^3} f^{k}_{2}\nonumber\\
                     &\times\int \frac{d\Omega}{4\pi} \sigma_{kj} |\vec{v}_1-\vec{v}_2|(f^{k}_{3}+1)(f^{j}_{4}+1),\label{gamma22}
\end{align}
where $\sigma_{kj}=(1+\delta_{kj})4\pi a_{kj}^2$ is the cross section, $\vec{v}_1$ and $\vec{v}_2$ are the initial velocities of atoms $j$ and $k$ respectively, $\Omega$ specifies the solid angle of the final relative velocity $\vec{v}_4-\vec{v}_3$. 

For collisions between condensate and non-condensate atoms, we first look at the $C^{kj}_{12}$ process (for both $k=j$ and $k\neq j$) which is present even in a single-component Bose gas. We specifically evaluate the ``out'' collision rate that represents the scattering of a non-condensed atom from a condensate to produce two non-condensed atoms,
\begin{equation}
  \Gamma_{12}^{kj,{\rm out}} = \int\frac{d{\bf p}_2}{(2\pi\hbar)^3}f^{k}_{2}\,n_{c,j}\,\sigma_{kj}v^{\text{out}}_{r}\int\frac{d\Omega}{4\pi}(1+ f^{k}_{3}+f^{j}_{4}), \label{gamma12}
\end{equation}
where $v^{\text{out}}_r = \sqrt{|\vec{v}_{c,j}-\vec{v}_2|^2-2(U_{\text{n}}^j-\mu_c^j)/m_{kj}}$ is the relative speed of the initial states corrected to take into account the local conservation of energy. 
The reverse process, where two non-condensed atoms collide such that one of them goes into a condensate, is given by the ``in'' rate as
\begin{align}
  \Gamma_{12}^{kj,{\rm in}} = \int\frac{d{\bf p}_4}{(2\pi\hbar)^3}f^{j}_{4}\frac{\,n_{c,j}\,\sigma_{kj}(m_k/m_{kj})^3}{4\pi(1+m_j/m_k)|\vec{v}_r^{\rm in}|}\int d\tilde{\vec{v}} f_3^k,\label{eq:gamma12_in}
\end{align}
where $\vec{v}_r^{\rm in}=\vec{v}_4^j-\vec{v}_c^j$ is the velocity of thermal atom $j$ relative to the local condensate velocity while the second integral is a two-dimensional integral over the velocity vector $\tilde{\vec{v}}$ which is in a plane normal to $\vec{v}_r^{\rm in}$. The velocity of the other incoming thermal atom $\vec{v}_3^k$ is then given by
\begin{align}
  \vec{v}_3^k = \vec{v}_c^j + \frac{(1-m_j/m_k)}{2}\vec{v}_r^{\rm in} + \tilde{\vec{v}} + \frac{(U_{\rm n}^j-\mu_c^j)\hat{\vec{v}}_r^{\rm in}}{m_j|\vec{v}_r^{\rm in}|}
\end{align}
and the outgoing velocity of the thermal atom is given by
\begin{align}
  \vec{v}_2^k = (m_j/m_k) \vec{v}_r^{\rm in} + \vec{v}_3^k.
\end{align}
Note that we follow~\cite{jackson_zaremba_2002} and drop the cubic term $f_2 f_3 f_4$ in numerical simulations as it cancels exactly between the ``in'' and ``out'' rates.

Finally, we consider the exchange collisions $\mathds{C}^{kj}_{12}$ ($k\neq j$ only) novel to our treatment of the binary Bose gas, which describes a process whereby one condensate atom (say atom $k$) collides with a non-condensed atom $j$ and are then scattered into a thermal (atom $k$) and condensed (atom $j$) state. The collision rate is
\begin{align}
\Gamma^{\rm out}_{\mathds{C}}=\sigma_{kj}\left(\frac{\mathcal{M}_{kj}}{m_{kj}}\right)^2 n_{c,k}\,n_{c,j} \tilde{v}_r\int\frac{d\Omega}{4\pi} f^{j}_{2}(f^{k}_{1}+1),\label{gamma_xc}
\end{align}
where $\mathcal{M}_{kj}^{-1}=m_k^{-1}-m_j^{-1}$ plays the role of an effective reduced mass while the effective relative speed is 
\begin{equation}
\tilde{v}_r{=}\sqrt{|\vec{v}_{c,j}{-}\vec{v}_{c,k}|^2{-}2([U_{\text{n}}^k{-}\mu_c^k]{-}[U_{\text{n}}^j{-}\mu_c^j])/\mathcal{M}_{kj}}. 
\end{equation}
Equations~\eqref{gamma22}, \eqref{gamma12} and \eqref{gamma_xc} are the key quantities computed in our simulations once the equilibrium densities are obtained. The first two are evaluated using Monte Carlo sampling of the integrals while an exact expression (see Sect.~\ref{subsect:temp_g12} for detailed discussions) can be obtained for Eq.~\eqref{gamma_xc} because $\vec{v}_{c,j}=\vec{v}_{c,k}=0$ at equilibrium. 

Examples of these spatially-resolved collision rates are shown in the middle ($\Gamma_{12}$ and $\Gamma_{\mathds{C}}$) and bottom ($\Gamma_{22}$) panels of Figs.~\ref{rbk_T21_density} and \ref{rbrb_T21_density}. Note that $\Gamma_{12}$ are drawn on a larger scale compared to $\Gamma_{22}$. Typically, the rates between condensed and non-condensed atoms (both $\Gamma_{12}$ and $\Gamma_{\mathds{C}}$) feature localized peaks at the condensate edges where the thermal cloud and condensate overlap the most, while the rates between non-condensed atoms ($\Gamma_{22}$) follow closely the shape of the thermal density profiles. This is because $\Gamma_{12}$ and $\Gamma_{\mathds{C}}$ are approximately proportional to the product of condensate densities and thermal cloud densities while $\Gamma_{22}$ is approximately proportional to the product of two thermal cloud densities. These observations are important to understand the variation of these collision rates with respect to the trap frequency; see Sec.~\ref{sec:varytrap}.

On the other hand, comparison among the $C_{22}$ or $C_{12}$ processes shows that the relative peak values of $\Gamma$ can be estimated by the relevant cross section $\sigma\propto a_s^2$. For example, $^{87}$Rb intraspecies collisions (red dashes) and the $^{87}$Rb-$^{41}$K interspecies collisions (black dots and black dash-dots) have comparable peak heights in the immiscible case (Fig.~\ref{rbk_T21_density}) because of similar cross sections, whereas the $^{85}$Rb intraspecies collisions (blue thick dashes) dominates over both $^{87}$Rb-$^{85}$Rb interspecies collisions (black dots and black dash-dots) and $^{87}$Rb intraspecies collisions (red dashes) in the miscible case (Fig.~\ref{rbrb_T21_density}) because of the large $^{85}$Rb scattering length $a_{\rm Rb85}=900a_0$. 

Finally, for the case of a $^{87}$Rb-$^{85}$Rb mixture, the sharp peak for the interspecies exchange collision $\Gamma_{\mathds{C}}$ (green curves in the right middle panels of Fig.~\ref{rbrb_T21_density}) is a consequence of the small mass difference between the two different atomic species. This is most easily seen if we consider a small spatial region around a critical radius $r_c$, at which $\tilde{v}_r=0$. In this case, we can approximate $\tilde{v}_r=\sqrt{\mathcal{C}(r-r_c)/\mathcal{M}_{kj}}$ for some constant $\mathcal{C}$ and substitute this into Eq.~\eqref{gamma_xc} to show that the spatial width of $\Gamma_{\mathds{C}}$ is proportional to $\mathcal{M}_{kj}^{-1}$ . For small mass difference, while $\Gamma_{\mathds{C}}$ is sharply-peaked in space, it is nevertheless possible to obtain the total number of interspecies exchange collisions, a physically meaningful and experimentally relevant quantity, by integrating $\Gamma_{\mathds{C}}$ over the full cloud volume. 

\subsection{\label{sec:hydrodynamic}Hydrodynamic analysis}
From the collision rates, we can further extract the mean free time $\tau$ as~\cite{griffin_nikuni_2009}
\begin{align}
  \frac{1}{\tau} = \frac{ 1}{ N_{\rm coll}} \int d\vec{r} \ \Gamma(\vec{r}), \label{eq:tau}
\end{align}
where $N_{\rm coll}$ is the relevant number of available non-condensed atoms taking part in collisions for each process. For example, with $\Gamma_{12}^{kj}$ in Eq.~\eqref{gamma12}, 
\begin{align}
  N_{\rm coll} = \int \frac{d\vec{r} \ d\vec{p}}{(2\pi\hbar)^3} f^k(\vec{p},\vec{r},t) = \int d\vec{r} \ \tilde{n}^k(\vec{r},t).
\end{align}
In the case of thermal-thermal collisions ($C_{22}$ processes) that involve two different components, $N_{\rm coll}$ refers to the number of non-condensed $^{87}$Rb atoms. This choice has no significant impact as, for our simulation parameters, the condensate fraction of $^{87}$Rb differs from the fraction of the other component ($^{41}$K or $^{85}$Rb) by less than 10\%, except for the miscible $^{87}$Rb-$^{85}$Rb mixture due to the strong mean-field corrections to the $^{85}$Rb fraction, see Fig.~\ref{con_frac}. 

We would like to mention that one could also use an alternative time scale defined by $\tilde{\tau}$~\cite{nikuni_zaremba_1999}
\begin{align}
  \frac{1}{\tilde{\tau}} = \frac{ 1}{ N_{\rm con}} \int d\vec{r} \ \Gamma(\vec{r}), \label{eq:tau_tilde}
\end{align}
where $N_{\rm con}$ is the relevant number of condensed atoms. The key difference lies in $N_{\rm coll}$ and $N_{\rm con}$, which simply reflects our interest with respect to the change in the number of either non-condensed or condensed atoms. These two time scales therefore differ by the order of condensate fraction.

\begin{figure*}[!htbp]
  \includegraphics[width=0.8\textwidth]{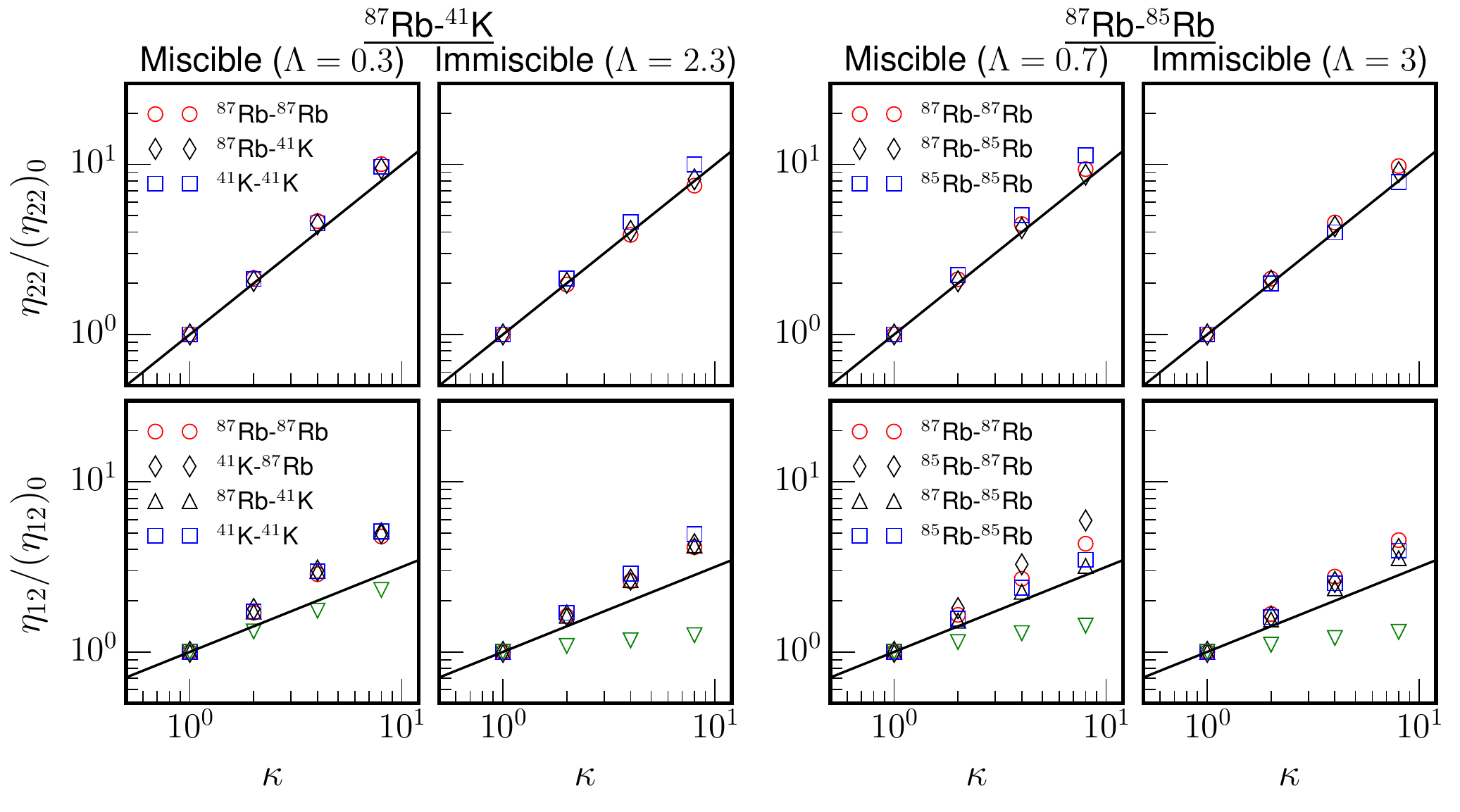}\caption{\label{fig:rbk_rbrb_trap_freq}(Color online) Variation of hydrodynamic parameters $\eta=1/(\omega\tau)$ of $^{87}$Rb-$^{41}$K (left) and $^{87}$Rb-$^{85}$Rb (right) mixtures with respect to the scaling of trap frequency and temperature, $(\omega,T)=\kappa(\omega_0,T_0)$ for $C_{22}$ (top), $C_{12}$ (bottom) and $\mathds{C}_{12}$ (bottom, green inverted triangles) processes. Scattering lengths and atom numbers are the same as Fig.~\ref{rbk_T21_density} and \ref{rbrb_T21_density}. The reference frequency and temperature are $\omega_0=2\pi\times20$Hz and $T_0=25$nK. Each data set of $\eta$ has been normalized by its value $\eta_0$ at $\kappa=1$. The solid lines give our predictions~\eqref{eq:alpha_scaling} for $\alpha=1$ (top) and $\alpha=1/2$ (bottom). $1/(\omega\tau_{\mathds{C}})$ departs from our prediction ($\alpha=0$) in the miscible $^{87}$Rb-$^{41}$K case (first column bottom) because our assumption of localised $\mathds{C}_{12}$ process is no longer valid; see Fig.~\ref{rbk_T21_density} for an example.}
\end{figure*}

When $\tau$ is compared with the trap frequency, which governs the oscillation frequency of a collisionless classical particle in the harmonic trap, we obtain the dimensionless hydrodynamic parameter 
\begin{align}
\eta=\frac{1}{\omega \tau}.
\end{align}
If $\eta>1$, a non-condensed atom will experience on average at least one collision before completing an oscillation in the harmonic trap, hence the system is in a hydrodynamic regime. Otherwise, the system is in the collisionless regime. These hydrodynamic parameters are highly relevant to cold-atom experiments as they determine the thermalisation rates~\cite{wu_foot_1996,delannoy_murdoch_2001,mosk_kraft_2001,anderlini_ciampini_2005}. In particular, the interspecies hydrodynamic parameters are crucial to the efficiency of sympathetic cooling~\cite{delannoy_murdoch_2001}. Understanding these parameters can therefore help to optimise future studies of the various cooling stages.

In the following subsections, we analyse the collisional processes using the hydrodynamic parameter given by Eq. \eqref{eq:tau}.
 
\subsection{\label{sec:varytrap}Trap frequency variation}

Our previous work~\cite{edmonds_lee_2015} has shown that the hydrodynamic parameter $\eta_{\mathds{C}}=1/(\omega \tau_{\mathds{C}})$ of the exchange process can be one to two orders of magnitude larger than the corresponding parameters of the $C_{12}$ and $C_{22}$ processes. In this subsection, we show that by varying the isotropic trap frequency, it is possible to bring $\eta$ of the various processes closer in magnitude. We provide a further explanation based on the scaling of length, energy and condensate densities.

In order to make meaningful comparisons, we scale the trap frequency $\omega$ and the temperature $T$ simultaneously by the same factor $\kappa$ such that the condensate fractions of the binary mixtures remain approximately the same as $\omega$ and $T$ are varied. This can be easily understood for the non-interacting Bose gas, where the single-particle energies appear as multiples of $\hbar\omega$ and the thermal occupation (determined by the ratio $\hbar\omega/k_BT$) thus remain unchanged. 

We use $\omega_0 = 2\pi\times 20$Hz and $T_0=25$nK as references for trap frequency and temperature and consider four different sets of isotropic trap frequency and temperature, $(\omega,T)=\kappa\times(\omega_0,T_0), \kappa\in\{1,2,4,8\}$. The numerically-obtained hydrodynamic parameters are shown in Fig.~\ref{fig:rbk_rbrb_trap_freq}, which clearly shows that 
\begin{align}
  \frac{\eta}{\eta_0}=\kappa^\alpha,
\end{align}
where
\begin{align}
  \alpha \approx\left\{\begin{array}{ll}
                                    1 & (C_{22})\\
                                    1/2 & (C_{12})\\
                                    0 & (\mathds{C}_{12}),
                                 \end{array}\right.\label{eq:alpha_scaling}
\end{align}
and $\eta_0 = 1/(\omega_0\tau_0)$ is the reference hydrodynamic parameter (different numerical values for different processes) while $\tau_0$ is the mean free time of the various processes at $\kappa=1$. In order to understand Eq.~\eqref{eq:alpha_scaling}, we rewrite Eqs.~\eqref{gamma22}, \eqref{gamma12} and \eqref{gamma_xc} in terms of dimensionless variables for position $\bar{\vec{r}} = \vec{r}/\ell$, momentum $\bar{\vec{p}} = \vec{p}\ell/\hbar$ and velocity $\bar{\vec{v}} = \vec{v}/(\ell \omega)$, where $\ell=\sqrt{\hbar/(m_{\rm Rb}\omega)}$ is the harmonic length for $^{87}$Rb atoms and we choose the mass of $^{87}$Rb here simply as a reference. For the $C_{22}$ processes, we have
\begin{align}
  \frac{1}{\omega\tau^{kj}_{22}}& = \frac{1}{\ell^2 N_{\rm coll}}\int \frac{d\bar{\vec{r}}\ d\bar{\vec{p}}_1\ d\bar{\vec{p}}_1}{(2\pi)^6} \nonumber\\
  & \times \int \frac{d\Omega}{4\pi} \sigma_{kj} |\bar{\vec{v}}_1-\bar{\vec{v}}_2| f_1^j f_2^k (f_3^k+1)(f_4^j+1).
\end{align}
Since the phase-space distribution $f$ as a function of $\bar{\vec{r}}$ and $\bar{\vec{p}}$ remains approximately unchanged as we vary $\kappa$, the sole dependence on $\kappa$ in $1/(\omega\tau_{22})$ comes from the prefactor $1/\ell^2\propto\omega\propto\kappa$ and thus $\alpha=1$, a result consistent with those from~\cite{delannoy_murdoch_2001}.

We perform the same transformations on the $C_{12}$ collisional process and arrive at
\begin{align}\nonumber
\frac{1}{\omega\tau^{kj}_{12}}&=\frac{1}{\ell^2N_{\rm coll}}\int\frac{d\bar{\vec{r}} \ d\bar{\vec{p}}_2}{(2\pi)^3}f^{k}_{2}\,\bar{n}_{c,j}\,\sigma_{kj}\bar{v}^{\text{out}}_{r}
\\&\times\int\frac{d\Omega}{4\pi}(1{+}f^{k}_{3}{+}f^{j}_{4}).\label{eq:c12_hydro}
\end{align}
It is now important to realise that $\Gamma^{kj}_{12}(\vec{r})$ is strongly-peaked around the condensate edges, hence it is sufficient to consider the scaling of the dimensionless condensate density $\bar{n}_{c,j}=n_{c,j}\ell^3$ in this region. To obtain a quantitative estimate, we approximate the condensate by a Thomas-Fermi profile and use the density at a healing length from the Thomas-Fermi radius as a reference to conclude that $\bar{n}_{c,j}\propto \ell$ as $\kappa$ varies. The net result is that $1/(\omega\tau^{kj}_{12})\propto 1/\ell\propto\sqrt{\omega}\propto \sqrt{\kappa}$, hence $\alpha=1/2$.

It is now straight forward to see that $1/(\omega\tau_{\mathds{C}})$ does not scale with $\ell$ because of the product $n_{c,j} n_{c,k}$ in Eq.~\eqref{gamma_xc}, hence $\alpha=0$. For the miscible Rb-K mixture in Fig.~\ref{fig:rbk_rbrb_trap_freq}, this prediction breaks down because the assumption that $\Gamma_{\mathds{C}}(\vec{r})$ is localised in space is not longer valid; see the left middle panel of Fig.~\ref{rbk_T21_density}.

While the scaled hydrodynamic parameter $\eta/\eta_0$ appears to be small for the exchange collisional process $\mathds{C}_{12}$ when compared to other $C_{12}$ and $C_{22}$ processes, the actual numerical values of $\eta$ are in fact large, hence $\mathds{C}_{12}$ remains a dominant process in the situation considered by Fig.~\ref{fig:rbk_rbrb_trap_freq}.

\subsection{\label{sec:trap_geometry}Trap geometry}

Besides variation of the isotropic trap frequency, a highly-relevant possibility, both experimentally and theoretically, is the variation of the trap aspect ratio $\lambda_{\rm trap}=\omega_z/\omega_\perp$,  where $\omega_z$ ($\omega_\perp$) is the axial (radial) trap frequency, so as to probe the physics of reduced dimensionality. If $\lambda_{\rm trap}<1$, we have a cigar-shaped condensate that can be used to study, e.g. solitons~\cite{burger_bongs_1999,becker_stellmer_2008} and solitonic vortices~\cite{donadello_serafini_2014}; if instead, $\lambda_{\rm trap}> 1$, the condensate cloud is pancake-shaped and it is commonly used to study vortices~\cite{matthews_anderson_1999,madison_cheby_2001,abo_shaeer_raman_2001}.

\begin{figure*}
  \includegraphics[width=0.72\textwidth]{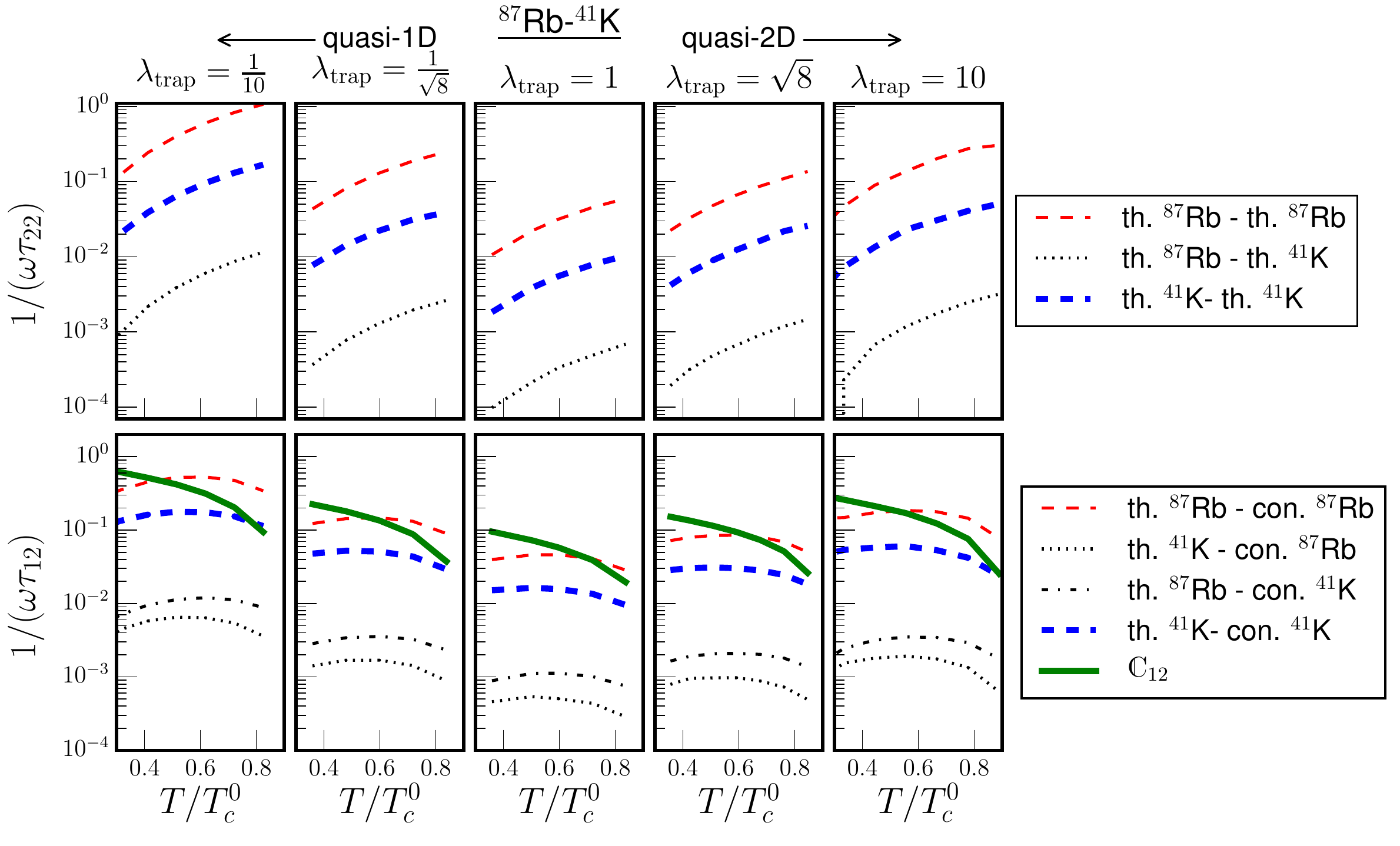}\caption{\label{fig:trap_geometry_rbk}(Color online) Variation of the hydrodynamic parameters $1/(\omega\tau)$ with increasing aspect ratio $\lambda_{\rm trap}=\omega_z/\omega_\perp$ (left to right) for miscible $^{87}$Rb-$^{41}$K mixtures. The scattering lengths are the same as Fig.~\ref{rbk_T21_density}. The reference frequency is $\omega=2\pi\times 20$Hz and the axial (radial) trap frequency $\omega_z$ ($\omega_\perp$) is equal to $\omega$ for quasi-1D (quasi-2D) system. For the ease of comparison, $T_c^0$ is chosen to be the critical temperature for the non-interacting single-component trapped gas.}
\end{figure*}

\begin{figure*}
  \includegraphics[width=0.72\textwidth]{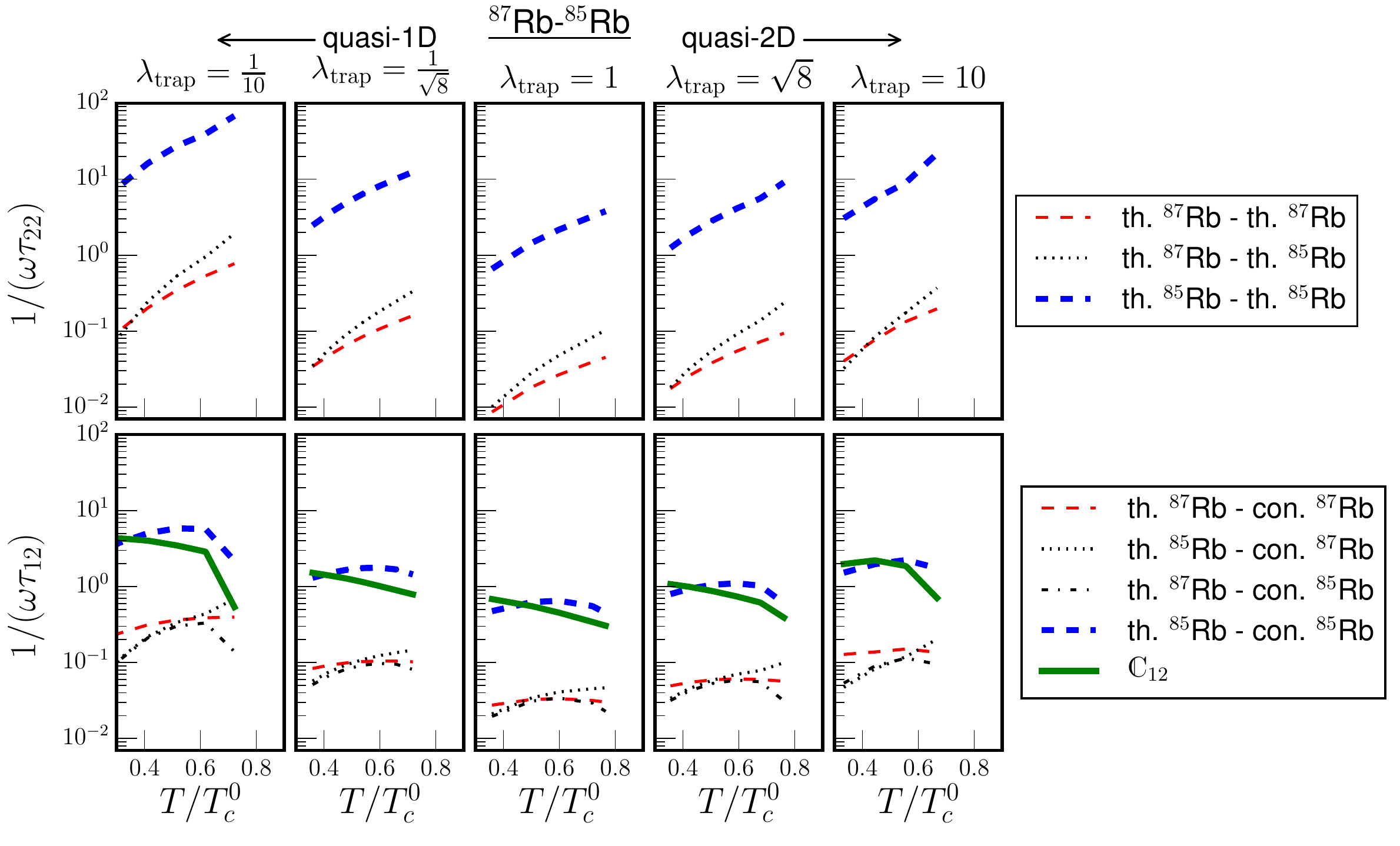}\caption{\label{fig:trap_geometry_rbrb}(Color online) Same as Fig.~\ref{fig:trap_geometry_rbk} but computed for $^{87}$Rb-$^{85}$Rb mixtures. The scattering lengths are the same as Fig.~\ref{rbrb_T21_density}}
\end{figure*}

In the following, we choose a reference frequency $\omega=2\pi\times20$Hz and fix $\omega_z(\omega_\perp)=\omega$ for cigar (pancake) clouds and consider miscible mixtures with two different aspect ratios for each trap geometry: $1/\lambda_{\rm trap}=\sqrt{8},10$ for a cigar shaped cloud (column 1,2 in Figs.~\ref{fig:trap_geometry_rbk} and \ref{fig:trap_geometry_rbrb}) and $\lambda_{\rm trap}=\sqrt{8},10$ for a pancake shaped cloud (column 4,5 in Figs.~\ref{fig:trap_geometry_rbk} and \ref{fig:trap_geometry_rbrb}). We have also checked that our conclusions are applicable to immiscible mixtures (not shown).

Figures~\ref{fig:trap_geometry_rbk} shows the variation of the hydrodynamic parameters of $^{87}$Rb-$^{41}$K mixtures for various collisional processes as a function of $T/T^0_c$, where $T^0_c$ is chosen to be the critical temperature of the non-interacting single-component Bose gas for the convenience of comparison. For each binary mixture, in going from left to right, the trap geometry changes from a quasi-1D geometry to an isotropic system, then to a quasi-2D geometry. As the geometry changes, all hydrodynamic parameters increase like $\lambda^{-1}_{\rm trap}$ for the cigar-shaped cloud, and like $\sqrt{\lambda_{\rm trap}}$ for the pancake-shaped cloud, meaning that the collisional time scale is mainly determined by the tighter trap frequency. This can be understood as a consequence that atoms are confined to a smaller region in space with a tighter trap, hence an increased probability of collisions. Despite the change in the collisional time scales, the relative magnitudes of $\eta$ when compared among the different processes remain roughly unchanged. In other words, if the $^{87}$Rb intraspecies scattering is the dominant $C_{22}$ process in an isotropic trap (third column of top panels in Fig.~\ref{fig:trap_geometry_rbk}), it remains so even if we tighten either the radial or axial trap frequency. 
It also means that, the $\mathds{C}_{12}$ collisional process (bottom green solid lines) remains the dominant interspecies collisional process (others are indicated by black dots and dash-dots) when the aspect ratio is changed. Similar observations can be made on $^{87}$Rb-$^{85}$Rb mixtures (Fig.~\ref{fig:trap_geometry_rbrb}). However, a comparison between Fig.~\ref{fig:trap_geometry_rbk} and Fig.~\ref{fig:trap_geometry_rbrb} reveals another important feature: the relative magnitudes of $\eta$, when compared between intraspecies and interspecies collisional processes, are largely determined by the scattering lengths, as we have already noted in our analysis of the spatial collisional rates (see the end of Sec.~\ref{sec:collrate}). For this reason, $^{85}$Rb intraspecies collisions (blue dashes in the right panels of Fig.~\ref{fig:trap_geometry_rbrb}) dominates both the $C_{12}$ and $C_{22}$ processes.

We would like to caution the reader that our numerical results are obtained by assuming the non-condensed atoms behave like classical particles in three-dimensional space. This is certainly not true when the confining trap frequency is much larger than the thermal energy, $\hbar\omega_\perp \gg k_BT$ (cigar-shaped cloud) or $\hbar\omega_z \gg k_BT$ (pancaked-shaped cloud). Our results here serve only as a general guide in changing trap geometry and should not be extended to extremely large or small aspect ratios.

\subsection{\label{subsect:temp_g12}Temperature-dependence of $\Gamma_{\mathds{C}}$}
\begin{figure*}
\includegraphics[width=0.75\textwidth]{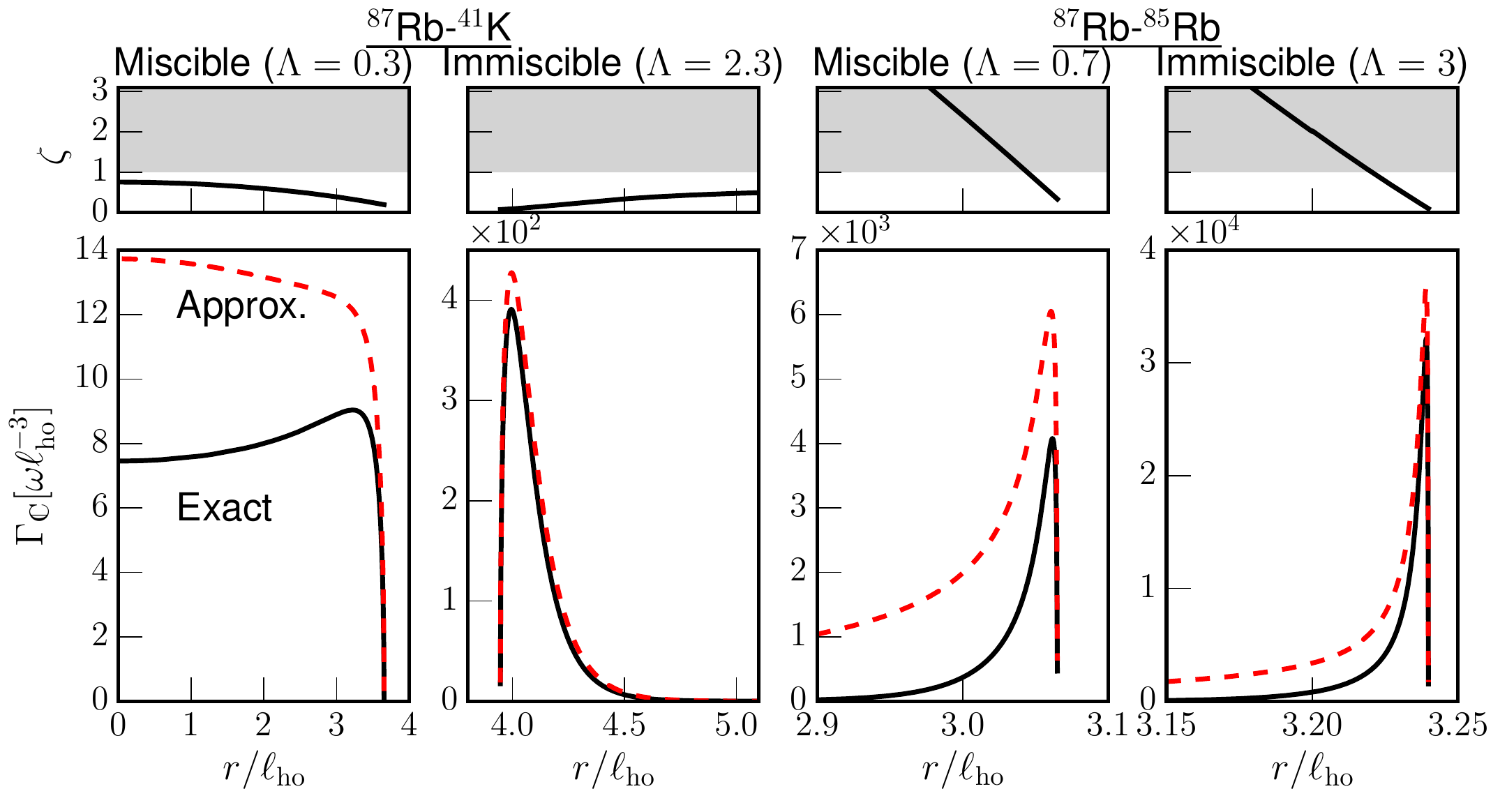}\caption{\label{fig:gamma_xc_T21}(Color online) The estimation parameter $\zeta=(p^2/2m_j+U^j_{\rm n}-\mu_c^j)/(k_BT)$ (top) and the exchange collision rate $\Gamma_{\mathds{C}}$~\eqref{gamma_xc_eq} evaluated with the exact phase-space distribution~\eqref{eq:eff_ph_sp} (bottom black solid) or a phase-space distribution expanded to leading order in $1/T$ (bottom red dashed) as a function of distance $r$ from the trap centre. Shaded regions at the top panels give the regime where the high-temperature expansion is inapplicable ($\zeta>1$). Simulation parameters of $^{87}$Rb-$^{41}$K (left) and $^{87}$Rb-$^{85}$Rb (right) mixtures are the same as Figs.~\ref{rbk_T21_density} and \ref{rbrb_T21_density} at a temperature $T=21$nK$\sim0.5T_c$.}
\end{figure*} 
The final question that we would like to address with our equilibrium simulations concerns the temperature dependence of the exchange collisional process $\mathds{C}_{12}$. At equilibrium, the spatially-resolved collision rate~\eqref{gamma_xc} can be simplified to 
\begin{align}
\Gamma_{\mathds{C}}=\sigma_{kj}\left(\frac{\mathcal{M}_{kj}}{m_{kj}}\right)^2 n_{c,k}\,n_{c,j} \tilde{v}'_r f'(f'+1),\label{gamma_xc_eq}
\end{align}
where $\tilde{v}'_r=\sqrt{2([U_{\text{n}}^j-\mu_c^j]-[U_{\text{n}}^k-\mu_c^k])/\mathcal{M}_{kj}}$ is the effective relative speed and the phase-space distribution is
\begin{align}
f'=&\frac{1}{\exp[(\frac{p^2}{2m_j}+U_{\rm n}^j-\mu_c^j)/(k_BT)]-1}\quad(j=a,b)\label{eq:eff_ph_sp}
\end{align}
with $p$ chosen such that Eq.~\eqref{eq:eff_ph_sp} holds for both $j=a,b$.

Since the $\mathds{C}_{12}$ collisional process is localised around the condensate edge, where $U^j_{\rm n}-\mu^j_c$ tends to be small, it is tempting to assume that the spatially-resolved collision rate $\Gamma_{\mathds{C}}(\vec{r})$ lies within the high-temperature region ($p^2/2m_j+U^j_{\rm n}-\mu_c^j \ll k_BT$). In this case, Taylor expansion of $f'$ in orders of $1/(k_BT)$ then leads to a temperature-dependent rate 
\begin{align}
  \Gamma_{\mathds{C}}\sim \mathcal{A}(c_1 T + c_2 T^2),
\end{align}
where $\mathcal{A}$, $c_1$ and $c_2$ are factors to be determined (see Appendix \ref{appendix:gamma_calc}).

In Fig.~\ref{fig:gamma_xc_T21}, we show $\Gamma_{\mathds{C}}$ (bottom solid) and the estimation parameter (top)
\begin{align}\label{zeta}
  \zeta=(p^2/2m_j+U^j_{\rm n}-\mu_c^j)/(k_BT)
\end{align}
as a function of distance $r$ from the trap centre. We have $^{87}$Rb-$^{41}$K (left) and $^{87}$Rb-$^{85}$Rb (right) mixtures in an isotropic harmonic trap (frequency $\omega=2\pi\times20$Hz) at a temperature $T=21$nK$\approx0.5T_c$. For better comparison and a somewhat indirect link to approaches based on ``classical'' distribution functions~\cite{bradley_blakie_2014,De2014a,Su2013a,liu_pattinson_2014}, we also expand the phase-space distribution to leading order in $1/T$, i.e. $f'\approx k_BT / (\frac{p^2}{2m_j}+U_{\rm n}^j-\mu_c^j)$ and plot the approximated $\Gamma_{\mathds{C}}$ as dashed lines in the bottom panels. For each mixture, the left and right columns show data for the miscible and immiscible phases respectively. The bottom panels clearly indicate that the high-temperature expansion is valid for the immiscible but not for the miscible phase. In the latter case, this is mainly because $\Gamma_{\mathds{C}}$ extends over a broader region in space.

\section{\label{sec:2fhydro}Two-fluid Hydrodynamics}
In this section we derive the hydrodynamic equations for the normal components of the multi-component system. One of the key theoretical successes of the single component ZNG theory is its agreement with the hydrodynamic (Landau-Khalatnikov) equations representing the interaction of the condensed and non-condensed components of the system \cite{nikuni_griffin_2001,nikuni_zaremba_1999}. This coupling of the two fluids has recently been explored for a two component Bose system \cite{armaitis_stoof_2015}, as well as for spin-orbit coupled thermal Bose gases \cite{hou_yu_2015}. The hydrodynamic equations for the condensate are obtained using the Madelung transformation along with Eq. \eqref{dschro1}, yielding
\begin{equation}
\frac{\partial}{\partial t}n_{c,j}+\nabla\cdot({n_{c,j}\bf v}_{c,j})=-\left(\Gamma^{jj}_{12}+\Gamma^{kj}_{12}+\Gamma^{kj}_{\mathds{C}}\right),\label{hydro_c1}
\end{equation}
where ${\bf v}_{c,j}=(\hbar/m_j)\nabla\theta_j$ defines the superfluid velocity of component $j$. Equation \eqref{hydro_c1} above defines the continuity equation. The Euler equation for component $j$ takes the form
\begin{equation}
m_{j}\left(\frac{\partial}{\partial t}{\bf v}_{c,j}+\frac{1}{2}\nabla{\bf v}_{c,j}^{2}\right)=-\nabla\mu_{c}^{j},
\label{hydro_c2}
\end{equation}
where $\mu_{c}^{j}=\hbar^2(\nabla^2\sqrt{n_{c,j}})/(2m\sqrt{n_{c,j}})+U^{j}_{c}$ is the non-equilibrium chemical potential for component $j$. To obtain the corresponding equations for the non-condensate, we take moments with respect to the powers of the momentum {\bf p}, i.e. $\int d{\bf p}\ {\bf p}^{n}$ (where $n=0,1,2$) of the kinetic equation, Eq. \eqref{qbe}. This leads to a set of three coupled nonlinear equations for the non-condensate that can be used to describe the limit where collisions dominate, i.e. the hydrodynamic regime. The first of these describes the conservation of mass of component $j$,
\begin{equation}
\frac{\partial}{\partial t}\tilde{n}_j+\nabla\cdot(\tilde{n}_{j}{\bf v}_{\text{n},j})=\Gamma^{jj}_{12}+\Gamma^{kj}_{12}+\Gamma^{kj}_{\mathds{C}},
\label{nc_hydro1}
\end{equation}
where the velocity of component $j$ is defined as ${\bf v}_{\text{n},j}({\bf r},t)$, while the non-condensate density is $\tilde{n}_{j}({\bf r},t)$. The velocity of component $j$ of the normal fluid is
\begin{align}
{\bf v}_{\text{n},j}({\bf r},t)\equiv&\int\frac{d{\bf p}}{(2\pi\hbar)^3}\frac{\bf p}{m_{j}}\frac{f^{j}({\bf r},{\bf p},t)}{\tilde{n}_{j}({\bf r},t)},
\end{align}
and $\tilde{n}_j({\bf r},t)$ has been defined previously by Eq.~\eqref{eq:tilden}. The corresponding conservation law for the momentum of component $j$ (or Navier-Stokes equation) appears as
\begin{align}\nonumber
&m_{j}\tilde{n}_j\bigg(\frac{\partial}{\partial t}+{\bf v}_{\text{n},j}\cdot\nabla\bigg){\bf v}_{\text{n}\mu,j}=-\frac{\partial}{\partial x_{\nu}}P_{\mu\nu,j}-\tilde{n}_{j}\frac{\partial U^{j}_{\text{n}}}{\partial x_{\mu}}\\&+\int\frac{d{\bf p}}{(2\pi\hbar)^3}\bigg({\bf p}-m_{j}{\bf v}_{\text{n}\mu,j}\bigg)\bigg(C^{jj}_{12}+C^{kj}_{12}+\mathds{C}^{kj}_{12}\bigg),\label{nc_hydro2}
\end{align}
and ($\nu,\mu$) subscripts refer to spatial components here and in what follows. Meanwhile, the conservation law for the energy density of component $j$, $\epsilon_{j}({\bf r},t)$ is written
\begin{align}\nonumber
&\frac{\partial\epsilon_{j}}{\partial t}+\nabla\cdot(\epsilon_{j}{\bf v}_{\text{n},j})=-\nabla\cdot{\bf Q}_{j}-D_{\mu\nu,j}P_{\mu\nu,j}\\&+\int\frac{d{\bf p}}{(2\pi\hbar)^3}\frac{({\bf p}-m_{j}{\bf v}_{\text{n}\mu,j})^2}{2m_{j}}\bigg(C^{jj}_{12}+C^{kj}_{12}+\mathds{C}^{kj}_{12}\bigg).\label{nc_hydro3}
\end{align}
The symmetric rate-of-strain tensor appearing in Eq. \eqref{nc_hydro3} above is defined as 
\begin{align}
  D_{\mu\nu,j}({\bf r},t)=\frac{1}{2}\bigg(\frac{\partial{\bf v}_{\text{n}\mu,j}}{\partial x_{\nu}}+\frac{\partial{\bf v}_{\text{n}\nu,j}}{\partial x_{\mu}}\bigg),
\end{align}
and obeys $\sum_{\nu}D_{\nu\nu,j}=\nabla\cdot{\bf v_{\text{n},j}}$. The set of equations, Eqs. \eqref{nc_hydro1},\eqref{nc_hydro2} and \eqref{nc_hydro3} introduce the three important local hydrodynamic quantities; the stress tensor $P_{\mu\nu,j}({\bf r},t)$, the heat current ${\bf Q}_{j}({\bf r},t)$ and the energy density $\epsilon_{j}({\bf r},t)$ defined respectively for component $j$ as \cite{griffin_nikuni_2009}
\begin{align}
P_{\mu\nu,j}&{\equiv}m_{j}\int\frac{d{\bf p}}{(2\pi\hbar)^3}\bigg[\frac{p_{\mu}}{m_{j}}{-}{\bf v}_{\text{n}\mu,j}\bigg]\bigg[\frac{p_{\nu}}{m_{j}}{-}{\bf v}_{\text{n}\nu,j}\bigg]f^{j},\\{\bf Q}_{j}&\equiv\frac{m_{j}}{2}\int\frac{d{\bf p}}{(2\pi\hbar)^3}\bigg[\frac{\bf p}{m_{j}}{-}{\bf v}_{\text{n},j}\bigg]^2\bigg[\frac{\bf p}{m_{j}}{-}{\bf v}_{\text{n},j}\bigg]f^{j},\\\epsilon_{j}&\equiv\int\frac{d{\bf p}}{(2\pi\hbar)^3}\frac{1}{2m_{j}}\bigg[{\bf p}-m_{j}{\bf v}_{\text{n},j}\bigg]^2f^{j}.
\end{align}
Note that there is no dependence on the thermal-thermal collisional rates $\Gamma^{kj}_{22}$ in these equations, a consequence of the conservation of number, energy and momentum of the thermal atoms. To demonstrate this, we can calculate the (time-dependent) number of condensate and thermal atoms in component $j$ as
\begin{align}
N_{c,j}(t)&=\int d{\bf r}\ n_{c,j}({\bf r},t),\\
\tilde{N}_{j}(t)&=\int d{\bf r}\int d{\bf p}\ f^{j}({\bf p},{\bf r},t),
\end{align}
respectively. Using eqs. \eqref{hydro_c1} and \eqref{nc_hydro1} one finds in turn that
\begin{equation}  
\frac{\partial}{\partial t}\left(N_{c,j}(t)+\tilde{N}_{j}(t)\right)=0.
\end{equation}
The conservation of energy and momentum is slightly different, as these quantities are conserved over both components, which is reflected in the fact that the delta functions appearing throughout the kinetic theory depend generally on both the $j$ and $k$ indices. 

Equations \eqref{hydro_c1}-\eqref{hydro_c2} for the condensates and \eqref{nc_hydro1}, \eqref{nc_hydro2} and \eqref{nc_hydro3} for the non-condensates are a direct generalization of the equivalent expressions (see Ref. \cite{nikuni_griffin_2001}) for the single-component case, albeit now for a dynamically coupled system comprising two condensates and two thermal clouds. Due to their inherent complexity, the study of these coupled equations lies beyond the scope of the present work.    

It is anticipated in future works that the hydrodynamic equations will yield novel physics, particularly for the case of the full Landau-Khalatnikov theory, where the entropy of the normal component will cause additional effects not present in single component thermal Bose gases. 

\section{Comparison of schemes}
Here, we present a brief overview of the different approaches used to describe the dynamical evolution of coupled multi-component condensates. In our scheme, we have, as usual, separated the slowly evolving degrees of freedom from those evolving on more rapid time scales. In particular, having identified the condensate and (diagonal) non-condensate density as the slowly varying ``mean-field" quantities, we  obtained a coupled kinetic theory for both condensate and non-condensate that includes all relevant scattering channels. An important point here, similar to some extent to Ref.~\cite{bradley_blakie_2014} is that we have treated diagonal elements of the normal pair density (corresponds to thermal population) separately from their off-diagonal elements (corresponds to ``coherences''). Such a rationale is valid only in the absence of optical couplings, where collisions are expected to be the dominant process, which is certainly the case for mixtures of different species, such as $^{87}$Rb-$^{85}$Rb and $^{87}$Rb-$^{41}$K considered here, where no interconversion is possible. Indeed, for the physically-distinct case of condensates with internal spin degrees of freedom, such an adiabatic treatment has to be modified in order to self-consistently account also for the (internal) coherent coupling of spin degrees of freedom \cite{nikuni_williams_2003,endo_nikuni_2011}. 

There are several other complementary approaches to tackling the coupled dynamics of multi-component condensates, including the number-conserving approach~\cite{mason_gardiner_2014}, the stochastic Gross-Pitaevskii formalism~\cite{bradley_blakie_2014,De2014a,Su2013a,liu_pattinson_15}, as well as classical field~\cite{pattinson_proukakis_14} and truncated Wigner~\cite{Sabbatini2011a,Sabbatini2012a,Swislocki2013a} treatments. 

The starting point of the number-conserving method is the Penrose-Onsager criterion, in which the single-particle density matrix is written in terms of quantum field operators. One then expands the field operators with the Beliaev decomposition explicitly, maintaining their operator form and condensate-noncondensate orthogonality, in order to identify the small parameter of the theory, namely the (number) ratio of non-condensate to condensate population. This in turn allows a set of dynamical equations to be extracted which describe the condensate through a set of generalized Gross-Pitaevskii equations (GGP) and the non-condensate with modified Bogoliubov-de Gennes equations. However, the purely-dynamical single-component case~\cite{billam_gardiner_2012} has yet to be extended to the multi-component setting.

In contrast, stochastic treatments of condensate dynamics are appropriate for describing the high-temperature regime, where the number of atoms in the non-condensate fraction is large compared to those in the condensate, such that the energetic parameter $\zeta$ defined by \eqref{zeta} is less than one. Interestingly, the nature of multi-component systems means that novel transport processes, for example spin-changing collisions will contribute to both damping processes as well as noise for these systems. Both the stochastic (projected) GPE and the closely-related classical field~\cite{pattinson_proukakis_14} and truncated Wigner~\cite{Sabbatini2011a,Sabbatini2012a,Swislocki2013a} approaches have an explicit cutoff in energy, with all higher-lying  (pure thermal) atoms treated stationary. In contrast to these, the ZNG formalism explicitly treats all modes dynamically and self-consistently, but it does not include the effect of fluctuations of the phase of the non-condensate atoms, which in turn limits its applicability to temperatures not too close to the transition.

It is interesting to note that both the multi-component number conserving work of \cite{mason_gardiner_2014} as well as the stochastic treatment of \cite{bradley_blakie_2014} have independently identified a condensate to non-condensate ``exchange'' collisional event, physically analogous to that presented in this work by $\mathds{C}^{kj}_{12}$. In the former case such terms appear as in the present work as off-diagonal normal pair averages of fluctuation operators in the corresponding generalized Gross-Piteavskii equation, with such averages however defined in terms of number-conserving operators, for example see Eq. (66a) in \cite{mason_gardiner_2014}. In the stochastic treatment, the exchange process enters as a novel scattering amplitude representing an extension to the ``scattering term'' of single-component stochastic GPE, which however involves mutlimode classical field populations, as opposed to those of the single-mode condensates arising within our current treatment, see Eqs. (73) and (74) in \cite{bradley_blakie_2014}. Clearly, each of these non-equilibrium theories is quite different in origin, assumption and applicability, yet all three nonetheless demonstrate universal aspects of quantum transport theory in low temperature multi-component Bose gases.

One should also explicitly comment on the link of our present approach to the earlier works of Nikuni et al.~\cite{nikuni_williams_2003,endo_nikuni_2011}. Both works were aimed at discussing spinor condensates, for which the Hamiltonian contains additional terms explicitly maintaining the coupling between the two (spin-$\frac{1}{2}$) or three (spin-$1$) different states of the system. Clearly, in the case of explicit coupling between different states, which enables inter-conversions (i.e. particles from one state transferred to another state through external coupling), our fundamental assumption of treating the off-diagonal normal pair averages $\langle \delta_j^\dag \delta_k \rangle$ ($j\neq k$) differently from $\langle \delta_j^\dag \delta_j \rangle$ could break down. This in turn would imply that we should revisit our ``slowly-varying master'' variables, and include $\langle \delta_j^\dag \delta_k \rangle$ at the same footing as $\langle \delta_j^\dag \delta_j \rangle$ and $|\phi_j|^2$. This is precisely what has been shown in Refs.~\cite{nikuni_williams_2003,endo_nikuni_2011}, and would presumably apply to any single-species multi-component condensates $|F,m_F\rangle$, in identical $F$ and different $m_F$ states, under the presence of internal or external coupling between those states. Due to the added complexity of dealing with an off-diagonal Wigner distribution operators, such a model has however never been numerically analysed, remaining nonetheless an impressive analytical work for such immensely complicated systems.

In contrast, our present multi-component treatment is intended for mixtures of two systems of different species, such as a the case of $^{87}$Rb-$^{85}$Rb and $^{87}$Rb-$^{41}$K we have analysed here, with the full dynamical treatment pending.

\section{Conclusions}
We have demonstrated how a finite temperature theory describing the out-of-equilibrium dynamics of binary Bose gases can be derived using quantum kinetic theory. In particular, it was demonstrated how dissipative Gross-Pitaevskii equations for the binary system feature three types of collision exchange terms with the non-condensate atoms of the multi-component system. The non-condensates on the other hand are modeled with quantum Boltzmann equations coupled to collision integrals which describe atomic scattering between the condensate and non-condensate atoms. It was shown in detail how all of these transport processes are derived, including the important ``exchange'' term as well as the triplet correlation functions, which are explicitly computed for the multi-component system.

We also presented results from numerical simulations of various binary condensate systems in both miscible and immiscible regimes. In particular the condensate fractions were estimated for mixtures consisting of $^{87}$Rb-$^{41}$K and $^{87}$Rb-$^{85}$Rb, which showed a slight deviation from the estimations based on the single-component Bose gas due to mean-field effects. The role of time scales on collisions was elaborated on, in particular the collisionless and hydrodynamic regimes were studied through the hydrodynamic parameters of the various collisional processes. 



Our numerical results demonstrate the interesting possibility to access different hydrodynamic regimes. For example, thermal-thermal collisions and thermal-condensate collisions can occur on comparable or vastly distinct time scales through scaling the trap frequency and the temperature by the same factor,  because the various collisional integrals obey different scaling laws. On the other hand,  the intraspecies collisions can dominate over the interspecies collisions because of the large intraspecies scattering lengths, as demonstrated by the miscible $^{87}$Rb-$^{85}$Rb mixture, such that Feshbach resonances within and between components could prove useful in investigating different regimes. It is also possible to increase the hydrodynamic parameters of all processes by changing the trap geometry. However, it is important to note that this only changes overall values, rather than the relative estimations of different collisional processes, which remain unaffected with the $\mathds{C}_{12}$ process remaining as the dominant one. We have also investigated the extent to which a commonly-implemented phase-space expansion to leading order in $1/T$ is appropriate, finding it to be a rather poor estimation in the miscible case.

The possibility of controlling the hydrodynamicity allows us to explore the interplay of the various collisional processes. The hydrodynamic equations developed in Sec.~\ref{sec:2fhydro} shows the added complexity due to the extra scattering channels present in the binary system. With the presence of eight collisional processes (three $C_{22}$ processes, four $C_{12}$ processes and a $\mathds{C}_{12}$ process) in a binary system compared to two collisional processes (a $C_{22}$ process and a $C_{12}$ process) in a single-component Bose gas, it is not a priori clear how an out-of-equilibrium binary mixture would relax to its final state and at what time scales, especially if several collisional processes were close to the hydrodynamic regime and were competing with each other. A definitive answer requires careful numerical simulations of the full non-equilibrium dynamics, which is a subject under our active investigation. 

\section{Acknowledgments}
We acknowledge stimulating discussions with J. Arlt, S.L. Cornish, S.A. Gardiner, P. Mason and E. Zaremba and support from EPSRC [Doctoral Prize Fellowship (MJE), Grant No.~EP/K03250X/1~(KLL, NPP)].

\appendix
\begin{widetext}
\section{Perturbing Hamiltonian}\label{app_b}
The purpose of this appendix is to detail the steps required to arrive at the perturbing Hamiltonian. After applying Wick's theorem
to $\hat{H}_{3}$ and $\hat{H}_{4}$, it can be shown that the non-quadratic terms appearing in the full system Hamiltonian $\hat{H}$
can be recast in the form $\hat{H}_3\rightarrow\delta\hat{H}_{1}$ and $\hat{H}_{4}\rightarrow\delta\hat{H}_{2}+\delta{H}_{0}$ respectively, where
\begin{subequations}
\begin{align}\label{dh0}
\delta H_{0}=&-\int d{\bf r}\bigg\{\sum_{j}\frac{g_{jj}}{2}\bigg[2\tilde{n}_{jj}^{2}+|\tilde{m}_{jj}|^2\bigg]+\sum_{k\neq j}g_{kj}\bigg[\tilde{n}_{jj}\tilde{n}_{kk}+\tilde{n}_{jk}\tilde{n}_{kj}+|\tilde{m}_{jk}|^2\bigg]\bigg\},\\\nonumber
\delta\hat{H}_{1}=&\int d{\bf r}\bigg\{\sum_{j}g_{jj}\bigg(\phi_{j}^{*}\bigg[2\tilde{n}_{jj}
\hat{\delta}_{j}+\tilde{m}_{jj}\hat{\delta}^{\dagger}_{j}\bigg]+\text{h.c.}\bigg)+\sum_{k\neq j}g_{kj}\bigg(\phi_{j}^{*}\bigg[\tilde{n}_{kk}\hat{\delta}_{j}+\tilde{n}_{jk}\hat{\delta}_{k}+
\tilde{m}_{kj}\hat{\delta}^{\dagger}_{k}\bigg]\\\label{dh1}
&\hspace{8cm}+\phi_{k}^{*}\bigg[\tilde{n}_{jj}
\hat{\delta}_{k}+\tilde{n}_{kj}\hat{\delta}_{j}+\tilde{m}_{jk}\hat{\delta}^{\dagger}_{j}\bigg]+\text{h.c.}\bigg)\bigg\},\\
\delta\hat{H}_{2}=&\int d{\bf r}\bigg\{\sum_{j}\frac{g_{jj}}{2}\bigg(4\tilde{n}_{jj}
\hat{\delta}^{\dagger}_{j}\hat{\delta}_{j}{+}\bigg[\tilde{m}_{jj}\hat{\delta}^{\dagger}_{j}\hat{\delta}^{\dagger}_{j}{+}\text{h.c.}\bigg]\bigg)\label{dh2}{+}\sum_{k\neq j}g_{kj}\bigg(\tilde{n}_{jj}\hat{\delta}^{\dagger}_{k}\hat{\delta}_{k}{+}\tilde{n}_{k}\hat{\delta}^{\dagger}_{j}\hat{\delta}_{j}\nonumber\\&\hspace{8cm}{+}\bigg[\tilde{n}_{kj}\hat{\delta}^{\dagger}_{k}\hat{\delta}_{j}{+}\tilde{m}_{kj}\hat{\delta}^{\dagger}_{k}\hat{\delta}^{\dagger}_{j}{+}\text{h.c.}\bigg]\bigg)\bigg\},
\end{align}
\end{subequations}
where $\tilde{n}_{kj}=\langle\hat{\delta}^{\dagger}_{j}\hat{\delta}_{k}\rangle$ and $\tilde{m}_{kj}=\langle\hat{\delta}_{j}\hat{\delta}_{k}\rangle$. Here Eqs. \eqref{dh0}-\eqref{dh2} are generated by inserting Eqs. \eqref{wick1} and \eqref{wick2} into Eqs. \eqref{h3} and \eqref{h4}. The term given by Eq. \eqref{dh0} constitutes a mean-field shift to the chemical potential of the system. However, as we follow the usual prescription of keeping interaction effects within the chemical potential to linear order in the scattering length \cite{nikuni_zaremba_1999}, these terms need not be considered any further. 
With the definitions of Eq. \eqref{dh0}-\eqref{dh2}, one can show that the resulting perturbing Hamiltonian
\begin{equation}
\hat{H}'(t)=\hat{H}-\hat{H}_{\text{MF}}
\end{equation}
can be broken into contributions comprising different numbers of fluctuation operators. Using Eq. \eqref{ham_mf}, the perturbing Hamiltonian can be expressed in the form \cite{nikuni_zaremba_1999,griffin_nikuni_2009} 
\begin{equation}
\hat{H}'(t)=\hat{H}_{1}'(t)+\hat{H}_{2}'(t)+\hat{H}_{3}'(t)+\hat{H}_{4}'(t),
\end{equation}
where
\begin{subequations}
\begin{align}\nonumber
\hat{H}_{1}'(t)&={-}\delta\hat{H}_{1}
={-}{\int}d{\bf r}\bigg\{\sum_{j}g_{jj}\bigg(\phi_{j}^{*}\bigg[2\tilde{n}_{jj}\hat{\delta}_{j}+\tilde{m}_{jj}\hat{\delta}^{\dagger}_{j}\bigg]{+}\text{h.c.}\bigg){+}\sum_{k\neq j}g_{kj}\bigg(\phi_{j}^{*} \bigg[\tilde{n}_{kk}\hat{\delta}_{j}+\tilde{n}_{jk}\hat{\delta}_{k}+\tilde{m}_{kj}\hat{\delta}^{\dagger}_{k}\bigg]
\\
&\label{ham_pet1}\hspace{8.5cm}+\phi_{k}^{*}\bigg[\tilde{n}_{jj}\hat{\delta}_{k}+\tilde{n}_{kj}\hat{\delta}_{j}
+\tilde{m}_{jk}\hat{\delta}_{j}^{\dagger}\bigg]+\text{h.c.}\bigg)\bigg\},\\
\hat{H}_{2}'(t)&=\hat{H}_{2}-\hat{H}^{\text{diag}}_{2}
=\int d{\bf r}\bigg\{\sum_{j}\frac{g_{jj}}{2}\bigg(\phi_{j}^{2}\hat{\delta}^{\dagger}_{j}\hat{\delta}^{\dagger}_{j}+\text{h.c.}\bigg)
\label{ham_pet2}+\sum_{k\neq j}g_{kj}\left(\phi^{*}_{j}\phi^{*}_{k}\hat{\delta}_{k}\hat{\delta}_{j}+\phi^{*}_{j}\phi_{k}\hat{\delta}^{\dagger}_{k}
\hat{\delta}_{j}+\text{h.c.}\right)\bigg\},\\
\hat{H}_{3}'(t)&=\hat{H}_{3}
=\int d{\bf r}\bigg\{\sum_{j}g_{jj}\left(\phi^{*}_{j}\hat{\delta}^{\dagger}_{j}\hat{\delta}_{j}\hat{\delta}_{j}+\text{h.c.}\right)\label{ham_pet3}+\sum_{k\neq j}g_{kj}\bigg(\phi^{*}_{j}\hat{\delta}^{\dagger}_{k}\hat{\delta}_{k}\hat{\delta}_{j}+\phi_{k}^{*}\hat{\delta}^{\dagger}_{j}\hat{\delta}_{j}\hat{\delta}_{k}+\text{h.c.}\bigg)\bigg\},\\
\hat{H}_{4}'(t)&=\hat{H}_{4}-\delta\hat{H}^{\text{diag}}_{2}
=\int d{\bf r}\bigg\{\sum_{j}\frac{g_{jj}}{2}\bigg(\hat{\delta}^{\dagger}_{j}\hat{\delta}^{\dagger}_{j}\hat{\delta}_{j}\hat{\delta}_{j}-4\tilde{n}_{jj}
\hat{\delta}^{\dagger}_{j}\hat{\delta}_{j}\bigg)\label{ham_pet4}+\sum_{k\neq j}g_{kj}\left(\hat{\delta}^{\dagger}_{j}\hat{\delta}^{\dagger}_{k}\hat{\delta}_{k}\hat{\delta}_{j}-
\tilde{n}_{jk}\hat{\delta}^{\dagger}_{j}\hat{\delta}_{k}-\tilde{n}_{kj}\hat{\delta}^{\dagger}_{k}\hat{\delta}_{j}\right)\bigg\}.
\end{align}
\end{subequations}
The full expression for the Fourier transformed perturbing Hamiltonian becomes 
\begin{align}
\hat{H}'(t)=\hat{H}'_{1}(t)+\hat{H}'_{2}(t)+\hat{H}'_{3}(t)+\hat{H}'_{4}(t),
\label{eqn:decomp}
\end{align}
which are used to derive the collision integrals in the body of the text. Next, each term in Eq. \eqref{eqn:decomp} is decomposed into an intra- $\hat{H}_{n,j}'(t)$ and inter- $\hat{H}_{n,kj}'(t)$ component contribution so that for $n{=}1{-}4$ one has the decomposition $\hat{H}_{n}'(t)=\hat{H}_{n,j}'(t)+\hat{H}_{n,kj}'(t)$. Then using the Fourier expansion of the fluctuation operator $\hat{\delta}_{j}({\bf r},t_0)$
one obtains
\begin{subequations}
\begin{align}
\label{hpf_1j}\hat{H}_{1,j}'(t)=&-\sqrt{V}\sum_{j}\sum_{\bf p}g_{jj}\bigg\{\sqrt{n_{c,j}}e^{-i(\theta_j-\varepsilon^{j}_{c}(t'-t)/\hbar-{\bf p}^{j}_{c}\cdot{\bf r}/\hbar)}\bigg[2\delta_{{\bf p},{\bf p}^{j}_{c}}\tilde{n}_{jj}\hat{a}_{j,{\bf p}}+
\delta_{{\bf p},-{\bf p}^{j}_{c}}\tilde{m}_{jj}\hat{a}^{\dagger}_{j,{\bf p}}\bigg]+\text{h.c.}\bigg\},
\\\nonumber\hat{H}_{1,kj}'(t)=&-\sqrt{V}\sum_{k\neq j}\sum_{\bf p}g_{kj}\bigg\{\sqrt{n_{c,j}}e^{-i(\theta_j-\varepsilon^{j}_{c}(t'-t)/\hbar-{\bf p}^{j}_{c}\cdot{\bf r}/\hbar)}\bigg[\delta_{{\bf p},{\bf p}^{j}_{c}}\tilde{n}_{kk}\hat{a}_{j,{\bf p}}+\delta_{{\bf p},{\bf p}^{j}_{c}}\tilde{n}_{jk}\hat{a}_{k,{\bf p}}+\delta_{{\bf p},-{\bf p}^{j}_{c}}\tilde{m}_{kj}\hat{a}^{\dagger}_{k,{\bf p}}\bigg]\\\label{hpf_1kj}&+\sqrt{n_{c,k}}e^{-i(\theta_{k}-\varepsilon_{c}^{k}(t'-t)/\hbar-{\bf p}_{c}^{k}\cdot{\bf r}/\hbar)}\bigg[\delta_{{\bf p},{\bf p}_{c}^{k}}\tilde{n}_{jj}\hat{a}_{k,{\bf p}}+\delta_{{\bf p},{\bf p}_{c}^{k}}\tilde{n}_{kj}\hat{a}_{j,{\bf p}}+\delta_{{\bf p},-{\bf p}_{c}^{k}}\tilde{m}_{jk}\hat{a}^{\dagger}_{j,{\bf p}}\bigg]+\text{h.c.}\bigg\},
\\\label{hpf_2j}\hat{H}_{2,j}'(t)=&\frac{1}{2}\sum_{j}g_{jj}\sum_{{\bf p}_1,{\bf p}_2}\bigg\{\delta_{{\bf p}_1+{\bf p}_2,2{\bf p}^{j}_{c}}n_{c,j}e^{2i(\theta_j-\varepsilon^{j}_{c}(t'-t)/\hbar-{\bf p}^{j}_{c}\cdot{\bf r}/\hbar)}\hat{a}^{\dagger}_{j,{\bf p}_1}\hat{a}_{j,{\bf p}_2}^{\dagger}+\text{h.c.}\bigg\},
\\\nonumber\hat{H}_{2,kj}'(t)=&\sum_{k\neq j}\sum_{{\bf p}_1,{\bf p}_2}g_{kj}\sqrt{n_{c,j}n_{c,k}}\bigg\{\delta_{{\bf p}_{1}+{\bf p}_{2},{\bf p}^{j}_{c}+{\bf p}^{k}_{c}}e^{-i((\theta_j+\theta_k)-(\varepsilon^{j}_{c}+\varepsilon^{k}_{c})(t'-t)/\hbar-({\bf p}^{j}_{c}+{\bf p}^{k}_{c})\cdot{\bf r}/\hbar)}\hat{a}_{k,{\bf p}_1}\hat{a}_{j,{\bf p}_2}\\\label{hpf_2kj}&+\delta_{{\bf p}_1+{\bf p}^{j}_{c},{\bf p}_2+{\bf p}^{k}_{c}}e^{-i((\theta_j-\theta_k)-(\varepsilon^{j}_{c}-\varepsilon^{k}_{c})(t'-t)/\hbar-({\bf p}^{c}_{j}-{\bf p}^{k}_{c})\cdot{\bf r}/\hbar)}\hat{a}^{\dagger}_{k,{\bf p}_1}\hat{a}_{j,{\bf p}_2}+\text{h.c.}\bigg\},
\\\label{hpf_3j}\hat{H}_{3,j}'(t)=&\frac{1}{\sqrt{V}}\sum_{j}\sum_{{\bf p}_2,{\bf p}_3,{\bf p}_4}g_{jj}\sqrt{n_{c,j}}\bigg\{\delta_{{\bf p}^{j}_{c}+{\bf p}_{2},{\bf p}_{3}+{\bf p}_{4}}e^{-i(\theta_j-\varepsilon^{j}_{c}(t'-t)/\hbar-{\bf p}^{j}_{c}\cdot{\bf r}/\hbar)}\hat{a}^{\dagger}_{j,{\bf p}_{2}}\hat{a}_{j,{\bf p}_{3}}\hat{a}_{j,{\bf p}_{4}}+\text{h.c.}\bigg\},
\\\label{hpf_3kj}\hat{H}_{3,kj}'(t)=&\frac{1}{\sqrt{V}}\sum_{k\neq j}\sum_{{\bf p}_2,{\bf p}_3,{\bf p}_4}g_{kj}\sqrt{n_{c,j}}\bigg\{\delta_{{\bf p}^{j}_{c}+{\bf p}_{2},{\bf p}_{3}+{\bf p}_{4}}e^{-i(\theta_j-\varepsilon^{j}_{c}(t'-t)/\hbar-{\bf p}^{j}_{c}\cdot{\bf r}/\hbar)}\hat{a}^{\dagger}_{k,{\bf p}_{2}}\hat{a}_{k,{\bf p}_{3}}\hat{a}_{j,{\bf p}_{4}}+\text{h.c.}\bigg\},
\\\label{hpf_4j}\hat{H}_{4,j}'(t)=&\frac{1}{V}\sum_{j}\frac{g_{jj}}{2}\bigg\{\sum_{{\bf p}_1,{\bf p}_2,{\bf p}_3,{\bf p}_4}\delta_{{\bf p}_1+{\bf p}_2,{\bf p}_3+{\bf p}_4}\hat{a}^{\dagger}_{j,{\bf p}_1}\hat{a}^{\dagger}_{j,{\bf p}_2}\hat{a}_{j,{\bf p}_3}\hat{a}_{j,{\bf p}_4}-4\tilde{n}_{jj}\sum_{{\bf p}_1,{\bf p}_2}\delta_{{\bf p}_1,{\bf p}_2}\hat{a}^{\dagger}_{j,{\bf p}_1}\hat{a}_{j,{\bf p}_2}\bigg\},
\\\nonumber\label{hpf_4kj}\hat{H}_{4,kj}'(t)=&\frac{1}{V}\sum_{k\neq j}g_{kj}\bigg\{\sum_{{\bf p}_1,{\bf p}_2,{\bf p}_3,{\bf p}_4}\delta_{{\bf p}_1+{\bf p}_2,{\bf p}_3+{\bf p}_4}\hat{a}^{\dagger}_{j,{\bf p}_1}\hat{a}^{\dagger}_{k,{\bf p}_2}\hat{a}_{k,{\bf p}_3}\hat{a}_{j,{\bf p}_4}-\tilde{n}_{jk}\sum_{{\bf p}_1,{\bf p}_2}\delta_{{\bf p}_1,{\bf p}_2}\hat{a}^{\dagger}_{j,{\bf p}_1}\hat{a}_{k,{\bf p}_2}\\&-\tilde{n}_{kj}\sum_{{\bf p}_1,{\bf p}_2}\delta_{{\bf p}_1,{\bf p}_2}\hat{a}^{\dagger}_{k,{\bf p}_1}\hat{a}_{j,{\bf p}_2}\bigg\}.
\end{align}
\end{subequations}
The expressions given above by Eqs. \eqref{hpf_1j}-\eqref{hpf_4kj} allow us to derive the collisional integrals in the body of the text. 

\section{Anomalous Pair Correlation Functions}\label{app_a}
In this appendix the anomalous pair correlation functions of the form $\langle\hat{\delta}_{j}\hat{\delta}_{j}\rangle$ and $\langle\hat{\delta}_{k}\hat{\delta}_{j}\rangle$ appearing in Eq. \eqref{eom_exp} are calculated. 
Following the steps detailed in the body of the paper, one can shown that the pair anomalous terms take the form
\begin{align}
&\langle\hat{\delta}_{k}\hat{\delta}_{j}\rangle=-i\pi g_{kj}\phi_{k}\phi_{j}(1{+}\delta_{kj})\frac{1}{2}\sum_{{\bf p}_1,{\bf p}_2}\delta(\varepsilon^{j}_{c}{+}\varepsilon^{k}_{c}{-}\varepsilon^{k}_{1}{-}\varepsilon^{j}_{2})\delta_{{\bf p}^{j}_{c}+{\bf p}^{k}_{c},{\bf p}_1+{\bf p}_2}\bigg[(f^{k}_{1}+1)(f^{j}_{2}+1)-f^{k}_{1}f^{j}_{2}\bigg].\label{anom_pairdkj}
\end{align}

By taking the continuum limit, one finds that Eq. \eqref{anom_pairdkj} becomes the collisional integral
\begin{align}
&\langle\hat{\delta}_{k}\hat{\delta}_{j}\rangle=-\frac{ig_{kj}\phi_j\phi_k}{4(2\pi)^2\hbar^3}(1{+}\delta_{kj})\int{d{\bf p}_1}\int{d{\bf p}_2}\delta({{\bf p}^{j}_{c}}+{{\bf p}^{k}_{c}}-{{\bf p}_1}-{{\bf p}_2})\delta({\varepsilon^{j}_{c}}+{\varepsilon^{k}_{c}}-{\varepsilon^{k}_{1}}-{\varepsilon^{j}_{2}})\bigg[(f^{k}_{1}{+}1)(f^{j}_{2}{+}1){-}f^{k}_{1}f^{j}_{2}\bigg].\label{anom_pair}
\end{align}
valid for both $j{=}k$ and $j\neq k$ This integral describes a collision process whereby a pair of condensate particles collide to give two thermal particles and it's inverse process. Such terms have not been included in our theory as they violate energy conservation. For this reason, these terms are conventionally dropped due to the Popov approximation \cite{griffin_1996}. Figure \ref{fig1} shows schematically the different kinetic scattering processes represented by Eq. \ref{anom_pair}. Squares represent condensate and circles represent non-condensate (thermal) atoms, while the blue and red colors represent the $a$ and $b$ components respectively. 
\begin{figure}[h]
\centering
\includegraphics[scale=1]{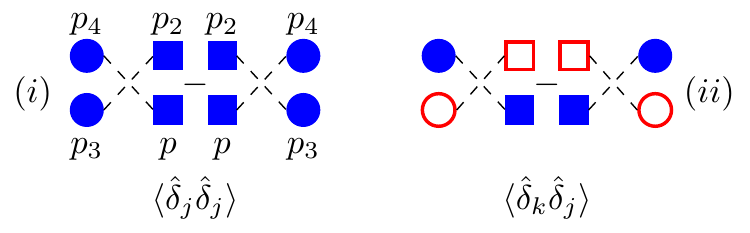}
\caption{(Color online) The two diagrams represent kinetic: (i) {\it intra}-component $\langle\hat{\delta}_{a}\hat{\delta}_{a}\rangle$; (ii) inter-component: $\langle\hat{\delta}_{a}\hat{\delta}_{b}\rangle$ pair anomalous scattering terms of Eq. \eqref{anom_pair}. Corresponding diagrams for the $b$ component are obtained by interchanging the (open) red and (closed) blue colors appearing above.}
\label{fig1}
\end{figure}

\section{\label{appendix:avgs}Evaluation of Non-Equilibrium averages}
Here, the form of various non-equilibrium averages used in the intermediate steps of deriving the full collisional theory are listed. We begin by writing out in full the triplet anomalous averages which are used in the definitions of the condensate growth terms $R^{jj}$ and $R^{kj}$ as
\begin{equation}  
\langle\hat{\delta}^{\dagger}_{k}\hat{\delta}_{k}\hat{\delta}_{j}\rangle_{\text{(1)}}{=}{-}\frac{i}{\hbar}\int\limits^{t}_{t_0}{dt'}\langle\hat{S}^{\dagger}(t',t_0)[\hat{S}^{\dagger}(t,t')\hat{\delta}^{\dagger}_{k}\hat{\delta}_{k}\hat{\delta}_{j}\hat{S}(t,t'),\hat{H}_{1,kj}'(t)]\hat{S}(t',t_0)\rangle\label{sing},
\end{equation}
and
\begin{equation}
\langle\hat{\delta}^{\dagger}_{k}\hat{\delta}_{k}\hat{\delta}_{j}\rangle_{\text{(3)}}{=}{-}\frac{i}{\hbar}\int\limits^{t}_{t_0}{dt'}\langle\hat{S}^{\dagger}(t',t_0)[\hat{S}^{\dagger}(t,t')\hat{\delta}^{\dagger}_{k}\hat{\delta}_{k}\hat{\delta}_{j}\hat{S}(t,t'),\hat{H}_{3,kj}'(t)]\hat{S}(t',t_0)\rangle.\label{trip}
\end{equation}
with similar expressions for $j=k$. Meanwhile, the condensate exchange terms $\mathds{R}^{kj}$ are calculated as (for $j \neq k$)
\begin{equation}
\langle\hat{\delta}^{\dagger}_{k}\hat{\delta}_{j}\rangle{=}{-}\frac{i}{\hbar}\int\limits^{t}_{t_0} dt'\langle\hat{S}^{\dagger}(t',t_0)[\hat{S}^{\dagger}(t,t')\hat{\delta}^{\dagger}_{k}\hat{\delta}_{j}\hat{S}(t,t'),\hat{H}_{2,kj}'(t)]\hat{S}(t',t_0)\rangle.\label{pair}
\end{equation}
This expression, based on symmetry-breaking did not appear in previous works for spinor gases~\cite{nikuni_williams_2003,endo_nikuni_2011}, but does formally appear in the related c-field approach of Ref.~\cite{bradley_blakie_2014}.

Next, we give the form of the quantum Boltzmann contributions $C^{jj}_{12}$ and $C^{kj}_{12}$ required to give the final form of the associated collision integrals. These are
\begin{equation}
C^{jj}_{12}=\frac{1}{i\hbar}\Tr\tilde{\rho}(t,t_0)[\hat{f}^{j}({\bf r},{\bf p},t_0),\hat{H}_{3,j}'(t)],\label{c12_jj1}
\end{equation}
and
\begin{equation}
C^{kj}_{12}=\frac{1}{i\hbar}\Tr\tilde{\rho}(t,t_0)[\hat{f}^{j}({\bf r},{\bf p},t_0),\hat{H}_{3,kj}'(t)].\label{c12_def1}
\end{equation}
Using the Fourier transform of the Wigner operator
\begin{equation}
\hat{f}^{j}({\bf r},{\bf p},t_0)=\sum_{\bf q}\hat{a}^{\dagger}_{j,{\bf p}-{\bf q}/2}\hat{a}_{j,{\bf p}+{\bf q}/2}e^{i{\bf r}\cdot{\bf q}/\hbar},\label{wig_mom1}
\end{equation}
along with $\hat{H}_{3,kj}'(t)$ (Eq.~\eqref{hpf_3kj}), it can be shown that the right hand side of Eq. \eqref{c12_def1} can be written as
\begin{align}\nonumber
&C^{kj}_{12}=\frac{g_{kj}}{i\hbar\sqrt{V}}\sum_{\bf q}e^{i{\bf q}\cdot{\bf r}/\hbar}\sum_{{\bf p}_2,{\bf p}_3,{\bf p}_4}\bigg\{\delta_{{\bf p}^{k}_{c}+{\bf p}_2,{\bf p}_3+{\bf p}_4}\phi^{*}_{k}\bigg(\delta_{{\bf p}_2,{\bf p}+{\bf q}/2}\langle\hat{a}^{\dagger}_{j,{\bf p}-{\bf q}/2}\hat{a}_{j,{\bf p}_3}\hat{a}_{k,{\bf p}_4}\rangle-\delta_{{\bf p}_3,{\bf p}-{\bf q}/2}\\&\times\langle\hat{a}^{\dagger}_{j,{\bf p}_2}\hat{a}_{j,{\bf p}+{\bf q}/2}\hat{a}_{k,{\bf p}_4}\rangle\bigg)-\delta_{{\bf p}^{j}_{c}+{\bf p}_2,{\bf p}_3+{\bf p}_4}\phi^{*}_{j}\delta_{{\bf p}_4,{\bf p}-{\bf q}/2}\langle\hat{a}^{\dagger}_{k,{\bf p}_2}\hat{a}_{k,{\bf p}_3}\hat{a}_{j,{\bf p}+{\bf q}/2}\rangle-\text{h.c.}\bigg\}.\label{c12_calc}
\end{align}
The expression given above contains two distinct types of three-field correlation functions, both of which can be computed using the definition of the non-equilibrium average along with the relevant contribution from the perturbing Hamiltonian, Eq. \eqref{hpf_3kj}. One can in particular show that
\begin{align}
&\langle\hat{a}^{\dagger}_{j,{\bf p}_2}\hat{a}_{j,{\bf p}_3}\hat{a}_{l,{\bf p}_4}\rangle=\frac{\pi}{i\sqrt{V}}g_{lj}\phi_{l}(1{+}\delta_{lj})\delta(\varepsilon^{l}_{c}{+}\varepsilon^{j}_{p_2}{-}\varepsilon^{j}_{p_3}{-}\varepsilon^{l}_{p_4})\delta_{{\bf p}^{l}_{c}+{\bf p}_2,{\bf p}_3+{\bf p}_4}\bigg[f^{j}_{2}(f^{j}_{3}{+}1)(f^{l}_{4}{+}1){-}(f^{j}_{2}{+}1)f^{j}_{3}f^{l}_{4}\bigg],\label{3corr}
\end{align}
and the corresponding expression for $\langle\hat{a}^{\dagger}_{l,{\bf p}_4}\hat{a}^{\dagger}_{j,{\bf p}_3}\hat{a}_{j,{\bf p}_2}\rangle$ can be obtained by taking the Hermitian conjugate of Eq. \eqref{3corr}. To simplify the first summation over the center of mass momentum ${\bf q}$ appearing in Eq. \eqref{c12_calc}, we can expand the sum and note that terms with ${\bf q}\neq0$ make zero overall contribution. After which, upon inserting Eq. \eqref{3corr} into Eq. \eqref{c12_calc} results in the final expression given by Eq. \eqref{c12jk} of Sec. \ref{sec:c12}.

Meanwhile, the exchange collisional term $\mathds{C}^{kj}_{12}$ is computed as (for $j\neq k$)
\begin{align}\label{eqn:exC12}
\mathds{C}^{kj}_{12}=\frac{1}{i\hbar}\Tr\tilde{\rho}(t,t_0)[\hat{f}^{j}({\bf r},{\bf p},t_0),\hat{H}_{2,kj}'(t)],
\end{align}
which after using the definition $\hat{H}_{2,kj}'(t)$ given by Eq.~\eqref{hpf_2kj} gives
\begin{align}
&\mathds{C}^{kj}_{12}=\frac{2g_{kj}}{i\hbar}\sum_{{\bf p}_1,{\bf p}_2}\delta_{{\bf p}_{c}^{j}+{\bf p}_{1},{\bf p}^{k}_{c}+{\bf p}_{2}}\bigg(\phi_{j}\phi_{k}^{*}\langle\hat{a}^{\dagger}_{j,{\bf p}_2}\hat{a}_{k,{\bf p}_1}\rangle-\phi_{j}^{*}\phi_{k}\langle\hat{a}^{\dagger}_{k,{\bf p}_1}\hat{a}_{j,{\bf p}_2}\rangle\bigg).\label{cC12kj}
\end{align}
Then, two quadratic correlation functions are required in order to write down a final expression for the collision integral, Eq. \eqref{cC12kj}. As before we can use the expression for the non-equilibrium average, Eq. \eqref{avg_approx} along with the relevant part of the perturbing Hamiltonian, Eq. \eqref{hpf_2kj}. This gives
\begin{align}
&\langle\hat{a}^{\dagger}_{k,{\bf p}_{1}}\hat{a}_{j,{\bf p}_{2}}\rangle=-i\pi g_{kj}\phi_{j}\phi^{*}_{k}\delta(\varepsilon^{j}_{c}+\varepsilon^{k}_{p_1}-\varepsilon^{k}_{c}-\varepsilon^{j}_{p_2})\delta_{{\bf p}^{j}_{c}+{\bf p}_{1},{\bf p}^{k}_{c}+{\bf p}_{2}}\bigg[(f^{j}_{2}+1)f^{k}_{1}-f^{j}_{2}(f^{k}_{1}+1)\bigg].\label{2corr}
\end{align}
Inserting Eq. \eqref{2corr} into \eqref{cC12kj} above allows us to write down the discrete form of the exchange collision integral, Eq. \eqref{c12jk2} of Sec. \ref{sec:c12e} of the text.

The final non-equilibrium average that is required are those terms describing scattering exclusively between thermal atoms, $C^{jj}_{22}$ and $C^{kj}_{22}$, which is given by the expressions
\begin{equation}
C^{jj}_{22}=\frac{1}{i\hbar}\Tr\tilde{\rho}(t,t_0)[\hat{f}^{j}({\bf r},{\bf p},t_0),\hat{H}_{4,j}'(t)],\label{c22_j1}
\end{equation}
and
\begin{equation}
C^{kj}_{22}=\frac{1}{i\hbar}\Tr\tilde{\rho}(t,t_0)[\hat{f}^{j}({\bf r},{\bf p},t_0),\hat{H}_{4,kj}'(t)]\label{c22_kj1}.
\end{equation}
Proceeding, we can write down an expression for $C^{kj}_{22}$ using $\hat{H}_{4,kj}'(t)$ (Eq.~\eqref{hpf_4kj}) which takes the form
\begin{align}
&C_{22}^{kj}={\frac{g_{kj}}{2i\hbar V}}\sum_{{\bf p}_1{\bf p}_2{\bf p}_3{\bf p}_4}\delta_{{\bf p}_1+{\bf p}_2,{\bf p}_3+{\bf p}_4}\bigg([\delta_{{\bf p}_1,{\bf p}}{-}\delta_{{\bf p}_4,{\bf p}}]\langle\hat{a}^{\dagger}_{j,{\bf p}_1}\hat{a}_{k,{\bf p}_2}\hat{a}_{k,{\bf p}_3}\hat{a}_{j,{\bf p}_4}\rangle{+}[\delta_{{\bf p}_2,{\bf p}}{-}\delta_{{\bf p}_3,{\bf p}}]\langle\hat{a}^{\dagger}_{k,{\bf p}_1}\hat{a}^{\dagger}_{j,{\bf p}_2}\hat{a}_{j,{\bf p}_3}\hat{a}_{k,{\bf p}_4}\rangle\bigg).
\label{c22jkd}
\end{align}
Again to simplify Eq. \eqref{c22jkd}, we require the quartic correlation function comprised of equal numbers of creation and annihilation operators. This is computed using the definition given by Eq. \eqref{avg_approx} along with $\hat{H}_{4,kj}'(t)$ as defined by Eq. \eqref{hpf_4kj}. This gives 
\begin{equation}
\langle\hat{a}^{\dagger}_{j,{\bf p}_1}\hat{a}^{\dagger}_{k,{\bf p}_2}\hat{a}_{k,{\bf p}_3}\hat{a}_{j,{\bf p}_4}\rangle=\frac{\pi g_{kj}}{iV}(1{+}\delta_{kj})\delta(\varepsilon^{j}_{p_1}{+}\varepsilon^{k}_{p_2}{-}\varepsilon^{k}_{p_3}{-}\varepsilon^{j}_{p_4})\delta_{{\bf p}_1+{\bf p}_2,{\bf p}_3+{\bf p}_4}[f^{j}_{1}f^{k}_{2}(f^{k}_{3}{+}1)(f^{j}_{4}{+}1){-}(f^{j}_{1}{+}1)(f^{k}_{2}{+}1)f^{k}_{3}f^{j}_{4}].\label{4corr}
\end{equation} 
Interchanging the dummy summation variables over momentum in Eq. \eqref{c22jkd}, and inserting the expression given by Eq. \eqref{4corr} leads to Eq. \eqref{c22jk} of Sec. \ref{sec:c22} of the main text. 

To obtain all of the collision integrals appearing in the body of the text, one also requires the following approximations 
\begin{equation}
\hat{S}^{\dagger}(t,t')\hat{a}_{n,{\bf p}}\hat{S}(t,t')\simeq\hat{a}_{n,{\bf p}}\exp(-i\varepsilon^{n}_{\bf p}(t-t')/\hbar)
\end{equation}
as well as
\begin{equation}
\langle\hat{a}^{\dagger}_{n,{\bf p}}\hat{a}_{n',{\bf p}'}\rangle_{t_0}=\delta_{n,n'}\delta_{{\bf p},{\bf p}'}f^{n}({\bf r},{\bf p},t),
\end{equation}
and taking the continuum limit requires replacing the summations with
\begin{equation}
\frac{1}{V}\sum_{\bf p}\rightarrow\int\frac{d{\bf p}}{(2\pi\hbar)^3}\hspace{1cm}\text{and}\hspace{1cm}V\delta_{{\bf p},{\bf 0}}\rightarrow(2\pi\hbar)^3\delta({\bf p}).
\end{equation}

Finally, in evaluating the integrals over time appearing in all of the collisional integrals, the following identity has been used
\begin{equation}
\frac{1}{\hbar}\int\limits^{t}_{-\infty}dt' e^{i(\varepsilon^{j}_{p_1}+\varepsilon^{k}_{p_2}-\varepsilon^{l}_{p_3}-\varepsilon^{m}_{p_4})(t-t')/\hbar}=\pi\delta(\varepsilon^{j}_{p_1}+\varepsilon^{k}_{p_2}-\varepsilon^{l}_{p_3}-\varepsilon^{m}_{p_4})+i\mathcal{P}\bigg(\frac{1}{\varepsilon^{j}_{p_1}+\varepsilon^{k}_{p_2}-\varepsilon^{l}_{p_3}-\varepsilon^{m}_{p_4}}\bigg),\label{eqn:cpv}
\end{equation}
where $\mathcal{P}(\dots)$ represents the Cauchy principle value in Eq. \eqref{eqn:cpv} above, and is conventially dropped \cite{nikuni_williams_2003,endo_nikuni_2011}. 

\section{\label{appendix:com_transform}Transformation between lab frame and centre-of-mass frame}
In the numerical evaluation of the collisional rates~\eqref{gamma22}, \eqref{gamma12} and \eqref{eq:gamma12_in}, it is useful to transform the momenta from the lab frame to the centre-of-mass frame. To this end, we define the center-of-mass and relative momenta, ($\vec{P}$, $\vec{p}_r$) and ($\vec{P}'$, $\vec{p}'_r$), such that 
\begin{subequations}
\label{com_transform}
\begin{align}
  \begin{pmatrix}\vec{P} \\ \vec{p}_r\end{pmatrix} = &\begin{pmatrix}1 & 1 \\ \frac{m_k}{m_j+m_k} & -\frac{m_j}{m_j+m_k}\end{pmatrix} \begin{pmatrix} \vec{p}_1 \\ \vec{p}_2 \end{pmatrix},\label{com_transform_a}\\
    \begin{pmatrix}\vec{P}' \\ \vec{p}'_r\end{pmatrix} = &\begin{pmatrix}1 & 1 \\ \frac{m_k}{m_j+m_k} & -\frac{m_j}{m_j+m_k}\end{pmatrix} \begin{pmatrix} \vec{p}_4 \\ \vec{p}_3 \end{pmatrix}.\label{com_transform_b}
\end{align}
\end{subequations}
The two Dirac delta functions then enforce $\vec{P}=\vec{P}'$, $|\vec{p}_r|=|\vec{p}'_r|$ because of energy and momentum conservation. 

The collision rates between non-condensed atoms (for both $k=j$ and $k\neq j$) is
\begin{equation}
\Gamma^{kj,{\rm out}}_{22}=\frac{(1{+}\delta_{kj})g_{kj}^2}{(2\pi)^8\hbar^{10}}\int d\vec{p}_1 \int d{\bf p}_2\int d{\bf p}_3\int d{\bf p}_4\delta({\bf p}_1{+}{\bf p}_2{-}{\bf p}_3{-}{\bf p}_4)\delta(\varepsilon^{j}_{p_1}{+}\varepsilon^{k}_{p_2}{-}\varepsilon^{k}_{p_3}{-}\varepsilon^{j}_{p_4})f_1^{j}f^{k}_{2}(f^{k}_{3}{+}1)(f^{j}_{4}{+}1),\label{orig_gamma22}
\end{equation}
which is conveniently evaluated with Eq. \eqref{com_transform} to obtain
\begin{equation}
\Gamma^{kj,{\rm out}}_{22}=\int \frac{d\vec{p}_1}{(2\pi\hbar)^3} f_1^{j} \int \frac{d{\bf p}_2}{(2\pi\hbar)^3} f^{k}_{2}\int \frac{d\Omega}{4\pi} \sigma_{kj} |\vec{v}_1-\vec{v}_2|(f^{k}_{3}+1)(f^{j}_{4}+1)
\end{equation}
where $\sigma_{kj}=(1+\delta_{kj})4\pi a_{kj}^2$ is the cross section, $\vec{v}_1$ and $\vec{v}_2$ are the initial velocities of atoms $j$ and $k$ respectively, $\Omega$ specifies the solid angle of the final relative velocity $\vec{v}_4-\vec{v}_3$. 

For collisions between condensate and non-condensate atoms, we first consider the $C^{kj}_{12}$ process (for both $k=j$ and $k\neq j$), the ``out'' collision rate that represents the scattering of a non-condensed atom from a condensate to produce two non-condensed atoms is given by
\begin{equation}
 \Gamma_{12}^{kj,{\rm out}}=\frac{(1{+}\delta_{kj})g_{kj}^{2}}{(2\pi)^5\hbar^7}n_{c,j}\ \int d{\bf p}_{2}\int d{\bf p}_{3}\int d{\bf p}_{4}\delta({\bf p}_{c}^{j}{+}{\bf p}_{2}{-}{\bf p}_{3}{-}{\bf p}_{4})\delta(\varepsilon^{j}_{c}{+}\varepsilon^{k}_{p_2}{-}\varepsilon^{k}_{p_3}{-}\varepsilon^{j}_{p_4}) f^{k}_{2}(f^{k}_{3}{+}1)(f^{j}_{4}{+}1).\label{orig_gamma12}
\end{equation}
We again use \eqref{com_transform} to obtain
\begin{equation}
  \Gamma_{12}^{kj,{\rm out}} = \int\frac{d{\bf p}_2}{(2\pi\hbar)^3}f^{k}_{2}\,n_{c,j}\,\sigma_{kj}v^{\text{out}}_{r}\int\frac{d\Omega}{4\pi}(1+ f^{k}_{3}+f^{j}_{4}),
\end{equation}
where $v^{\text{out}}_r = \sqrt{|\vec{v}_{c,j}-\vec{v}_2|^2-2(U_{\text{n}}^j-\mu_c^j)/m_{kj}}$ is the relative speed of the initial states corrected to take into account the local conservation of energy. 

The reverse process, where two non-condensed atoms collide such that one of them goes into a condensate, is given by the ``in'' rate as
\begin{equation}
  \Gamma_{12}^{kj,{\rm in}}=\frac{(1{+}\delta_{kj})g_{kj}^{2}}{(2\pi)^5\hbar^7}n_{c,j}\ \int d{\bf p}_{2}\int d{\bf p}_{3}\int d{\bf p}_{4}\delta({\bf p}_{c}^{j}{+}{\bf p}_{2}{-}{\bf p}_{3}{-}{\bf p}_{4})\delta(\varepsilon^{j}_{c}{+}\varepsilon^{k}_{p_2}{-}\varepsilon^{k}_{p_3}{-}\varepsilon^{j}_{p_4})(1{+}f^{k}_{2})f^{k}_{3}f^{j}_{4}. \label{eq:orig_gamma12_in}
\end{equation}
In order to reduce Eq.~\eqref{eq:orig_gamma12_in} into a useful form for Monte Carlo sampling as well as dynamical simulations, we follow the approach of Jackson and Zaremba~\cite{jackson_zaremba_2002} and arrive at
\begin{align}
  \Gamma_{12}^{kj,{\rm in}} = \int\frac{d{\bf p}_4}{(2\pi\hbar)^3}f^{j}_{4}\frac{\,n_{c,j}\,\sigma_{kj}(m_k/m_{kj})^3}{4\pi(1+m_j/m_k)|\vec{v}_r^{\rm in}|}\int d\tilde{\vec{v}} f_3^k,
\end{align}
where $\vec{v}_r^{\rm in}=\vec{v}_4^j-\vec{v}_c^j$ is the velocity of thermal atom $j$ relative to the local condensate velocity while the second integral is a two-dimensional integral over the velocity vector $\tilde{\vec{v}}$ which is in a plane normal to $\vec{v}_r^{\rm in}$. The velocity of the other incoming thermal atom $\vec{v}_3^k$ is then given by
\begin{align}
  \vec{v}_3^k = \vec{v}_c^j + \frac{(1-m_j/m_k)}{2}\vec{v}_r^{\rm in} + \tilde{\vec{v}} + \frac{(U_{\rm n}^j-\mu_c^j)\hat{\vec{v}}_r^{\rm in}}{m_j|\vec{v}_r^{\rm in}|},
\end{align}
and the outgoing velocity of the thermal atom is given by
\begin{align}
  \vec{v}_2^k = (m_j/m_k) \vec{v}_r^{\rm in} + \vec{v}_3^k.
\end{align}
Note that we follow~\cite{jackson_zaremba_2002} and drop the cubic term $f_2 f_3 f_4$ in numerical simulations because it cancels exactly between the ``in'' and ``out'' rates.

For the exchange collisions~\eqref{gamma_xc}, instead of the usual center-of-mass transformation~\eqref{com_transform}, we use an alternative transformation
\begin{subequations}
\label{effcom_transform}
\begin{align}
  \begin{pmatrix}\tilde{\vec{P}} \\ \tilde{\vec{p}}_r\end{pmatrix} = &\begin{pmatrix}1 & -1 \\ \frac{m_j}{m_j-m_k} & -\frac{m_k}{m_j-m_k}\end{pmatrix} \begin{pmatrix} \vec{p}_1 \\ \vec{p}_2 \end{pmatrix},\label{effcom_transform_a}\\
    \begin{pmatrix}\tilde{\vec{P}}' \\ \tilde{\vec{p}}'_r\end{pmatrix} = &\begin{pmatrix}1 & -1 \\ \frac{m_j}{m_j-m_k} & -\frac{m_k}{m_j-m_k}\end{pmatrix} \begin{pmatrix} \vec{p}_c^k \\ \vec{p}_c^j \end{pmatrix}.\label{effcom_transform_b}
\end{align}
\end{subequations}

The exchange collision rate is therefore
\begin{align}
\Gamma^{\rm out}_{\mathds{C}}=&\frac{g_{kj}^{2}}{(2\pi)^2\hbar^4}\ n_{c,k}\,n_{c,j}\,\int d{\bf p}_{1}\int d{\bf p}_{2} \delta({\bf p}_{c}^{j}{+}{\bf p}_{1}{-}{\bf p}_{c}^{k}{-}{\bf p}_{2})\delta(\varepsilon^{j}_{c}{+}\varepsilon^{k}_{p_1}{-}\varepsilon^{k}_{c}{-}\varepsilon^{j}_{p_2}
)f^{j}_{2}(f^{k}_{1}+1)\nonumber\\
=&\sigma_{kj}\left(\frac{\mathcal{M}_{kj}}{m_{kj}}\right)^2 n_{c,k}\,n_{c,j} \tilde{v}_r\int\frac{d\Omega}{4\pi} f^{j}_{2}(f^{k}_{1}+1),\label{eqn:gamma_xc}
\end{align}
where $\mathcal{M}_{kj}^{-1}=m_k^{-1}-m_j^{-1}$ plays the role of an effective reduced mass while the effective relative speed is 
\begin{equation}
\tilde{v}_r{=}\sqrt{|\vec{v}_{c,j}{-}\vec{v}_{c,k}|^2{-}2([U_{\text{n}}^k{-}\mu_c^k]{-}[U_{\text{n}}^j{-}\mu_c^j])/\mathcal{M}_{kj}}. 
\end{equation}

\section{\label{appendix:gamma_calc}Equilibrium evaluation of $\Gamma_{\mathds{C}}^{kj}$}
In general the collision integrals cannot be evaluated analytically. However, the exchange rate, Eq. \eqref{eqn:gamma_xc} can be calculated in the limit $\beta_0(\frac{p^2}{2m_j}+U_{c}^{j}-\mu^{j}_{c})\ll1$. Using the momentum transformations given by \eqref{effcom_transform_a} and \eqref{effcom_transform_b}, the scattering rate $\Gamma^{\rm out}_{\mathds{C}}$ becomes
\begin{align}
\Gamma^{\rm out}_{\mathds{C}}&=\frac{2\mathcal{M}_{kj}g_{kj}^{2}}{(2\pi)^2\hbar^4}n_{c,j}n_{c,k}{p}^{\rm out}\int d\Omega(f^{k}_{1}+1)f^{j}_{2},\simeq\frac{4\pi\mathcal{M}_{kj}g_{kj}^{2}}{(2\pi)^2\hbar^4}n_{c,j}n_{c,k}p^{\rm out}(c_1T+c_2T^2),
\label{appendixD:G}
\end{align}
where the constants $c_i$ are defined as
\begin{align}\label{appendixD:c1}
c_1&=\frac{k_{B}}{\delta}\ln\bigg[\frac{\gamma^{j}+\delta+g_{jj}n_{c,j}}{\gamma^{j}-\delta+g_{jj}n_{c,j}}\bigg],\\
c_2&=\frac{k_{B}^{2}}{\delta}\frac{\ln\bigg(\frac{[\gamma^{j}-\delta+g_{jj}n_{c,j}][\gamma^{k}+\delta+g_{kk}n_{c,k}]}{[\gamma^{j}+\delta+g_{jj}n_{c,j}][\gamma^{k}-\delta+g_{kk}n_{c,k}]}\bigg)}{\gamma^{j}-\gamma^{k}+g_{jj}n_{c,j}-g_{kk}n_{c,k}}.\label{appendixD:c2}
\end{align}
The constants $\gamma^{\alpha}$ and $\delta$ are defined in terms of the momenta as
\begin{align}
\gamma^{\alpha}&=\frac{{p^{\text{out}}}^2}{2m_{\alpha}}+\frac{m_\alpha}{(m_j-m_k)^2}\frac{p^{2}_{0}}{2},\\
\delta&=\frac{1}{m_j-m_k}|p^{\text{out}}||p_0|.
\end{align}
Finally, the `relative' momentum is defined as $p^{\text{out}}=\sqrt{p^2+2\mathcal{M}_{kj}(g_{jj}n_{c,j}-g_{kk}n_{c,k})}$. Equation \eqref{appendixD:G} shows the temperature dependence of $\Gamma^{kj}_{\mathds{C}}$, while the constants $c_1$ and $c_2$ defined by Eq. \eqref{appendixD:c1} and \eqref{appendixD:c2} give respectively the general form of the coefficients of $T^2$ and $T$.
\end{widetext}

\bibliography{2comp_kinetic_long_041115}

\begin{thebibliography}{156}%
\makeatletter
\providecommand \@ifxundefined [1]{%
 \@ifx{#1\undefined}
}%
\providecommand \@ifnum [1]{%
 \ifnum #1\expandafter \@firstoftwo
 \else \expandafter \@secondoftwo
 \fi
}%
\providecommand \@ifx [1]{%
 \ifx #1\expandafter \@firstoftwo
 \else \expandafter \@secondoftwo
 \fi
}%
\providecommand \natexlab [1]{#1}%
\providecommand \enquote  [1]{``#1''}%
\providecommand \bibnamefont  [1]{#1}%
\providecommand \bibfnamefont [1]{#1}%
\providecommand \citenamefont [1]{#1}%
\providecommand \href@noop [0]{\@secondoftwo}%
\providecommand \href [0]{\begingroup \@sanitize@url \@href}%
\providecommand \@href[1]{\@@startlink{#1}\@@href}%
\providecommand \@@href[1]{\endgroup#1\@@endlink}%
\providecommand \@sanitize@url [0]{\catcode `\\12\catcode `\$12\catcode
  `\&12\catcode `\#12\catcode `\^12\catcode `\_12\catcode `\%12\relax}%
\providecommand \@@startlink[1]{}%
\providecommand \@@endlink[0]{}%
\providecommand \url  [0]{\begingroup\@sanitize@url \@url }%
\providecommand \@url [1]{\endgroup\@href {#1}{\urlprefix }}%
\providecommand \urlprefix  [0]{URL }%
\providecommand \Eprint [0]{\href }%
\providecommand \doibase [0]{http://dx.doi.org/}%
\providecommand \selectlanguage [0]{\@gobble}%
\providecommand \bibinfo  [0]{\@secondoftwo}%
\providecommand \bibfield  [0]{\@secondoftwo}%
\providecommand \translation [1]{[#1]}%
\providecommand \BibitemOpen [0]{}%
\providecommand \bibitemStop [0]{}%
\providecommand \bibitemNoStop [0]{.\EOS\space}%
\providecommand \EOS [0]{\spacefactor3000\relax}%
\providecommand \BibitemShut  [1]{\csname bibitem#1\endcsname}%
\let\auto@bib@innerbib\@empty
\bibitem [{\citenamefont {Weiner}\ \emph {et~al.}(1999)\citenamefont {Weiner},
  \citenamefont {Bagnato}, \citenamefont {Zilio},\ and\ \citenamefont
  {Julienne}}]{weiner_bagnato_1999}%
  \BibitemOpen
  \bibfield  {author} {\bibinfo {author} {\bibfnamefont {J.}~\bibnamefont
  {Weiner}}, \bibinfo {author} {\bibfnamefont {V.~S.}\ \bibnamefont {Bagnato}},
  \bibinfo {author} {\bibfnamefont {S.}~\bibnamefont {Zilio}}, \ and\ \bibinfo
  {author} {\bibfnamefont {P.~S.}\ \bibnamefont {Julienne}},\ }\href@noop {}
  {\bibfield  {journal} {\bibinfo  {journal} {Rev. Mod. Phys.}\ }\textbf
  {\bibinfo {volume} {71}},\ \bibinfo {pages} {1} (\bibinfo {year}
  {1999})}\BibitemShut {NoStop}%
\bibitem [{\citenamefont {Leggett}(2001)}]{leggett_2001}%
  \BibitemOpen
  \bibfield  {author} {\bibinfo {author} {\bibfnamefont {A.~J.}\ \bibnamefont
  {Leggett}},\ }\href@noop {} {\bibfield  {journal} {\bibinfo  {journal} {Rev.
  Mod. Phys.}\ }\textbf {\bibinfo {volume} {73}},\ \bibinfo {pages} {307}
  (\bibinfo {year} {2001})}\BibitemShut {NoStop}%
\bibitem [{\citenamefont {{Lewenstein}}\ \emph {et~al.}(2007)\citenamefont
  {{Lewenstein}}, \citenamefont {{Sanpera}}, \citenamefont {{Ahufinger}},
  \citenamefont {{Damski}}, \citenamefont {{Sen}},\ and\ \citenamefont
  {{Sen}}}]{lewenstein_sanpera_2007}%
  \BibitemOpen
  \bibfield  {author} {\bibinfo {author} {\bibfnamefont {M.}~\bibnamefont
  {{Lewenstein}}}, \bibinfo {author} {\bibfnamefont {A.}~\bibnamefont
  {{Sanpera}}}, \bibinfo {author} {\bibfnamefont {V.}~\bibnamefont
  {{Ahufinger}}}, \bibinfo {author} {\bibfnamefont {B.}~\bibnamefont
  {{Damski}}}, \bibinfo {author} {\bibfnamefont {A.}~\bibnamefont {{Sen}}}, \
  and\ \bibinfo {author} {\bibfnamefont {U.}~\bibnamefont {{Sen}}},\
  }\href@noop {} {\bibfield  {journal} {\bibinfo  {journal} {Adv. Phys.}\
  }\textbf {\bibinfo {volume} {56}},\ \bibinfo {pages} {243} (\bibinfo {year}
  {2007})}\BibitemShut {NoStop}%
\bibitem [{\citenamefont {Giorgini}\ \emph {et~al.}(2008)\citenamefont
  {Giorgini}, \citenamefont {Pitaevskii},\ and\ \citenamefont
  {Stringari}}]{giorgini_pitaevskii_2008}%
  \BibitemOpen
  \bibfield  {author} {\bibinfo {author} {\bibfnamefont {S.}~\bibnamefont
  {Giorgini}}, \bibinfo {author} {\bibfnamefont {L.~P.}\ \bibnamefont
  {Pitaevskii}}, \ and\ \bibinfo {author} {\bibfnamefont {S.}~\bibnamefont
  {Stringari}},\ }\href@noop {} {\bibfield  {journal} {\bibinfo  {journal}
  {Rev. Mod. Phys.}\ }\textbf {\bibinfo {volume} {80}},\ \bibinfo {pages}
  {1215} (\bibinfo {year} {2008})}\BibitemShut {NoStop}%
\bibitem [{\citenamefont {Bloch}\ \emph {et~al.}(2008)\citenamefont {Bloch},
  \citenamefont {Dalibard},\ and\ \citenamefont
  {Zwerger}}]{bloch_dalibard_2008}%
  \BibitemOpen
  \bibfield  {author} {\bibinfo {author} {\bibfnamefont {I.}~\bibnamefont
  {Bloch}}, \bibinfo {author} {\bibfnamefont {J.}~\bibnamefont {Dalibard}}, \
  and\ \bibinfo {author} {\bibfnamefont {W.}~\bibnamefont {Zwerger}},\ }\href
  {\doibase 10.1103/RevModPhys.80.885} {\bibfield  {journal} {\bibinfo
  {journal} {Rev. Mod. Phys.}\ }\textbf {\bibinfo {volume} {80}},\ \bibinfo
  {pages} {885} (\bibinfo {year} {2008})}\BibitemShut {NoStop}%
\bibitem [{\citenamefont {Bloch}(2005)}]{bloch_2005}%
  \BibitemOpen
  \bibfield  {author} {\bibinfo {author} {\bibfnamefont {I.}~\bibnamefont
  {Bloch}},\ }\href {http://stacks.iop.org/0953-4075/38/i=9/a=013} {\bibfield
  {journal} {\bibinfo  {journal} {Journal of Physics B: Atomic, Molecular and
  Optical Physics}\ }\textbf {\bibinfo {volume} {38}},\ \bibinfo {pages} {S629}
  (\bibinfo {year} {2005})}\BibitemShut {NoStop}%
\bibitem [{\citenamefont {Morsch}\ and\ \citenamefont
  {Oberthaler}(2006)}]{morsch_oberthaler_2006}%
  \BibitemOpen
  \bibfield  {author} {\bibinfo {author} {\bibfnamefont {O.}~\bibnamefont
  {Morsch}}\ and\ \bibinfo {author} {\bibfnamefont {M.}~\bibnamefont
  {Oberthaler}},\ }\href@noop {} {\bibfield  {journal} {\bibinfo  {journal}
  {Rev. Mod. Phys.}\ }\textbf {\bibinfo {volume} {78}},\ \bibinfo {pages} {179}
  (\bibinfo {year} {2006})}\BibitemShut {NoStop}%
\bibitem [{\citenamefont {Brennen}\ \emph {et~al.}(1999)\citenamefont
  {Brennen}, \citenamefont {Caves}, \citenamefont {Jessen},\ and\ \citenamefont
  {Deutsch}}]{brennen_caves_99}%
  \BibitemOpen
  \bibfield  {author} {\bibinfo {author} {\bibfnamefont {G.~K.}\ \bibnamefont
  {Brennen}}, \bibinfo {author} {\bibfnamefont {C.~M.}\ \bibnamefont {Caves}},
  \bibinfo {author} {\bibfnamefont {P.~S.}\ \bibnamefont {Jessen}}, \ and\
  \bibinfo {author} {\bibfnamefont {I.~H.}\ \bibnamefont {Deutsch}},\ }\href
  {\doibase 10.1103/PhysRevLett.82.1060} {\bibfield  {journal} {\bibinfo
  {journal} {Phys. Rev. Lett.}\ }\textbf {\bibinfo {volume} {82}},\ \bibinfo
  {pages} {1060} (\bibinfo {year} {1999})}\BibitemShut {NoStop}%
\bibitem [{\citenamefont {Jaksch}\ \emph {et~al.}(2000)\citenamefont {Jaksch},
  \citenamefont {Cirac}, \citenamefont {Zoller}, \citenamefont {Rolston},
  \citenamefont {C\^ot\'e},\ and\ \citenamefont {Lukin}}]{jaksch_cirac_00}%
  \BibitemOpen
  \bibfield  {author} {\bibinfo {author} {\bibfnamefont {D.}~\bibnamefont
  {Jaksch}}, \bibinfo {author} {\bibfnamefont {J.~I.}\ \bibnamefont {Cirac}},
  \bibinfo {author} {\bibfnamefont {P.}~\bibnamefont {Zoller}}, \bibinfo
  {author} {\bibfnamefont {S.~L.}\ \bibnamefont {Rolston}}, \bibinfo {author}
  {\bibfnamefont {R.}~\bibnamefont {C\^ot\'e}}, \ and\ \bibinfo {author}
  {\bibfnamefont {M.~D.}\ \bibnamefont {Lukin}},\ }\href {\doibase
  10.1103/PhysRevLett.85.2208} {\bibfield  {journal} {\bibinfo  {journal}
  {Phys. Rev. Lett.}\ }\textbf {\bibinfo {volume} {85}},\ \bibinfo {pages}
  {2208} (\bibinfo {year} {2000})}\BibitemShut {NoStop}%
\bibitem [{\citenamefont {Schmiedmayer}\ \emph {et~al.}(2002)\citenamefont
  {Schmiedmayer}, \citenamefont {Folman},\ and\ \citenamefont
  {Calarco}}]{schmiedmayer_folman_02}%
  \BibitemOpen
  \bibfield  {author} {\bibinfo {author} {\bibfnamefont {J.}~\bibnamefont
  {Schmiedmayer}}, \bibinfo {author} {\bibfnamefont {R.}~\bibnamefont
  {Folman}}, \ and\ \bibinfo {author} {\bibfnamefont {T.}~\bibnamefont
  {Calarco}},\ }\href {\doibase 10.1080/09500340110111077} {\bibfield
  {journal} {\bibinfo  {journal} {Journal of Modern Optics}\ }\textbf {\bibinfo
  {volume} {49}},\ \bibinfo {pages} {1375} (\bibinfo {year}
  {2002})}\BibitemShut {NoStop}%
\bibitem [{\citenamefont {Hall}\ \emph {et~al.}(1998)\citenamefont {Hall},
  \citenamefont {Matthews}, \citenamefont {Ensher}, \citenamefont {Wieman},\
  and\ \citenamefont {Cornell}}]{hall_matthews_1998}%
  \BibitemOpen
  \bibfield  {author} {\bibinfo {author} {\bibfnamefont {D.~S.}\ \bibnamefont
  {Hall}}, \bibinfo {author} {\bibfnamefont {M.~R.}\ \bibnamefont {Matthews}},
  \bibinfo {author} {\bibfnamefont {J.~R.}\ \bibnamefont {Ensher}}, \bibinfo
  {author} {\bibfnamefont {C.~E.}\ \bibnamefont {Wieman}}, \ and\ \bibinfo
  {author} {\bibfnamefont {E.~A.}\ \bibnamefont {Cornell}},\ }\href {\doibase
  10.1103/PhysRevLett.81.1539} {\bibfield  {journal} {\bibinfo  {journal}
  {Phys. Rev. Lett.}\ }\textbf {\bibinfo {volume} {81}},\ \bibinfo {pages}
  {1539} (\bibinfo {year} {1998})}\BibitemShut {NoStop}%
\bibitem [{\citenamefont {Matthews}\ \emph {et~al.}(1999)\citenamefont
  {Matthews}, \citenamefont {Anderson}, \citenamefont {Haljan}, \citenamefont
  {Hall}, \citenamefont {Wieman},\ and\ \citenamefont
  {Cornell}}]{matthews_anderson_1999}%
  \BibitemOpen
  \bibfield  {author} {\bibinfo {author} {\bibfnamefont {M.~R.}\ \bibnamefont
  {Matthews}}, \bibinfo {author} {\bibfnamefont {B.~P.}\ \bibnamefont
  {Anderson}}, \bibinfo {author} {\bibfnamefont {P.~C.}\ \bibnamefont
  {Haljan}}, \bibinfo {author} {\bibfnamefont {D.~S.}\ \bibnamefont {Hall}},
  \bibinfo {author} {\bibfnamefont {C.~E.}\ \bibnamefont {Wieman}}, \ and\
  \bibinfo {author} {\bibfnamefont {E.~A.}\ \bibnamefont {Cornell}},\ }\href
  {\doibase 10.1103/PhysRevLett.83.2498} {\bibfield  {journal} {\bibinfo
  {journal} {Phys. Rev. Lett.}\ }\textbf {\bibinfo {volume} {83}},\ \bibinfo
  {pages} {2498} (\bibinfo {year} {1999})}\BibitemShut {NoStop}%
\bibitem [{\citenamefont {Maddaloni}\ \emph {et~al.}(2000)\citenamefont
  {Maddaloni}, \citenamefont {Modugno}, \citenamefont {Fort}, \citenamefont
  {Minardi},\ and\ \citenamefont {Inguscio}}]{maddaloni_modugno_2000}%
  \BibitemOpen
  \bibfield  {author} {\bibinfo {author} {\bibfnamefont {P.}~\bibnamefont
  {Maddaloni}}, \bibinfo {author} {\bibfnamefont {M.}~\bibnamefont {Modugno}},
  \bibinfo {author} {\bibfnamefont {C.}~\bibnamefont {Fort}}, \bibinfo {author}
  {\bibfnamefont {F.}~\bibnamefont {Minardi}}, \ and\ \bibinfo {author}
  {\bibfnamefont {M.}~\bibnamefont {Inguscio}},\ }\href {\doibase
  10.1103/PhysRevLett.85.2413} {\bibfield  {journal} {\bibinfo  {journal}
  {Phys. Rev. Lett.}\ }\textbf {\bibinfo {volume} {85}},\ \bibinfo {pages}
  {2413} (\bibinfo {year} {2000})}\BibitemShut {NoStop}%
\bibitem [{\citenamefont {Modugno}\ \emph {et~al.}(2002)\citenamefont
  {Modugno}, \citenamefont {Modugno}, \citenamefont {Riboli}, \citenamefont
  {Roati},\ and\ \citenamefont {Inguscio}}]{modugno_modugno_2002}%
  \BibitemOpen
  \bibfield  {author} {\bibinfo {author} {\bibfnamefont {G.}~\bibnamefont
  {Modugno}}, \bibinfo {author} {\bibfnamefont {M.}~\bibnamefont {Modugno}},
  \bibinfo {author} {\bibfnamefont {F.}~\bibnamefont {Riboli}}, \bibinfo
  {author} {\bibfnamefont {G.}~\bibnamefont {Roati}}, \ and\ \bibinfo {author}
  {\bibfnamefont {M.}~\bibnamefont {Inguscio}},\ }\href {\doibase
  10.1103/PhysRevLett.89.190404} {\bibfield  {journal} {\bibinfo  {journal}
  {Phys. Rev. Lett.}\ }\textbf {\bibinfo {volume} {89}},\ \bibinfo {pages}
  {190404} (\bibinfo {year} {2002})}\BibitemShut {NoStop}%
\bibitem [{\citenamefont {Simoni}\ \emph {et~al.}(2003)\citenamefont {Simoni},
  \citenamefont {Ferlaino}, \citenamefont {Roati}, \citenamefont {Modugno},\
  and\ \citenamefont {Inguscio}}]{simoni_ferlaino_2003}%
  \BibitemOpen
  \bibfield  {author} {\bibinfo {author} {\bibfnamefont {A.}~\bibnamefont
  {Simoni}}, \bibinfo {author} {\bibfnamefont {F.}~\bibnamefont {Ferlaino}},
  \bibinfo {author} {\bibfnamefont {G.}~\bibnamefont {Roati}}, \bibinfo
  {author} {\bibfnamefont {G.}~\bibnamefont {Modugno}}, \ and\ \bibinfo
  {author} {\bibfnamefont {M.}~\bibnamefont {Inguscio}},\ }\href {\doibase
  10.1103/PhysRevLett.90.163202} {\bibfield  {journal} {\bibinfo  {journal}
  {Phys. Rev. Lett.}\ }\textbf {\bibinfo {volume} {90}},\ \bibinfo {pages}
  {163202} (\bibinfo {year} {2003})}\BibitemShut {NoStop}%
\bibitem [{\citenamefont {Papp}\ \emph {et~al.}(2008)\citenamefont {Papp},
  \citenamefont {Pino},\ and\ \citenamefont {Wieman}}]{papp_pino_2008}%
  \BibitemOpen
  \bibfield  {author} {\bibinfo {author} {\bibfnamefont {S.~B.}\ \bibnamefont
  {Papp}}, \bibinfo {author} {\bibfnamefont {J.~M.}\ \bibnamefont {Pino}}, \
  and\ \bibinfo {author} {\bibfnamefont {C.~E.}\ \bibnamefont {Wieman}},\
  }\href {\doibase 10.1103/PhysRevLett.101.040402} {\bibfield  {journal}
  {\bibinfo  {journal} {Phys. Rev. Lett.}\ }\textbf {\bibinfo {volume} {101}},\
  \bibinfo {pages} {040402} (\bibinfo {year} {2008})}\BibitemShut {NoStop}%
\bibitem [{\citenamefont {Tojo}\ \emph {et~al.}(2010)\citenamefont {Tojo},
  \citenamefont {Taguchi}, \citenamefont {Masuyama}, \citenamefont {Hayashi},
  \citenamefont {Saito},\ and\ \citenamefont {Hirano}}]{tojo_taguchi_2010}%
  \BibitemOpen
  \bibfield  {author} {\bibinfo {author} {\bibfnamefont {S.}~\bibnamefont
  {Tojo}}, \bibinfo {author} {\bibfnamefont {Y.}~\bibnamefont {Taguchi}},
  \bibinfo {author} {\bibfnamefont {Y.}~\bibnamefont {Masuyama}}, \bibinfo
  {author} {\bibfnamefont {T.}~\bibnamefont {Hayashi}}, \bibinfo {author}
  {\bibfnamefont {H.}~\bibnamefont {Saito}}, \ and\ \bibinfo {author}
  {\bibfnamefont {T.}~\bibnamefont {Hirano}},\ }\href {\doibase
  10.1103/PhysRevA.82.033609} {\bibfield  {journal} {\bibinfo  {journal} {Phys.
  Rev. A}\ }\textbf {\bibinfo {volume} {82}},\ \bibinfo {pages} {033609}
  (\bibinfo {year} {2010})}\BibitemShut {NoStop}%
\bibitem [{\citenamefont {Sugawa}\ \emph {et~al.}(2011)\citenamefont {Sugawa},
  \citenamefont {Yamazaki}, \citenamefont {Taie},\ and\ \citenamefont
  {Takahashi}}]{sugawa_yamazaki_2011}%
  \BibitemOpen
  \bibfield  {author} {\bibinfo {author} {\bibfnamefont {S.}~\bibnamefont
  {Sugawa}}, \bibinfo {author} {\bibfnamefont {R.}~\bibnamefont {Yamazaki}},
  \bibinfo {author} {\bibfnamefont {S.}~\bibnamefont {Taie}}, \ and\ \bibinfo
  {author} {\bibfnamefont {Y.}~\bibnamefont {Takahashi}},\ }\href {\doibase
  10.1103/PhysRevA.84.011610} {\bibfield  {journal} {\bibinfo  {journal} {Phys.
  Rev. A}\ }\textbf {\bibinfo {volume} {84}},\ \bibinfo {pages} {011610}
  (\bibinfo {year} {2011})}\BibitemShut {NoStop}%
\bibitem [{\citenamefont {Lercher}\ \emph {et~al.}(2011)\citenamefont
  {Lercher}, \citenamefont {Takekoshi}, \citenamefont {Debatin}, \citenamefont
  {Schuster}, \citenamefont {Rameshan}, \citenamefont {Ferlaino}, \citenamefont
  {Grimm},\ and\ \citenamefont {N\"agerl}}]{lercher_takekoshi_2011}%
  \BibitemOpen
  \bibfield  {author} {\bibinfo {author} {\bibfnamefont {A.}~\bibnamefont
  {Lercher}}, \bibinfo {author} {\bibfnamefont {T.}~\bibnamefont {Takekoshi}},
  \bibinfo {author} {\bibfnamefont {M.}~\bibnamefont {Debatin}}, \bibinfo
  {author} {\bibfnamefont {B.}~\bibnamefont {Schuster}}, \bibinfo {author}
  {\bibfnamefont {R.}~\bibnamefont {Rameshan}}, \bibinfo {author}
  {\bibfnamefont {F.}~\bibnamefont {Ferlaino}}, \bibinfo {author}
  {\bibfnamefont {R.}~\bibnamefont {Grimm}}, \ and\ \bibinfo {author}
  {\bibfnamefont {H.-C.}\ \bibnamefont {N\"agerl}},\ }\href {\doibase
  10.1140/epjd/e2011-20015-6} {\bibfield  {journal} {\bibinfo  {journal} {Eur.
  Phys. J. D}\ }\textbf {\bibinfo {volume} {65}},\ \bibinfo {pages} {3}
  (\bibinfo {year} {2011})}\BibitemShut {NoStop}%
\bibitem [{\citenamefont {McCarron}\ \emph {et~al.}(2011)\citenamefont
  {McCarron}, \citenamefont {Cho}, \citenamefont {Jenkin}, \citenamefont
  {K\"oppinger},\ and\ \citenamefont {Cornish}}]{mccarron_cho_11}%
  \BibitemOpen
  \bibfield  {author} {\bibinfo {author} {\bibfnamefont {D.~J.}\ \bibnamefont
  {McCarron}}, \bibinfo {author} {\bibfnamefont {H.~W.}\ \bibnamefont {Cho}},
  \bibinfo {author} {\bibfnamefont {D.~L.}\ \bibnamefont {Jenkin}}, \bibinfo
  {author} {\bibfnamefont {M.~P.}\ \bibnamefont {K\"oppinger}}, \ and\ \bibinfo
  {author} {\bibfnamefont {S.~L.}\ \bibnamefont {Cornish}},\ }\href {\doibase
  10.1103/PhysRevA.84.011603} {\bibfield  {journal} {\bibinfo  {journal} {Phys.
  Rev. A}\ }\textbf {\bibinfo {volume} {84}},\ \bibinfo {pages} {011603}
  (\bibinfo {year} {2011})}\BibitemShut {NoStop}%
\bibitem [{\citenamefont {Pasquiou}\ \emph {et~al.}(2013)\citenamefont
  {Pasquiou}, \citenamefont {Bayerle}, \citenamefont {Tzanova}, \citenamefont
  {Stellmer}, \citenamefont {Szczepkowski}, \citenamefont {Parigger},
  \citenamefont {Grimm},\ and\ \citenamefont
  {Schreck}}]{pasquiou_bayerle_2013}%
  \BibitemOpen
  \bibfield  {author} {\bibinfo {author} {\bibfnamefont {B.}~\bibnamefont
  {Pasquiou}}, \bibinfo {author} {\bibfnamefont {A.}~\bibnamefont {Bayerle}},
  \bibinfo {author} {\bibfnamefont {S.~M.}\ \bibnamefont {Tzanova}}, \bibinfo
  {author} {\bibfnamefont {S.}~\bibnamefont {Stellmer}}, \bibinfo {author}
  {\bibfnamefont {J.}~\bibnamefont {Szczepkowski}}, \bibinfo {author}
  {\bibfnamefont {M.}~\bibnamefont {Parigger}}, \bibinfo {author}
  {\bibfnamefont {R.}~\bibnamefont {Grimm}}, \ and\ \bibinfo {author}
  {\bibfnamefont {F.}~\bibnamefont {Schreck}},\ }\href {\doibase
  10.1103/PhysRevA.88.023601} {\bibfield  {journal} {\bibinfo  {journal} {Phys.
  Rev. A}\ }\textbf {\bibinfo {volume} {88}},\ \bibinfo {pages} {023601}
  (\bibinfo {year} {2013})}\BibitemShut {NoStop}%
\bibitem [{\citenamefont {Xiong}\ \emph {et~al.}(2013)\citenamefont {Xiong},
  \citenamefont {Li}, \citenamefont {Wang},\ and\ \citenamefont
  {Wang}}]{xiong_li_2013}%
  \BibitemOpen
  \bibfield  {author} {\bibinfo {author} {\bibfnamefont {D.}~\bibnamefont
  {Xiong}}, \bibinfo {author} {\bibfnamefont {X.}~\bibnamefont {Li}}, \bibinfo
  {author} {\bibfnamefont {F.}~\bibnamefont {Wang}}, \ and\ \bibinfo {author}
  {\bibfnamefont {D.}~\bibnamefont {Wang}},\ }\href@noop {} {} (\bibinfo {year}
  {2013}),\ \Eprint {http://arxiv.org/abs/1305.7091 [cond-mat.quant-gas]}
  {arXiv:1305.7091 [cond-mat.quant-gas]} \BibitemShut {NoStop}%
\bibitem [{\citenamefont {Wacker}\ \emph {et~al.}(2015)\citenamefont {Wacker},
  \citenamefont {J{\o}rgensen}, \citenamefont {Birkmose}, \citenamefont
  {Horchani}, \citenamefont {Ertmer}, \citenamefont {Klempt}, \citenamefont
  {Winter}, \citenamefont {Sherson},\ and\ \citenamefont
  {Arlt}}]{wacker_jorgensen_2015}%
  \BibitemOpen
  \bibfield  {author} {\bibinfo {author} {\bibfnamefont {L.}~\bibnamefont
  {Wacker}}, \bibinfo {author} {\bibfnamefont {N.~B.}\ \bibnamefont
  {J{\o}rgensen}}, \bibinfo {author} {\bibfnamefont {D.}~\bibnamefont
  {Birkmose}}, \bibinfo {author} {\bibfnamefont {R.}~\bibnamefont {Horchani}},
  \bibinfo {author} {\bibfnamefont {W.}~\bibnamefont {Ertmer}}, \bibinfo
  {author} {\bibfnamefont {C.}~\bibnamefont {Klempt}}, \bibinfo {author}
  {\bibfnamefont {N.}~\bibnamefont {Winter}}, \bibinfo {author} {\bibfnamefont
  {J.}~\bibnamefont {Sherson}}, \ and\ \bibinfo {author} {\bibfnamefont
  {J.}~\bibnamefont {Arlt}},\ }\href@noop {} {} (\bibinfo {year} {2015}),\
  \Eprint {http://arxiv.org/abs/1505.07975 [cond-mat.quant-gas]}
  {arXiv:1505.07975 [cond-mat.quant-gas]} \BibitemShut {NoStop}%
\bibitem [{\citenamefont {Stamper-Kurn}\ and\ \citenamefont
  {Ueda}(2013)}]{stamperkurn_ueda_2013}%
  \BibitemOpen
  \bibfield  {author} {\bibinfo {author} {\bibfnamefont {D.~M.}\ \bibnamefont
  {Stamper-Kurn}}\ and\ \bibinfo {author} {\bibfnamefont {M.}~\bibnamefont
  {Ueda}},\ }\href {\doibase 10.1103/RevModPhys.85.1191} {\bibfield  {journal}
  {\bibinfo  {journal} {Rev. Mod. Phys.}\ }\textbf {\bibinfo {volume} {85}},\
  \bibinfo {pages} {1191} (\bibinfo {year} {2013})}\BibitemShut {NoStop}%
\bibitem [{\citenamefont {Stenger}\ \emph {et~al.}(1998)\citenamefont
  {Stenger}, \citenamefont {Inouye}, \citenamefont {Stamper-Kurn},
  \citenamefont {Miesner}, \citenamefont {Chikkatur},\ and\ \citenamefont
  {Ketterle}}]{stenger_inouye_1998}%
  \BibitemOpen
  \bibfield  {author} {\bibinfo {author} {\bibfnamefont {J.}~\bibnamefont
  {Stenger}}, \bibinfo {author} {\bibfnamefont {S.}~\bibnamefont {Inouye}},
  \bibinfo {author} {\bibfnamefont {D.~M.}\ \bibnamefont {Stamper-Kurn}},
  \bibinfo {author} {\bibfnamefont {H.-J.}\ \bibnamefont {Miesner}}, \bibinfo
  {author} {\bibfnamefont {A.~P.}\ \bibnamefont {Chikkatur}}, \ and\ \bibinfo
  {author} {\bibfnamefont {W.}~\bibnamefont {Ketterle}},\ }\href@noop {}
  {\bibfield  {journal} {\bibinfo  {journal} {Nature (London)}\ }\textbf
  {\bibinfo {volume} {396}},\ \bibinfo {pages} {345} (\bibinfo {year}
  {1998})}\BibitemShut {NoStop}%
\bibitem [{\citenamefont {Lewandowski}\ \emph {et~al.}(2003)\citenamefont
  {Lewandowski}, \citenamefont {McGuirk}, \citenamefont {Harber},\ and\
  \citenamefont {Cornell}}]{lewandowski_mcguirk_2003}%
  \BibitemOpen
  \bibfield  {author} {\bibinfo {author} {\bibfnamefont {H.~J.}\ \bibnamefont
  {Lewandowski}}, \bibinfo {author} {\bibfnamefont {J.~M.}\ \bibnamefont
  {McGuirk}}, \bibinfo {author} {\bibfnamefont {D.~M.}\ \bibnamefont {Harber}},
  \ and\ \bibinfo {author} {\bibfnamefont {E.~A.}\ \bibnamefont {Cornell}},\
  }\href {\doibase 10.1103/PhysRevLett.91.240404} {\bibfield  {journal}
  {\bibinfo  {journal} {Phys. Rev. Lett.}\ }\textbf {\bibinfo {volume} {91}},\
  \bibinfo {pages} {240404} (\bibinfo {year} {2003})}\BibitemShut {NoStop}%
\bibitem [{\citenamefont {Schweikhard}\ \emph {et~al.}(2004)\citenamefont
  {Schweikhard}, \citenamefont {Coddington}, \citenamefont {Engels},
  \citenamefont {Tung},\ and\ \citenamefont
  {Cornell}}]{schweikhard_coddington_2004}%
  \BibitemOpen
  \bibfield  {author} {\bibinfo {author} {\bibfnamefont {V.}~\bibnamefont
  {Schweikhard}}, \bibinfo {author} {\bibfnamefont {I.}~\bibnamefont
  {Coddington}}, \bibinfo {author} {\bibfnamefont {P.}~\bibnamefont {Engels}},
  \bibinfo {author} {\bibfnamefont {S.}~\bibnamefont {Tung}}, \ and\ \bibinfo
  {author} {\bibfnamefont {E.~A.}\ \bibnamefont {Cornell}},\ }\href {\doibase
  10.1103/PhysRevLett.93.210403} {\bibfield  {journal} {\bibinfo  {journal}
  {Phys. Rev. Lett.}\ }\textbf {\bibinfo {volume} {93}},\ \bibinfo {pages}
  {210403} (\bibinfo {year} {2004})}\BibitemShut {NoStop}%
\bibitem [{\citenamefont {Kronj{\"a}ger}\ \emph {et~al.}(2005)\citenamefont
  {Kronj{\"a}ger}, \citenamefont {Becker}, \citenamefont {Brinkmann},
  \citenamefont {Walser}, \citenamefont {Navez}, \citenamefont {Bongs},\ and\
  \citenamefont {Sengstock}}]{kronjager_becker_2005}%
  \BibitemOpen
  \bibfield  {author} {\bibinfo {author} {\bibfnamefont {J.}~\bibnamefont
  {Kronj{\"a}ger}}, \bibinfo {author} {\bibfnamefont {C.}~\bibnamefont
  {Becker}}, \bibinfo {author} {\bibfnamefont {M.}~\bibnamefont {Brinkmann}},
  \bibinfo {author} {\bibfnamefont {R.}~\bibnamefont {Walser}}, \bibinfo
  {author} {\bibfnamefont {P.}~\bibnamefont {Navez}}, \bibinfo {author}
  {\bibfnamefont {K.}~\bibnamefont {Bongs}}, \ and\ \bibinfo {author}
  {\bibfnamefont {K.}~\bibnamefont {Sengstock}},\ }\href@noop {} {\bibfield
  {journal} {\bibinfo  {journal} {Phys. Rev. A}\ }\textbf {\bibinfo {volume}
  {72}},\ \bibinfo {pages} {063619} (\bibinfo {year} {2005})}\BibitemShut
  {NoStop}%
\bibitem [{\citenamefont {Sadler}\ \emph {et~al.}(2006)\citenamefont {Sadler},
  \citenamefont {Higbie}, \citenamefont {Leslie}, \citenamefont
  {Vengalattore},\ and\ \citenamefont {Stamper-Kurn}}]{sadler_higbie_06}%
  \BibitemOpen
  \bibfield  {author} {\bibinfo {author} {\bibfnamefont {L.~E.}\ \bibnamefont
  {Sadler}}, \bibinfo {author} {\bibfnamefont {J.~M.}\ \bibnamefont {Higbie}},
  \bibinfo {author} {\bibfnamefont {S.~R.}\ \bibnamefont {Leslie}}, \bibinfo
  {author} {\bibfnamefont {M.}~\bibnamefont {Vengalattore}}, \ and\ \bibinfo
  {author} {\bibfnamefont {D.~M.}\ \bibnamefont {Stamper-Kurn}},\ }\href@noop
  {} {\bibfield  {journal} {\bibinfo  {journal} {Nature (London)}\ }\textbf
  {\bibinfo {volume} {443}},\ \bibinfo {pages} {312} (\bibinfo {year}
  {2006})}\BibitemShut {NoStop}%
\bibitem [{\citenamefont {Beattie}\ \emph {et~al.}(2013)\citenamefont
  {Beattie}, \citenamefont {Moulder}, \citenamefont {Fletcher},\ and\
  \citenamefont {Hadzibabic}}]{beattie_moulder_2013}%
  \BibitemOpen
  \bibfield  {author} {\bibinfo {author} {\bibfnamefont {S.}~\bibnamefont
  {Beattie}}, \bibinfo {author} {\bibfnamefont {S.}~\bibnamefont {Moulder}},
  \bibinfo {author} {\bibfnamefont {R.~J.}\ \bibnamefont {Fletcher}}, \ and\
  \bibinfo {author} {\bibfnamefont {Z.}~\bibnamefont {Hadzibabic}},\ }\href
  {\doibase 10.1103/PhysRevLett.110.025301} {\bibfield  {journal} {\bibinfo
  {journal} {Phys. Rev. Lett.}\ }\textbf {\bibinfo {volume} {110}},\ \bibinfo
  {pages} {025301} (\bibinfo {year} {2013})}\BibitemShut {NoStop}%
\bibitem [{\citenamefont {Ho}\ and\ \citenamefont
  {Shenoy}(1996)}]{ho_shenoy_1996}%
  \BibitemOpen
  \bibfield  {author} {\bibinfo {author} {\bibfnamefont {T.-L.}\ \bibnamefont
  {Ho}}\ and\ \bibinfo {author} {\bibfnamefont {V.~B.}\ \bibnamefont
  {Shenoy}},\ }\href {\doibase 10.1103/PhysRevLett.77.3276} {\bibfield
  {journal} {\bibinfo  {journal} {Phys. Rev. Lett.}\ }\textbf {\bibinfo
  {volume} {77}},\ \bibinfo {pages} {3276} (\bibinfo {year}
  {1996})}\BibitemShut {NoStop}%
\bibitem [{\citenamefont {Pu}\ and\ \citenamefont
  {Bigelow}(1998{\natexlab{a}})}]{pu_bigelow_1998}%
  \BibitemOpen
  \bibfield  {author} {\bibinfo {author} {\bibfnamefont {H.}~\bibnamefont
  {Pu}}\ and\ \bibinfo {author} {\bibfnamefont {N.~P.}\ \bibnamefont
  {Bigelow}},\ }\href {\doibase 10.1103/PhysRevLett.80.1130} {\bibfield
  {journal} {\bibinfo  {journal} {Phys. Rev. Lett.}\ }\textbf {\bibinfo
  {volume} {80}},\ \bibinfo {pages} {1130} (\bibinfo {year}
  {1998}{\natexlab{a}})}\BibitemShut {NoStop}%
\bibitem [{\citenamefont {Pu}\ and\ \citenamefont
  {Bigelow}(1998{\natexlab{b}})}]{pu_bigelow_1998a}%
  \BibitemOpen
  \bibfield  {author} {\bibinfo {author} {\bibfnamefont {H.}~\bibnamefont
  {Pu}}\ and\ \bibinfo {author} {\bibfnamefont {N.~P.}\ \bibnamefont
  {Bigelow}},\ }\href {\doibase 10.1103/PhysRevLett.80.1134} {\bibfield
  {journal} {\bibinfo  {journal} {Phys. Rev. Lett.}\ }\textbf {\bibinfo
  {volume} {80}},\ \bibinfo {pages} {1134} (\bibinfo {year}
  {1998}{\natexlab{b}})}\BibitemShut {NoStop}%
\bibitem [{\citenamefont {Dalibard}\ \emph {et~al.}(2011)\citenamefont
  {Dalibard}, \citenamefont {Gerbier}, \citenamefont
  {Juzeli\ifmmode~\bar{u}\else \={u}\fi{}nas},\ and\ \citenamefont
  {\"Ohberg}}]{dalibard_gerbier_2011}%
  \BibitemOpen
  \bibfield  {author} {\bibinfo {author} {\bibfnamefont {J.}~\bibnamefont
  {Dalibard}}, \bibinfo {author} {\bibfnamefont {F.}~\bibnamefont {Gerbier}},
  \bibinfo {author} {\bibfnamefont {G.}~\bibnamefont
  {Juzeli\ifmmode~\bar{u}\else \={u}\fi{}nas}}, \ and\ \bibinfo {author}
  {\bibfnamefont {P.}~\bibnamefont {\"Ohberg}},\ }\href@noop {} {\bibfield
  {journal} {\bibinfo  {journal} {Rev. Mod. Phys.}\ }\textbf {\bibinfo {volume}
  {83}},\ \bibinfo {pages} {1523} (\bibinfo {year} {2011})}\BibitemShut
  {NoStop}%
\bibitem [{\citenamefont {Goldman}\ \emph {et~al.}(2014)\citenamefont
  {Goldman}, \citenamefont {Juzeli\=unas}, \citenamefont {\"Ohberg},\ and\
  \citenamefont {Spielman}}]{goldman_juzeliunas_2014}%
  \BibitemOpen
  \bibfield  {author} {\bibinfo {author} {\bibfnamefont {N.}~\bibnamefont
  {Goldman}}, \bibinfo {author} {\bibfnamefont {G.}~\bibnamefont
  {Juzeli\=unas}}, \bibinfo {author} {\bibfnamefont {P.}~\bibnamefont
  {\"Ohberg}}, \ and\ \bibinfo {author} {\bibfnamefont {I.~B.}\ \bibnamefont
  {Spielman}},\ }\href@noop {} {\bibfield  {journal} {\bibinfo  {journal} {Rep.
  Prog. Phys.}\ }\textbf {\bibinfo {volume} {77}},\ \bibinfo {pages} {126401}
  (\bibinfo {year} {2014})}\BibitemShut {NoStop}%
\bibitem [{\citenamefont {Ferrier-Barbut}\ \emph {et~al.}(2014)\citenamefont
  {Ferrier-Barbut}, \citenamefont {Delehaye}, \citenamefont {Laurent},
  \citenamefont {Grier}, \citenamefont {Pierce}, \citenamefont {Rem},
  \citenamefont {Chevy},\ and\ \citenamefont {Salomon}}]{ferrier_delehaye_14}%
  \BibitemOpen
  \bibfield  {author} {\bibinfo {author} {\bibfnamefont {I.}~\bibnamefont
  {Ferrier-Barbut}}, \bibinfo {author} {\bibfnamefont {M.}~\bibnamefont
  {Delehaye}}, \bibinfo {author} {\bibfnamefont {S.}~\bibnamefont {Laurent}},
  \bibinfo {author} {\bibfnamefont {A.~T.}\ \bibnamefont {Grier}}, \bibinfo
  {author} {\bibfnamefont {M.}~\bibnamefont {Pierce}}, \bibinfo {author}
  {\bibfnamefont {B.~S.}\ \bibnamefont {Rem}}, \bibinfo {author} {\bibfnamefont
  {F.}~\bibnamefont {Chevy}}, \ and\ \bibinfo {author} {\bibfnamefont
  {C.}~\bibnamefont {Salomon}},\ }\href {\doibase 10.1126/science.1255380}
  {\bibfield  {journal} {\bibinfo  {journal} {Science}\ }\textbf {\bibinfo
  {volume} {345}},\ \bibinfo {pages} {1035} (\bibinfo {year}
  {2014})}\BibitemShut {NoStop}%
\bibitem [{\citenamefont {Proukakis}\ \emph {et~al.}(2013)\citenamefont
  {Proukakis}, \citenamefont {Gardiner}, \citenamefont {Davis},\ and\
  \citenamefont {Szymanska}}]{proukakis_gardiner_2013}%
  \BibitemOpen
  \bibfield  {author} {\bibinfo {author} {\bibfnamefont {N.~P.}\ \bibnamefont
  {Proukakis}}, \bibinfo {author} {\bibfnamefont {S.~A.}\ \bibnamefont
  {Gardiner}}, \bibinfo {author} {\bibfnamefont {M.~J.}\ \bibnamefont {Davis}},
  \ and\ \bibinfo {author} {\bibfnamefont {M.~H.}\ \bibnamefont {Szymanska}},\
  }\href@noop {} {\emph {\bibinfo {title} {Quantum Gases: Finite Temperature
  and Non-Equilibrium Dynamics}}}\ (\bibinfo  {publisher} {Imperial College
  Press},\ \bibinfo {year} {2013})\BibitemShut {NoStop}%
\bibitem [{\citenamefont {Proukakis}\ and\ \citenamefont
  {Jackson}(2008)}]{proukakis_jackson_2008}%
  \BibitemOpen
  \bibfield  {author} {\bibinfo {author} {\bibfnamefont {N.~P.}\ \bibnamefont
  {Proukakis}}\ and\ \bibinfo {author} {\bibfnamefont {B.}~\bibnamefont
  {Jackson}},\ }\href@noop {} {\bibfield  {journal} {\bibinfo  {journal} {J.
  Phys. B: At. Mol. Opt. Phys.}\ }\textbf {\bibinfo {volume} {41}},\ \bibinfo
  {pages} {203002} (\bibinfo {year} {2008})}\BibitemShut {NoStop}%
\bibitem [{\citenamefont {Berloff}\ \emph {et~al.}(2014)\citenamefont
  {Berloff}, \citenamefont {Brachet},\ and\ \citenamefont
  {Proukakis}}]{berloff_brachet_14}%
  \BibitemOpen
  \bibfield  {author} {\bibinfo {author} {\bibfnamefont {N.~G.}\ \bibnamefont
  {Berloff}}, \bibinfo {author} {\bibfnamefont {M.}~\bibnamefont {Brachet}}, \
  and\ \bibinfo {author} {\bibfnamefont {N.~P.}\ \bibnamefont {Proukakis}},\
  }\href {\doibase 10.1073/pnas.1312549111} {\bibfield  {journal} {\bibinfo
  {journal} {Proceedings of the National Academy of Sciences}\ }\textbf
  {\bibinfo {volume} {111}},\ \bibinfo {pages} {4675} (\bibinfo {year}
  {2014})}\BibitemShut {NoStop}%
\bibitem [{\citenamefont {Griffin}\ \emph {et~al.}(2009)\citenamefont
  {Griffin}, \citenamefont {Nikuni},\ and\ \citenamefont
  {Zaremba}}]{griffin_nikuni_2009}%
  \BibitemOpen
  \bibfield  {author} {\bibinfo {author} {\bibfnamefont {A.}~\bibnamefont
  {Griffin}}, \bibinfo {author} {\bibfnamefont {T.}~\bibnamefont {Nikuni}}, \
  and\ \bibinfo {author} {\bibfnamefont {E.}~\bibnamefont {Zaremba}},\
  }\href@noop {} {\emph {\bibinfo {title} {Bose-Condensed Gases at Finite
  Temperatures}}}\ (\bibinfo  {publisher} {Cambridge University Press},\
  \bibinfo {year} {2009})\BibitemShut {NoStop}%
\bibitem [{\citenamefont {Blakie}\ \emph {et~al.}(2008)\citenamefont {Blakie},
  \citenamefont {Bradley}, \citenamefont {Davis}, \citenamefont {Ballagh},\
  and\ \citenamefont {Gardiner}}]{blakie_bradley_08}%
  \BibitemOpen
  \bibfield  {author} {\bibinfo {author} {\bibfnamefont {P.~B.}\ \bibnamefont
  {Blakie}}, \bibinfo {author} {\bibfnamefont {A.~S.}\ \bibnamefont {Bradley}},
  \bibinfo {author} {\bibfnamefont {M.~J.}\ \bibnamefont {Davis}}, \bibinfo
  {author} {\bibfnamefont {R.~J.}\ \bibnamefont {Ballagh}}, \ and\ \bibinfo
  {author} {\bibfnamefont {C.~W.}\ \bibnamefont {Gardiner}},\ }\href@noop {}
  {\bibfield  {journal} {\bibinfo  {journal} {Adv. Phys.}\ }\textbf {\bibinfo
  {volume} {57}},\ \bibinfo {pages} {363} (\bibinfo {year} {2008})}\BibitemShut
  {NoStop}%
\bibitem [{\citenamefont {Brewczyk}\ \emph {et~al.}(2007)\citenamefont
  {Brewczyk}, \citenamefont {Gajda},\ and\ \citenamefont
  {Rz\c{a}\.zewski}}]{brewczyk_gajda_07}%
  \BibitemOpen
  \bibfield  {author} {\bibinfo {author} {\bibfnamefont {M.}~\bibnamefont
  {Brewczyk}}, \bibinfo {author} {\bibfnamefont {M.}~\bibnamefont {Gajda}}, \
  and\ \bibinfo {author} {\bibfnamefont {K.}~\bibnamefont {Rz\c{a}\.zewski}},\
  }\href@noop {} {\bibfield  {journal} {\bibinfo  {journal} {J. Phys. B: At.
  Mol. Opt.}\ }\textbf {\bibinfo {volume} {40}},\ \bibinfo {pages} {R1}
  (\bibinfo {year} {2007})}\BibitemShut {NoStop}%
\bibitem [{\citenamefont {Kagan}\ \emph
  {et~al.}(1992{\natexlab{a}})\citenamefont {Kagan}, \citenamefont
  {Svistunov},\ and\ \citenamefont {Shlyapnikov}}]{kagan_svistunov_92a}%
  \BibitemOpen
  \bibfield  {author} {\bibinfo {author} {\bibfnamefont {Y.}~\bibnamefont
  {Kagan}}, \bibinfo {author} {\bibfnamefont {B.}~\bibnamefont {Svistunov}}, \
  and\ \bibinfo {author} {\bibfnamefont {G.}~\bibnamefont {Shlyapnikov}},\
  }\href@noop {} {\bibfield  {journal} {\bibinfo  {journal} {Sov. Phys. JETP}\
  }\textbf {\bibinfo {volume} {74}},\ \bibinfo {pages} {279} (\bibinfo {year}
  {1992}{\natexlab{a}})}\BibitemShut {NoStop}%
\bibitem [{\citenamefont {Kagan}\ \emph
  {et~al.}(1992{\natexlab{b}})\citenamefont {Kagan}, \citenamefont
  {Svistunov},\ and\ \citenamefont {Shlyapnikov}}]{kagan_svistunov_92b}%
  \BibitemOpen
  \bibfield  {author} {\bibinfo {author} {\bibfnamefont {Y.}~\bibnamefont
  {Kagan}}, \bibinfo {author} {\bibfnamefont {B.}~\bibnamefont {Svistunov}}, \
  and\ \bibinfo {author} {\bibfnamefont {G.}~\bibnamefont {Shlyapnikov}},\
  }\href@noop {} {\bibfield  {journal} {\bibinfo  {journal} {Sov. Phys. JETP}\
  }\textbf {\bibinfo {volume} {75}},\ \bibinfo {pages} {387} (\bibinfo {year}
  {1992}{\natexlab{b}})}\BibitemShut {NoStop}%
\bibitem [{\citenamefont {Davis}\ \emph
  {et~al.}(2001{\natexlab{a}})\citenamefont {Davis}, \citenamefont {Morgan},\
  and\ \citenamefont {Burnett}}]{davis_morgan_01}%
  \BibitemOpen
  \bibfield  {author} {\bibinfo {author} {\bibfnamefont {M.~J.}\ \bibnamefont
  {Davis}}, \bibinfo {author} {\bibfnamefont {S.~A.}\ \bibnamefont {Morgan}}, \
  and\ \bibinfo {author} {\bibfnamefont {K.}~\bibnamefont {Burnett}},\
  }\href@noop {} {\bibfield  {journal} {\bibinfo  {journal} {Phys. Rev. Lett.}\
  }\textbf {\bibinfo {volume} {87}},\ \bibinfo {pages} {160402} (\bibinfo
  {year} {2001}{\natexlab{a}})}\BibitemShut {NoStop}%
\bibitem [{\citenamefont {Davis}\ \emph {et~al.}(2002)\citenamefont {Davis},
  \citenamefont {Morgan},\ and\ \citenamefont {Burnett}}]{davis_morgan_02}%
  \BibitemOpen
  \bibfield  {author} {\bibinfo {author} {\bibfnamefont {M.~J.}\ \bibnamefont
  {Davis}}, \bibinfo {author} {\bibfnamefont {S.~A.}\ \bibnamefont {Morgan}}, \
  and\ \bibinfo {author} {\bibfnamefont {K.}~\bibnamefont {Burnett}},\
  }\href@noop {} {\bibfield  {journal} {\bibinfo  {journal} {Phys.\ Rev.\ A}\
  }\textbf {\bibinfo {volume} {66}},\ \bibinfo {pages} {053618} (\bibinfo
  {year} {2002})}\BibitemShut {NoStop}%
\bibitem [{\citenamefont {Berloff}\ and\ \citenamefont
  {Svistunov}(2002)}]{berloff_svistunov_02}%
  \BibitemOpen
  \bibfield  {author} {\bibinfo {author} {\bibfnamefont {N.~G.}\ \bibnamefont
  {Berloff}}\ and\ \bibinfo {author} {\bibfnamefont {B.~V.}\ \bibnamefont
  {Svistunov}},\ }\href@noop {} {\bibfield  {journal} {\bibinfo  {journal}
  {Phys.\ Rev.\ A}\ }\textbf {\bibinfo {volume} {66}},\ \bibinfo {pages}
  {013603} (\bibinfo {year} {2002})}\BibitemShut {NoStop}%
\bibitem [{\citenamefont {Steel}\ \emph {et~al.}(1998)\citenamefont {Steel},
  \citenamefont {Olsen}, \citenamefont {Plimak}, \citenamefont {Drummond},
  \citenamefont {Tan}, \citenamefont {Collett}, \citenamefont {Walls},\ and\
  \citenamefont {Graham}}]{steel_olsen_98}%
  \BibitemOpen
  \bibfield  {author} {\bibinfo {author} {\bibfnamefont {M.~J.}\ \bibnamefont
  {Steel}}, \bibinfo {author} {\bibfnamefont {M.~K.}\ \bibnamefont {Olsen}},
  \bibinfo {author} {\bibfnamefont {L.~I.}\ \bibnamefont {Plimak}}, \bibinfo
  {author} {\bibfnamefont {P.~D.}\ \bibnamefont {Drummond}}, \bibinfo {author}
  {\bibfnamefont {S.~M.}\ \bibnamefont {Tan}}, \bibinfo {author} {\bibfnamefont
  {M.~J.}\ \bibnamefont {Collett}}, \bibinfo {author} {\bibfnamefont {D.~F.}\
  \bibnamefont {Walls}}, \ and\ \bibinfo {author} {\bibfnamefont
  {R.}~\bibnamefont {Graham}},\ }\href@noop {} {\bibfield  {journal} {\bibinfo
  {journal} {Phys. Rev. A}\ }\textbf {\bibinfo {volume} {58}},\ \bibinfo
  {pages} {4824} (\bibinfo {year} {1998})}\BibitemShut {NoStop}%
\bibitem [{\citenamefont {Sinatra}\ \emph {et~al.}(2002)\citenamefont
  {Sinatra}, \citenamefont {Lobo},\ and\ \citenamefont
  {Castin}}]{sinatra_lobo_02}%
  \BibitemOpen
  \bibfield  {author} {\bibinfo {author} {\bibfnamefont {A.}~\bibnamefont
  {Sinatra}}, \bibinfo {author} {\bibfnamefont {C.}~\bibnamefont {Lobo}}, \
  and\ \bibinfo {author} {\bibfnamefont {Y.}~\bibnamefont {Castin}},\
  }\href@noop {} {\bibfield  {journal} {\bibinfo  {journal} {J. Phys. B: At.
  Mol. Opt.}\ }\textbf {\bibinfo {volume} {35}},\ \bibinfo {pages} {3599}
  (\bibinfo {year} {2002})}\BibitemShut {NoStop}%
\bibitem [{\citenamefont {Stoof}\ and\ \citenamefont
  {Bijlsma}(2001)}]{stoof_bijlsma_00}%
  \BibitemOpen
  \bibfield  {author} {\bibinfo {author} {\bibfnamefont {H.~T.~C.}\
  \bibnamefont {Stoof}}\ and\ \bibinfo {author} {\bibfnamefont {M.~J.}\
  \bibnamefont {Bijlsma}},\ }\href@noop {} {\bibfield  {journal} {\bibinfo
  {journal} {J. Low Temp. Phys.}\ }\textbf {\bibinfo {volume} {124}},\ \bibinfo
  {pages} {431} (\bibinfo {year} {2001})}\BibitemShut {NoStop}%
\bibitem [{\citenamefont {Duine}\ and\ \citenamefont
  {Stoof}(2001)}]{duine_stoof_2001}%
  \BibitemOpen
  \bibfield  {author} {\bibinfo {author} {\bibfnamefont {R.~A.}\ \bibnamefont
  {Duine}}\ and\ \bibinfo {author} {\bibfnamefont {H.~T.~C.}\ \bibnamefont
  {Stoof}},\ }\href@noop {} {\bibfield  {journal} {\bibinfo  {journal} {Phys.
  Rev. A}\ }\textbf {\bibinfo {volume} {65}},\ \bibinfo {pages} {013603}
  (\bibinfo {year} {2001})}\BibitemShut {NoStop}%
\bibitem [{\citenamefont {Gardiner}\ and\ \citenamefont
  {Davis}(2003)}]{gardiner_davis_2003}%
  \BibitemOpen
  \bibfield  {author} {\bibinfo {author} {\bibfnamefont {C.~W.}\ \bibnamefont
  {Gardiner}}\ and\ \bibinfo {author} {\bibfnamefont {M.~J.}\ \bibnamefont
  {Davis}},\ }\href {http://stacks.iop.org/0953-4075/36/i=23/a=010} {\bibfield
  {journal} {\bibinfo  {journal} {Journal of Physics B: Atomic, Molecular and
  Optical Physics}\ }\textbf {\bibinfo {volume} {36}},\ \bibinfo {pages} {4731}
  (\bibinfo {year} {2003})}\BibitemShut {NoStop}%
\bibitem [{\citenamefont {Proukakis}(2003)}]{proukakis_03}%
  \BibitemOpen
  \bibfield  {author} {\bibinfo {author} {\bibfnamefont {N.}~\bibnamefont
  {Proukakis}},\ }\href@noop {} {\bibfield  {journal} {\bibinfo  {journal}
  {Laser Phys.}\ }\textbf {\bibinfo {volume} {13}},\ \bibinfo {pages} {527}
  (\bibinfo {year} {2003})}\BibitemShut {NoStop}%
\bibitem [{\citenamefont {Rooney}\ \emph {et~al.}(2012)\citenamefont {Rooney},
  \citenamefont {Blakie},\ and\ \citenamefont {Bradley}}]{rooney_blakie_12}%
  \BibitemOpen
  \bibfield  {author} {\bibinfo {author} {\bibfnamefont {S.~J.}\ \bibnamefont
  {Rooney}}, \bibinfo {author} {\bibfnamefont {P.~B.}\ \bibnamefont {Blakie}},
  \ and\ \bibinfo {author} {\bibfnamefont {A.~S.}\ \bibnamefont {Bradley}},\
  }\href {\doibase 10.1103/PhysRevA.86.053634} {\bibfield  {journal} {\bibinfo
  {journal} {Phys. Rev. A}\ }\textbf {\bibinfo {volume} {86}},\ \bibinfo
  {pages} {053634} (\bibinfo {year} {2012})}\BibitemShut {NoStop}%
\bibitem [{\citenamefont {Andersen}\ \emph {et~al.}(2002)\citenamefont
  {Andersen}, \citenamefont {Khawaja},\ and\ \citenamefont
  {Stoof}}]{andersen_alkhawaja_02}%
  \BibitemOpen
  \bibfield  {author} {\bibinfo {author} {\bibfnamefont {J.}~\bibnamefont
  {Andersen}}, \bibinfo {author} {\bibfnamefont {U.~A.}\ \bibnamefont
  {Khawaja}}, \ and\ \bibinfo {author} {\bibfnamefont {H.}~\bibnamefont
  {Stoof}},\ }\href@noop {} {\bibfield  {journal} {\bibinfo  {journal} {Phys.
  Rev. Lett.}\ }\textbf {\bibinfo {volume} {88}},\ \bibinfo {pages} {070407}
  (\bibinfo {year} {2002})}\BibitemShut {NoStop}%
\bibitem [{\citenamefont {Al~Khawaja}\ \emph {et~al.}(2002)\citenamefont
  {Al~Khawaja}, \citenamefont {Andersen}, \citenamefont {Proukakis},\ and\
  \citenamefont {Stoof}}]{alkhawaja_andersen_02b}%
  \BibitemOpen
  \bibfield  {author} {\bibinfo {author} {\bibfnamefont {U.}~\bibnamefont
  {Al~Khawaja}}, \bibinfo {author} {\bibfnamefont {J.~O.}\ \bibnamefont
  {Andersen}}, \bibinfo {author} {\bibfnamefont {N.~P.}\ \bibnamefont
  {Proukakis}}, \ and\ \bibinfo {author} {\bibfnamefont {H.~T.~C.}\
  \bibnamefont {Stoof}},\ }\href@noop {} {\bibfield  {journal} {\bibinfo
  {journal} {Phys. Rev. A}\ }\textbf {\bibinfo {volume} {66}},\ \bibinfo
  {pages} {059902} (\bibinfo {year} {2002})}\BibitemShut {NoStop}%
\bibitem [{\citenamefont {Cockburn}\ \emph
  {et~al.}(2011{\natexlab{a}})\citenamefont {Cockburn}, \citenamefont
  {Negretti}, \citenamefont {Proukakis},\ and\ \citenamefont
  {Henkel}}]{cockburn_negretti_11}%
  \BibitemOpen
  \bibfield  {author} {\bibinfo {author} {\bibfnamefont {S.~P.}\ \bibnamefont
  {Cockburn}}, \bibinfo {author} {\bibfnamefont {A.}~\bibnamefont {Negretti}},
  \bibinfo {author} {\bibfnamefont {N.~P.}\ \bibnamefont {Proukakis}}, \ and\
  \bibinfo {author} {\bibfnamefont {C.}~\bibnamefont {Henkel}},\ }\href@noop {}
  {\bibfield  {journal} {\bibinfo  {journal} {Phys. Rev. A}\ }\textbf {\bibinfo
  {volume} {83}},\ \bibinfo {pages} {043619} (\bibinfo {year}
  {2011}{\natexlab{a}})}\BibitemShut {NoStop}%
\bibitem [{\citenamefont {Cockburn}\ \emph
  {et~al.}(2011{\natexlab{b}})\citenamefont {Cockburn}, \citenamefont
  {Gallucci},\ and\ \citenamefont {Proukakis}}]{cockburn_gallucci_11}%
  \BibitemOpen
  \bibfield  {author} {\bibinfo {author} {\bibfnamefont {S.~P.}\ \bibnamefont
  {Cockburn}}, \bibinfo {author} {\bibfnamefont {D.}~\bibnamefont {Gallucci}},
  \ and\ \bibinfo {author} {\bibfnamefont {N.~P.}\ \bibnamefont {Proukakis}},\
  }\href@noop {} {\bibfield  {journal} {\bibinfo  {journal} {Phys. Rev. A}\
  }\textbf {\bibinfo {volume} {84}},\ \bibinfo {pages} {023613} (\bibinfo
  {year} {2011}{\natexlab{b}})}\BibitemShut {NoStop}%
\bibitem [{\citenamefont {Gallucci}\ \emph {et~al.}(2012)\citenamefont
  {Gallucci}, \citenamefont {Cockburn},\ and\ \citenamefont
  {Proukakis}}]{gallucci_cockburn_12}%
  \BibitemOpen
  \bibfield  {author} {\bibinfo {author} {\bibfnamefont {D.}~\bibnamefont
  {Gallucci}}, \bibinfo {author} {\bibfnamefont {S.~P.}\ \bibnamefont
  {Cockburn}}, \ and\ \bibinfo {author} {\bibfnamefont {N.~P.}\ \bibnamefont
  {Proukakis}},\ }\href@noop {} {\bibfield  {journal} {\bibinfo  {journal}
  {Preprint}\ } (\bibinfo {year} {2012})}\BibitemShut {NoStop}%
\bibitem [{\citenamefont {Cockburn}\ and\ \citenamefont
  {Proukakis}(2012)}]{cockburn_proukakis_2012}%
  \BibitemOpen
  \bibfield  {author} {\bibinfo {author} {\bibfnamefont {S.~P.}\ \bibnamefont
  {Cockburn}}\ and\ \bibinfo {author} {\bibfnamefont {N.~P.}\ \bibnamefont
  {Proukakis}},\ }\href {\doibase 10.1103/PhysRevA.86.033610} {\bibfield
  {journal} {\bibinfo  {journal} {Phys. Rev. A}\ }\textbf {\bibinfo {volume}
  {86}},\ \bibinfo {pages} {033610} (\bibinfo {year} {2012})}\BibitemShut
  {NoStop}%
\bibitem [{\citenamefont {Davis}\ \emph {et~al.}(2012)\citenamefont {Davis},
  \citenamefont {Blakie}, \citenamefont {van Amerongen}, \citenamefont {van
  Druten},\ and\ \citenamefont {Kheruntsyan}}]{davis_blakie_2012}%
  \BibitemOpen
  \bibfield  {author} {\bibinfo {author} {\bibfnamefont {M.~J.}\ \bibnamefont
  {Davis}}, \bibinfo {author} {\bibfnamefont {P.~B.}\ \bibnamefont {Blakie}},
  \bibinfo {author} {\bibfnamefont {A.~H.}\ \bibnamefont {van Amerongen}},
  \bibinfo {author} {\bibfnamefont {N.~J.}\ \bibnamefont {van Druten}}, \ and\
  \bibinfo {author} {\bibfnamefont {K.~V.}\ \bibnamefont {Kheruntsyan}},\
  }\href {\doibase 10.1103/PhysRevA.85.031604} {\bibfield  {journal} {\bibinfo
  {journal} {Phys. Rev. A}\ }\textbf {\bibinfo {volume} {85}},\ \bibinfo
  {pages} {031604} (\bibinfo {year} {2012})}\BibitemShut {NoStop}%
\bibitem [{\citenamefont {Weiler}\ \emph {et~al.}(2008)\citenamefont {Weiler},
  \citenamefont {Neely}, \citenamefont {Scherer}, \citenamefont {Bradley},
  \citenamefont {Davis},\ and\ \citenamefont {Anderson}}]{weiler_neely_08}%
  \BibitemOpen
  \bibfield  {author} {\bibinfo {author} {\bibfnamefont {C.~N.}\ \bibnamefont
  {Weiler}}, \bibinfo {author} {\bibfnamefont {T.~W.}\ \bibnamefont {Neely}},
  \bibinfo {author} {\bibfnamefont {D.~R.}\ \bibnamefont {Scherer}}, \bibinfo
  {author} {\bibfnamefont {A.~S.}\ \bibnamefont {Bradley}}, \bibinfo {author}
  {\bibfnamefont {M.~J.}\ \bibnamefont {Davis}}, \ and\ \bibinfo {author}
  {\bibfnamefont {B.~P.}\ \bibnamefont {Anderson}},\ }\href@noop {} {\bibfield
  {journal} {\bibinfo  {journal} {Nature}\ }\textbf {\bibinfo {volume} {455}},\
  \bibinfo {pages} {948} (\bibinfo {year} {2008})}\BibitemShut {NoStop}%
\bibitem [{\citenamefont {Proukakis}\ \emph {et~al.}(2006)\citenamefont
  {Proukakis}, \citenamefont {Schmiedmayer},\ and\ \citenamefont
  {Stoof}}]{proukakis_schmiedmayer_2006}%
  \BibitemOpen
  \bibfield  {author} {\bibinfo {author} {\bibfnamefont {N.~P.}\ \bibnamefont
  {Proukakis}}, \bibinfo {author} {\bibfnamefont {J.}~\bibnamefont
  {Schmiedmayer}}, \ and\ \bibinfo {author} {\bibfnamefont {H.~T.~C.}\
  \bibnamefont {Stoof}},\ }\href@noop {} {\bibfield  {journal} {\bibinfo
  {journal} {Phys. Rev. A}\ }\textbf {\bibinfo {volume} {73}},\ \bibinfo
  {pages} {053603} (\bibinfo {year} {2006})}\BibitemShut {NoStop}%
\bibitem [{\citenamefont {Cockburn}\ \emph {et~al.}(2010)\citenamefont
  {Cockburn}, \citenamefont {Nistazakis}, \citenamefont {Horikis},
  \citenamefont {Kevrekidis}, \citenamefont {Proukakis},\ and\ \citenamefont
  {Frantzeskakis}}]{cockburn_nistazakis_2010}%
  \BibitemOpen
  \bibfield  {author} {\bibinfo {author} {\bibfnamefont {S.~P.}\ \bibnamefont
  {Cockburn}}, \bibinfo {author} {\bibfnamefont {H.~E.}\ \bibnamefont
  {Nistazakis}}, \bibinfo {author} {\bibfnamefont {T.~P.}\ \bibnamefont
  {Horikis}}, \bibinfo {author} {\bibfnamefont {P.~G.}\ \bibnamefont
  {Kevrekidis}}, \bibinfo {author} {\bibfnamefont {N.~P.}\ \bibnamefont
  {Proukakis}}, \ and\ \bibinfo {author} {\bibfnamefont {D.~J.}\ \bibnamefont
  {Frantzeskakis}},\ }\href@noop {} {\bibfield  {journal} {\bibinfo  {journal}
  {Phys. Rev. Lett.}\ }\textbf {\bibinfo {volume} {104}},\ \bibinfo {pages}
  {174101} (\bibinfo {year} {2010})}\BibitemShut {NoStop}%
\bibitem [{\citenamefont {Damski}\ and\ \citenamefont
  {Zurek}(2010)}]{damski_zurek_10}%
  \BibitemOpen
  \bibfield  {author} {\bibinfo {author} {\bibfnamefont {B.}~\bibnamefont
  {Damski}}\ and\ \bibinfo {author} {\bibfnamefont {W.~H.}\ \bibnamefont
  {Zurek}},\ }\href@noop {} {\bibfield  {journal} {\bibinfo  {journal} {Phys.
  Rev. Lett.}\ }\textbf {\bibinfo {volume} {104}},\ \bibinfo {pages} {160404}
  (\bibinfo {year} {2010})}\BibitemShut {NoStop}%
\bibitem [{\citenamefont {Kirkpatrick}\ and\ \citenamefont
  {Dorfman}(1983)}]{kirkpatrick_dorfman_1983}%
  \BibitemOpen
  \bibfield  {author} {\bibinfo {author} {\bibfnamefont {T.~R.}\ \bibnamefont
  {Kirkpatrick}}\ and\ \bibinfo {author} {\bibfnamefont {J.~R.}\ \bibnamefont
  {Dorfman}},\ }\href {\doibase 10.1103/PhysRevA.28.2576} {\bibfield  {journal}
  {\bibinfo  {journal} {Phys. Rev. A}\ }\textbf {\bibinfo {volume} {28}},\
  \bibinfo {pages} {2576} (\bibinfo {year} {1983})}\BibitemShut {NoStop}%
\bibitem [{\citenamefont {Kirkpatrick}\ and\ \citenamefont
  {Dorfman}(1985{\natexlab{a}})}]{kirkpatrick_dorfman_1985}%
  \BibitemOpen
  \bibfield  {author} {\bibinfo {author} {\bibfnamefont {T.}~\bibnamefont
  {Kirkpatrick}}\ and\ \bibinfo {author} {\bibfnamefont {J.}~\bibnamefont
  {Dorfman}},\ }\href {\doibase 10.1007/BF00681309} {\bibfield  {journal}
  {\bibinfo  {journal} {Journal of Low Temperature Physics}\ }\textbf {\bibinfo
  {volume} {58}},\ \bibinfo {pages} {301} (\bibinfo {year}
  {1985}{\natexlab{a}})}\BibitemShut {NoStop}%
\bibitem [{\citenamefont {Kirkpatrick}\ and\ \citenamefont
  {Dorfman}(1985{\natexlab{b}})}]{kirkpatrick_dorfman_1985a}%
  \BibitemOpen
  \bibfield  {author} {\bibinfo {author} {\bibfnamefont {T.}~\bibnamefont
  {Kirkpatrick}}\ and\ \bibinfo {author} {\bibfnamefont {J.}~\bibnamefont
  {Dorfman}},\ }\href {\doibase 10.1007/BF00681133} {\bibfield  {journal}
  {\bibinfo  {journal} {Journal of Low Temperature Physics}\ }\textbf {\bibinfo
  {volume} {58}},\ \bibinfo {pages} {399} (\bibinfo {year}
  {1985}{\natexlab{b}})}\BibitemShut {NoStop}%
\bibitem [{\citenamefont {Zaremba}\ \emph {et~al.}(1999)\citenamefont
  {Zaremba}, \citenamefont {Nikuni}, \citenamefont {Griffin},\ and\
  \citenamefont {Griffin}}]{zaremba_nikuni_1999}%
  \BibitemOpen
  \bibfield  {author} {\bibinfo {author} {\bibfnamefont {E.}~\bibnamefont
  {Zaremba}}, \bibinfo {author} {\bibfnamefont {T.}~\bibnamefont {Nikuni}},
  \bibinfo {author} {\bibfnamefont {A.}~\bibnamefont {Griffin}}, \ and\
  \bibinfo {author} {\bibfnamefont {A.}~\bibnamefont {Griffin}},\ }\href@noop
  {} {\bibfield  {journal} {\bibinfo  {journal} {J. Low Temp. Phys.}\ }\textbf
  {\bibinfo {volume} {116}},\ \bibinfo {pages} {277} (\bibinfo {year}
  {1999})}\BibitemShut {NoStop}%
\bibitem [{\citenamefont {Walser}\ \emph {et~al.}(1999)\citenamefont {Walser},
  \citenamefont {Williams}, \citenamefont {Cooper},\ and\ \citenamefont
  {Holland}}]{walser_williams_1999}%
  \BibitemOpen
  \bibfield  {author} {\bibinfo {author} {\bibfnamefont {R.}~\bibnamefont
  {Walser}}, \bibinfo {author} {\bibfnamefont {J.}~\bibnamefont {Williams}},
  \bibinfo {author} {\bibfnamefont {J.}~\bibnamefont {Cooper}}, \ and\ \bibinfo
  {author} {\bibfnamefont {M.}~\bibnamefont {Holland}},\ }\href {\doibase
  10.1103/PhysRevA.59.3878} {\bibfield  {journal} {\bibinfo  {journal} {Phys.
  Rev. A}\ }\textbf {\bibinfo {volume} {59}},\ \bibinfo {pages} {3878}
  (\bibinfo {year} {1999})}\BibitemShut {NoStop}%
\bibitem [{\citenamefont {Walser}\ \emph {et~al.}(2001)\citenamefont {Walser},
  \citenamefont {Cooper},\ and\ \citenamefont {Holland}}]{walser_cooper_01}%
  \BibitemOpen
  \bibfield  {author} {\bibinfo {author} {\bibfnamefont {R.}~\bibnamefont
  {Walser}}, \bibinfo {author} {\bibfnamefont {J.}~\bibnamefont {Cooper}}, \
  and\ \bibinfo {author} {\bibfnamefont {M.}~\bibnamefont {Holland}},\
  }\href@noop {} {\bibfield  {journal} {\bibinfo  {journal} {Phys. Rev. A}\
  }\textbf {\bibinfo {volume} {63}},\ \bibinfo {pages} {013607} (\bibinfo
  {year} {2001})}\BibitemShut {NoStop}%
\bibitem [{\citenamefont {Wachter}\ \emph {et~al.}(2001)\citenamefont
  {Wachter}, \citenamefont {Walser}, \citenamefont {Cooper},\ and\
  \citenamefont {Holland}}]{wachter_walser_01a}%
  \BibitemOpen
  \bibfield  {author} {\bibinfo {author} {\bibfnamefont {J.}~\bibnamefont
  {Wachter}}, \bibinfo {author} {\bibfnamefont {R.}~\bibnamefont {Walser}},
  \bibinfo {author} {\bibfnamefont {J.}~\bibnamefont {Cooper}}, \ and\ \bibinfo
  {author} {\bibfnamefont {M.}~\bibnamefont {Holland}},\ }\href@noop {}
  {\bibfield  {journal} {\bibinfo  {journal} {Phys. Rev. A}\ }\textbf {\bibinfo
  {volume} {64}},\ \bibinfo {pages} {053612} (\bibinfo {year}
  {2001})}\BibitemShut {NoStop}%
\bibitem [{\citenamefont {Wachter}\ \emph {et~al.}(2002)\citenamefont
  {Wachter}, \citenamefont {Walser}, \citenamefont {Cooper},\ and\
  \citenamefont {Holland}}]{wachter_walser_01b}%
  \BibitemOpen
  \bibfield  {author} {\bibinfo {author} {\bibfnamefont {J.}~\bibnamefont
  {Wachter}}, \bibinfo {author} {\bibfnamefont {R.}~\bibnamefont {Walser}},
  \bibinfo {author} {\bibfnamefont {J.}~\bibnamefont {Cooper}}, \ and\ \bibinfo
  {author} {\bibfnamefont {M.}~\bibnamefont {Holland}},\ }\href@noop {}
  {\bibfield  {journal} {\bibinfo  {journal} {Phys. Rev. A}\ }\textbf {\bibinfo
  {volume} {65}},\ \bibinfo {pages} {039904} (\bibinfo {year}
  {2002})}\BibitemShut {NoStop}%
\bibitem [{\citenamefont {Proukakis}\ and\ \citenamefont
  {Burnett}(1996)}]{proukakis_burnett_1996}%
  \BibitemOpen
  \bibfield  {author} {\bibinfo {author} {\bibfnamefont {N.~P.}\ \bibnamefont
  {Proukakis}}\ and\ \bibinfo {author} {\bibfnamefont {K.}~\bibnamefont
  {Burnett}},\ }\href@noop {} {\bibfield  {journal} {\bibinfo  {journal} {J. of
  Research at the National Institute of Standards and Technology}\ }\textbf
  {\bibinfo {volume} {101}},\ \bibinfo {pages} {457} (\bibinfo {year}
  {1996})}\BibitemShut {NoStop}%
\bibitem [{\citenamefont {Proukakis}\ \emph {et~al.}(1998)\citenamefont
  {Proukakis}, \citenamefont {Burnett},\ and\ \citenamefont
  {Stoof}}]{proukakis_burnett_1998}%
  \BibitemOpen
  \bibfield  {author} {\bibinfo {author} {\bibfnamefont {N.~P.}\ \bibnamefont
  {Proukakis}}, \bibinfo {author} {\bibfnamefont {K.}~\bibnamefont {Burnett}},
  \ and\ \bibinfo {author} {\bibfnamefont {H.~T.~C.}\ \bibnamefont {Stoof}},\
  }\href {\doibase 10.1103/PhysRevA.57.1230} {\bibfield  {journal} {\bibinfo
  {journal} {Phys. Rev. A}\ }\textbf {\bibinfo {volume} {57}},\ \bibinfo
  {pages} {1230} (\bibinfo {year} {1998})}\BibitemShut {NoStop}%
\bibitem [{\citenamefont {Proukakis}(2001)}]{proukakis_2001a}%
  \BibitemOpen
  \bibfield  {author} {\bibinfo {author} {\bibfnamefont {N.~P.}\ \bibnamefont
  {Proukakis}},\ }\href@noop {} {\bibfield  {journal} {\bibinfo  {journal} {J.
  Phys. B: At. Mol. Opt. Phys.}\ }\textbf {\bibinfo {volume} {34}},\ \bibinfo
  {pages} {4737} (\bibinfo {year} {2001})}\BibitemShut {NoStop}%
\bibitem [{\citenamefont {Griffin}(1996)}]{griffin_1996}%
  \BibitemOpen
  \bibfield  {author} {\bibinfo {author} {\bibfnamefont {A.}~\bibnamefont
  {Griffin}},\ }\href {\doibase 10.1103/PhysRevB.53.9341} {\bibfield  {journal}
  {\bibinfo  {journal} {Phys. Rev. B}\ }\textbf {\bibinfo {volume} {53}},\
  \bibinfo {pages} {9341} (\bibinfo {year} {1996})}\BibitemShut {NoStop}%
\bibitem [{\citenamefont {Williams}\ and\ \citenamefont
  {Griffin}(2001)}]{williams_griffin_01}%
  \BibitemOpen
  \bibfield  {author} {\bibinfo {author} {\bibfnamefont {J.~E.}\ \bibnamefont
  {Williams}}\ and\ \bibinfo {author} {\bibfnamefont {A.}~\bibnamefont
  {Griffin}},\ }\href@noop {} {\bibfield  {journal} {\bibinfo  {journal} {Phys.
  Rev. A}\ }\textbf {\bibinfo {volume} {63}},\ \bibinfo {pages} {023612}
  (\bibinfo {year} {2001})}\BibitemShut {NoStop}%
\bibitem [{\citenamefont {Jackson}\ and\ \citenamefont
  {Zaremba}(2001{\natexlab{a}})}]{jackson_zaremba_01}%
  \BibitemOpen
  \bibfield  {author} {\bibinfo {author} {\bibfnamefont {B.}~\bibnamefont
  {Jackson}}\ and\ \bibinfo {author} {\bibfnamefont {E.}~\bibnamefont
  {Zaremba}},\ }\href@noop {} {\bibfield  {journal} {\bibinfo  {journal} {Phys.
  Rev. Lett.}\ }\textbf {\bibinfo {volume} {87}},\ \bibinfo {pages} {100404}
  (\bibinfo {year} {2001}{\natexlab{a}})}\BibitemShut {NoStop}%
\bibitem [{\citenamefont {Jackson}\ and\ \citenamefont
  {Zaremba}(2002{\natexlab{a}})}]{jackson_zaremba_02c}%
  \BibitemOpen
  \bibfield  {author} {\bibinfo {author} {\bibfnamefont {B.}~\bibnamefont
  {Jackson}}\ and\ \bibinfo {author} {\bibfnamefont {E.}~\bibnamefont
  {Zaremba}},\ }\href@noop {} {\bibfield  {journal} {\bibinfo  {journal} {Phys.
  Rev. Lett.}\ }\textbf {\bibinfo {volume} {88}},\ \bibinfo {pages} {180402}
  (\bibinfo {year} {2002}{\natexlab{a}})}\BibitemShut {NoStop}%
\bibitem [{\citenamefont {Jackson}\ and\ \citenamefont
  {Zaremba}(2003)}]{jackson_zaremba_03}%
  \BibitemOpen
  \bibfield  {author} {\bibinfo {author} {\bibfnamefont {B.}~\bibnamefont
  {Jackson}}\ and\ \bibinfo {author} {\bibfnamefont {E.}~\bibnamefont
  {Zaremba}},\ }\href@noop {} {\bibfield  {journal} {\bibinfo  {journal} {New
  J. Phys.}\ }\textbf {\bibinfo {volume} {5}},\ \bibinfo {pages} {88} (\bibinfo
  {year} {2003})}\BibitemShut {NoStop}%
\bibitem [{\citenamefont {Jackson}\ \emph {et~al.}(2007)\citenamefont
  {Jackson}, \citenamefont {Proukakis},\ and\ \citenamefont
  {Barenghi}}]{jackson_proukakis_2007}%
  \BibitemOpen
  \bibfield  {author} {\bibinfo {author} {\bibfnamefont {B.}~\bibnamefont
  {Jackson}}, \bibinfo {author} {\bibfnamefont {N.~P.}\ \bibnamefont
  {Proukakis}}, \ and\ \bibinfo {author} {\bibfnamefont {C.~F.}\ \bibnamefont
  {Barenghi}},\ }\href {\doibase 10.1103/PhysRevA.75.051601} {\bibfield
  {journal} {\bibinfo  {journal} {Phys. Rev. A}\ }\textbf {\bibinfo {volume}
  {75}},\ \bibinfo {pages} {051601} (\bibinfo {year} {2007})}\BibitemShut
  {NoStop}%
\bibitem [{\citenamefont {Jackson}\ \emph {et~al.}(2009)\citenamefont
  {Jackson}, \citenamefont {Proukakis}, \citenamefont {Barenghi},\ and\
  \citenamefont {Zaremba}}]{jackson_proukakis_09}%
  \BibitemOpen
  \bibfield  {author} {\bibinfo {author} {\bibfnamefont {B.}~\bibnamefont
  {Jackson}}, \bibinfo {author} {\bibfnamefont {N.~P.}\ \bibnamefont
  {Proukakis}}, \bibinfo {author} {\bibfnamefont {C.~F.}\ \bibnamefont
  {Barenghi}}, \ and\ \bibinfo {author} {\bibfnamefont {E.}~\bibnamefont
  {Zaremba}},\ }\href@noop {} {\bibfield  {journal} {\bibinfo  {journal} {Phys.
  Rev. A}\ }\textbf {\bibinfo {volume} {79}},\ \bibinfo {pages} {053615}
  (\bibinfo {year} {2009})}\BibitemShut {NoStop}%
\bibitem [{\citenamefont {Allen}\ \emph
  {et~al.}(2013{\natexlab{a}})\citenamefont {Allen}, \citenamefont {Zaremba},
  \citenamefont {Barenghi},\ and\ \citenamefont
  {Proukakis}}]{allen_zaremba_13}%
  \BibitemOpen
  \bibfield  {author} {\bibinfo {author} {\bibfnamefont {A.~J.}\ \bibnamefont
  {Allen}}, \bibinfo {author} {\bibfnamefont {E.}~\bibnamefont {Zaremba}},
  \bibinfo {author} {\bibfnamefont {C.~F.}\ \bibnamefont {Barenghi}}, \ and\
  \bibinfo {author} {\bibfnamefont {N.~P.}\ \bibnamefont {Proukakis}},\ }\href
  {\doibase 10.1103/PhysRevA.87.013630} {\bibfield  {journal} {\bibinfo
  {journal} {Phys. Rev. A}\ }\textbf {\bibinfo {volume} {87}},\ \bibinfo
  {pages} {013630} (\bibinfo {year} {2013}{\natexlab{a}})}\BibitemShut
  {NoStop}%
\bibitem [{\citenamefont {Bijlsma}\ \emph {et~al.}(2000)\citenamefont
  {Bijlsma}, \citenamefont {Zaremba},\ and\ \citenamefont
  {Stoof}}]{bijlsma_zaremba_2000}%
  \BibitemOpen
  \bibfield  {author} {\bibinfo {author} {\bibfnamefont {M.~J.}\ \bibnamefont
  {Bijlsma}}, \bibinfo {author} {\bibfnamefont {E.}~\bibnamefont {Zaremba}}, \
  and\ \bibinfo {author} {\bibfnamefont {H.~T.~C.}\ \bibnamefont {Stoof}},\
  }\href {\doibase 10.1103/PhysRevA.62.063609} {\bibfield  {journal} {\bibinfo
  {journal} {Phys. Rev. A}\ }\textbf {\bibinfo {volume} {62}},\ \bibinfo
  {pages} {063609} (\bibinfo {year} {2000})}\BibitemShut {NoStop}%
\bibitem [{\citenamefont {M\"arkle}\ \emph {et~al.}(2014)\citenamefont
  {M\"arkle}, \citenamefont {Allen}, \citenamefont {Federsel}, \citenamefont
  {Jetter}, \citenamefont {G\"unther}, \citenamefont {Fort\'agh}, \citenamefont
  {Proukakis},\ and\ \citenamefont {Judd}}]{markle_allen_14}%
  \BibitemOpen
  \bibfield  {author} {\bibinfo {author} {\bibfnamefont {J.}~\bibnamefont
  {M\"arkle}}, \bibinfo {author} {\bibfnamefont {A.~J.}\ \bibnamefont {Allen}},
  \bibinfo {author} {\bibfnamefont {P.}~\bibnamefont {Federsel}}, \bibinfo
  {author} {\bibfnamefont {B.}~\bibnamefont {Jetter}}, \bibinfo {author}
  {\bibfnamefont {A.}~\bibnamefont {G\"unther}}, \bibinfo {author}
  {\bibfnamefont {J.}~\bibnamefont {Fort\'agh}}, \bibinfo {author}
  {\bibfnamefont {N.~P.}\ \bibnamefont {Proukakis}}, \ and\ \bibinfo {author}
  {\bibfnamefont {T.~E.}\ \bibnamefont {Judd}},\ }\href {\doibase
  10.1103/PhysRevA.90.023614} {\bibfield  {journal} {\bibinfo  {journal} {Phys.
  Rev. A}\ }\textbf {\bibinfo {volume} {90}},\ \bibinfo {pages} {023614}
  (\bibinfo {year} {2014})}\BibitemShut {NoStop}%
\bibitem [{\citenamefont {Davis}\ \emph {et~al.}(2000)\citenamefont {Davis},
  \citenamefont {Gardiner},\ and\ \citenamefont {Ballagh}}]{davis_gardiner_00}%
  \BibitemOpen
  \bibfield  {author} {\bibinfo {author} {\bibfnamefont {M.~J.}\ \bibnamefont
  {Davis}}, \bibinfo {author} {\bibfnamefont {C.~W.}\ \bibnamefont {Gardiner}},
  \ and\ \bibinfo {author} {\bibfnamefont {R.~J.}\ \bibnamefont {Ballagh}},\
  }\href@noop {} {\bibfield  {journal} {\bibinfo  {journal} {Phys. Rev. A}\
  }\textbf {\bibinfo {volume} {62}},\ \bibinfo {pages} {063608} (\bibinfo
  {year} {2000})}\BibitemShut {NoStop}%
\bibitem [{\citenamefont {Gardiner}\ \emph {et~al.}(1998)\citenamefont
  {Gardiner}, \citenamefont {Lee}, \citenamefont {Ballagh}, \citenamefont
  {Davis},\ and\ \citenamefont {Zoller}}]{gardiner_lee_1998}%
  \BibitemOpen
  \bibfield  {author} {\bibinfo {author} {\bibfnamefont {C.~W.}\ \bibnamefont
  {Gardiner}}, \bibinfo {author} {\bibfnamefont {M.~D.}\ \bibnamefont {Lee}},
  \bibinfo {author} {\bibfnamefont {R.~J.}\ \bibnamefont {Ballagh}}, \bibinfo
  {author} {\bibfnamefont {M.~J.}\ \bibnamefont {Davis}}, \ and\ \bibinfo
  {author} {\bibfnamefont {P.}~\bibnamefont {Zoller}},\ }\href {\doibase
  10.1103/PhysRevLett.81.5266} {\bibfield  {journal} {\bibinfo  {journal}
  {Phys. Rev. Lett.}\ }\textbf {\bibinfo {volume} {81}},\ \bibinfo {pages}
  {5266} (\bibinfo {year} {1998})}\BibitemShut {NoStop}%
\bibitem [{\citenamefont {K\=ohl}\ \emph {et~al.}(2002)\citenamefont {K\=ohl},
  \citenamefont {Davis}, \citenamefont {Gardiner}, \citenamefont {H\=ansch},\
  and\ \citenamefont {Esslinger}}]{kohl_davis_2002}%
  \BibitemOpen
  \bibfield  {author} {\bibinfo {author} {\bibfnamefont {M.}~\bibnamefont
  {K\=ohl}}, \bibinfo {author} {\bibfnamefont {M.~J.}\ \bibnamefont {Davis}},
  \bibinfo {author} {\bibfnamefont {C.~W.}\ \bibnamefont {Gardiner}}, \bibinfo
  {author} {\bibfnamefont {T.~W.}\ \bibnamefont {H\=ansch}}, \ and\ \bibinfo
  {author} {\bibfnamefont {T.}~\bibnamefont {Esslinger}},\ }\href@noop {}
  {\bibfield  {journal} {\bibinfo  {journal} {Phys. Rev. Lett.}\ }\textbf
  {\bibinfo {volume} {88}},\ \bibinfo {pages} {080402} (\bibinfo {year}
  {2002})}\BibitemShut {NoStop}%
\bibitem [{\citenamefont {Hugbart}\ \emph {et~al.}(2007)\citenamefont
  {Hugbart}, \citenamefont {Retter}, \citenamefont {Var\'on}, \citenamefont
  {Bouyer}, \citenamefont {Aspect},\ and\ \citenamefont
  {Davis}}]{hugbart_retter_07}%
  \BibitemOpen
  \bibfield  {author} {\bibinfo {author} {\bibfnamefont {M.}~\bibnamefont
  {Hugbart}}, \bibinfo {author} {\bibfnamefont {J.~A.}\ \bibnamefont {Retter}},
  \bibinfo {author} {\bibfnamefont {A.~F.}\ \bibnamefont {Var\'on}}, \bibinfo
  {author} {\bibfnamefont {P.}~\bibnamefont {Bouyer}}, \bibinfo {author}
  {\bibfnamefont {A.}~\bibnamefont {Aspect}}, \ and\ \bibinfo {author}
  {\bibfnamefont {M.~J.}\ \bibnamefont {Davis}},\ }\href@noop {} {\bibfield
  {journal} {\bibinfo  {journal} {Phys. Rev. A}\ }\textbf {\bibinfo {volume}
  {75}},\ \bibinfo {pages} {011602} (\bibinfo {year} {2007})}\BibitemShut
  {NoStop}%
\bibitem [{\citenamefont {Gardiner}\ and\ \citenamefont
  {Morgan}(2007{\natexlab{a}})}]{gardiner_morgan_2007}%
  \BibitemOpen
  \bibfield  {author} {\bibinfo {author} {\bibfnamefont {S.~A.}\ \bibnamefont
  {Gardiner}}\ and\ \bibinfo {author} {\bibfnamefont {S.~A.}\ \bibnamefont
  {Morgan}},\ }\href {\doibase 10.1103/PhysRevA.75.043621} {\bibfield
  {journal} {\bibinfo  {journal} {Phys. Rev. A}\ }\textbf {\bibinfo {volume}
  {75}},\ \bibinfo {pages} {043621} (\bibinfo {year}
  {2007}{\natexlab{a}})}\BibitemShut {NoStop}%
\bibitem [{\citenamefont {Gardiner}\ and\ \citenamefont
  {Morgan}(2007{\natexlab{b}})}]{gardiner_morgan_2007err}%
  \BibitemOpen
  \bibfield  {author} {\bibinfo {author} {\bibfnamefont {S.~A.}\ \bibnamefont
  {Gardiner}}\ and\ \bibinfo {author} {\bibfnamefont {S.~A.}\ \bibnamefont
  {Morgan}},\ }\href {\doibase 10.1103/PhysRevA.76.029902} {\bibfield
  {journal} {\bibinfo  {journal} {Phys. Rev. A}\ }\textbf {\bibinfo {volume}
  {76}},\ \bibinfo {pages} {029902} (\bibinfo {year}
  {2007}{\natexlab{b}})}\BibitemShut {NoStop}%
\bibitem [{\citenamefont {{Billam}}\ and\ \citenamefont
  {{Gardiner}}(2012)}]{billam_gardiner_12}%
  \BibitemOpen
  \bibfield  {author} {\bibinfo {author} {\bibfnamefont {T.~P.}\ \bibnamefont
  {{Billam}}}\ and\ \bibinfo {author} {\bibfnamefont {S.~A.}\ \bibnamefont
  {{Gardiner}}},\ }\href@noop {} {\bibfield  {journal} {\bibinfo  {journal}
  {New J. Phys.}\ }\textbf {\bibinfo {volume} {14}},\ \bibinfo {pages} {013038}
  (\bibinfo {year} {2012})}\BibitemShut {NoStop}%
\bibitem [{\citenamefont {Billam}\ \emph {et~al.}(2013)\citenamefont {Billam},
  \citenamefont {Mason},\ and\ \citenamefont {Gardiner}}]{billam_mason_13}%
  \BibitemOpen
  \bibfield  {author} {\bibinfo {author} {\bibfnamefont {T.~P.}\ \bibnamefont
  {Billam}}, \bibinfo {author} {\bibfnamefont {P.}~\bibnamefont {Mason}}, \
  and\ \bibinfo {author} {\bibfnamefont {S.~A.}\ \bibnamefont {Gardiner}},\
  }\href {\doibase 10.1103/PhysRevA.87.033628} {\bibfield  {journal} {\bibinfo
  {journal} {Phys. Rev. A}\ }\textbf {\bibinfo {volume} {87}},\ \bibinfo
  {pages} {033628} (\bibinfo {year} {2013})}\BibitemShut {NoStop}%
\bibitem [{\citenamefont {Myatt}\ \emph {et~al.}(1997)\citenamefont {Myatt},
  \citenamefont {Burt}, \citenamefont {Ghrist}, \citenamefont {Cornell},\ and\
  \citenamefont {Wieman}}]{myatt_burt_1997}%
  \BibitemOpen
  \bibfield  {author} {\bibinfo {author} {\bibfnamefont {C.~J.}\ \bibnamefont
  {Myatt}}, \bibinfo {author} {\bibfnamefont {E.~A.}\ \bibnamefont {Burt}},
  \bibinfo {author} {\bibfnamefont {R.~W.}\ \bibnamefont {Ghrist}}, \bibinfo
  {author} {\bibfnamefont {E.~A.}\ \bibnamefont {Cornell}}, \ and\ \bibinfo
  {author} {\bibfnamefont {C.~E.}\ \bibnamefont {Wieman}},\ }\href {\doibase
  10.1103/PhysRevLett.78.586} {\bibfield  {journal} {\bibinfo  {journal} {Phys.
  Rev. Lett.}\ }\textbf {\bibinfo {volume} {78}},\ \bibinfo {pages} {586}
  (\bibinfo {year} {1997})}\BibitemShut {NoStop}%
\bibitem [{\citenamefont {Trippenbach}\ \emph {et~al.}(2000)\citenamefont
  {Trippenbach}, \citenamefont {Goral}, \citenamefont {Rzazewski},
  \citenamefont {Malomed},\ and\ \citenamefont {Band}}]{Trippenbach2000a}%
  \BibitemOpen
  \bibfield  {author} {\bibinfo {author} {\bibfnamefont {M.}~\bibnamefont
  {Trippenbach}}, \bibinfo {author} {\bibfnamefont {K.}~\bibnamefont {Goral}},
  \bibinfo {author} {\bibfnamefont {K.}~\bibnamefont {Rzazewski}}, \bibinfo
  {author} {\bibfnamefont {B.}~\bibnamefont {Malomed}}, \ and\ \bibinfo
  {author} {\bibfnamefont {Y.}~\bibnamefont {Band}},\ }\href {\doibase
  10.1088/0953-4075/33/19/314} {\bibfield  {journal} {\bibinfo  {journal} {J.
  Phys. B}\ }\textbf {\bibinfo {volume} {33}},\ \bibinfo {pages} {4017}
  (\bibinfo {year} {2000})}\BibitemShut {NoStop}%
\bibitem [{\citenamefont {Busch}\ and\ \citenamefont
  {Anglin}(2001)}]{Busch2001a}%
  \BibitemOpen
  \bibfield  {author} {\bibinfo {author} {\bibfnamefont {T.}~\bibnamefont
  {Busch}}\ and\ \bibinfo {author} {\bibfnamefont {J.~R.}\ \bibnamefont
  {Anglin}},\ }\href {\doibase 10.1103/PhysRevLett.87.010401} {\bibfield
  {journal} {\bibinfo  {journal} {Phys. Rev. Lett.}\ }\textbf {\bibinfo
  {volume} {87}},\ \bibinfo {pages} {010401} (\bibinfo {year}
  {2001})}\BibitemShut {NoStop}%
\bibitem [{\citenamefont {\"Ohberg}\ and\ \citenamefont
  {Santos}(2001)}]{Ohberg2001a}%
  \BibitemOpen
  \bibfield  {author} {\bibinfo {author} {\bibfnamefont {P.}~\bibnamefont
  {\"Ohberg}}\ and\ \bibinfo {author} {\bibfnamefont {L.}~\bibnamefont
  {Santos}},\ }\href {\doibase 10.1103/PhysRevLett.86.2918} {\bibfield
  {journal} {\bibinfo  {journal} {Phys. Rev. Lett.}\ }\textbf {\bibinfo
  {volume} {86}},\ \bibinfo {pages} {2918} (\bibinfo {year}
  {2001})}\BibitemShut {NoStop}%
\bibitem [{\citenamefont {Coen}\ and\ \citenamefont
  {Haelterman}(2001)}]{Coen2001a}%
  \BibitemOpen
  \bibfield  {author} {\bibinfo {author} {\bibfnamefont {S.}~\bibnamefont
  {Coen}}\ and\ \bibinfo {author} {\bibfnamefont {M.}~\bibnamefont
  {Haelterman}},\ }\href {\doibase 10.1103/PhysRevLett.87.140401} {\bibfield
  {journal} {\bibinfo  {journal} {Phys. Rev. Lett.}\ }\textbf {\bibinfo
  {volume} {87}},\ \bibinfo {pages} {140401} (\bibinfo {year}
  {2001})}\BibitemShut {NoStop}%
\bibitem [{\citenamefont {Berloff}(2005)}]{Berloff_05}%
  \BibitemOpen
  \bibfield  {author} {\bibinfo {author} {\bibfnamefont {N.~G.}\ \bibnamefont
  {Berloff}},\ }\href {\doibase 10.1103/PhysRevLett.94.120401} {\bibfield
  {journal} {\bibinfo  {journal} {Phys. Rev. Lett.}\ }\textbf {\bibinfo
  {volume} {94}},\ \bibinfo {pages} {120401} (\bibinfo {year}
  {2005})}\BibitemShut {NoStop}%
\bibitem [{\citenamefont {Kasamatsu}\ and\ \citenamefont
  {Tsubota}(2006)}]{Kasamatsu2006a}%
  \BibitemOpen
  \bibfield  {author} {\bibinfo {author} {\bibfnamefont {K.}~\bibnamefont
  {Kasamatsu}}\ and\ \bibinfo {author} {\bibfnamefont {M.}~\bibnamefont
  {Tsubota}},\ }\href {\doibase 10.1103/PhysRevA.74.013617} {\bibfield
  {journal} {\bibinfo  {journal} {Phys. Rev. A}\ }\textbf {\bibinfo {volume}
  {74}},\ \bibinfo {pages} {013617} (\bibinfo {year} {2006})}\BibitemShut
  {NoStop}%
\bibitem [{\citenamefont {Sasaki}\ \emph {et~al.}(2011)\citenamefont {Sasaki},
  \citenamefont {Suzuki},\ and\ \citenamefont {Saito}}]{Sasaki2011a}%
  \BibitemOpen
  \bibfield  {author} {\bibinfo {author} {\bibfnamefont {K.}~\bibnamefont
  {Sasaki}}, \bibinfo {author} {\bibfnamefont {N.}~\bibnamefont {Suzuki}}, \
  and\ \bibinfo {author} {\bibfnamefont {H.}~\bibnamefont {Saito}},\ }\href
  {\doibase 10.1103/PhysRevA.83.033602} {\bibfield  {journal} {\bibinfo
  {journal} {Phys. Rev. A}\ }\textbf {\bibinfo {volume} {83}},\ \bibinfo
  {pages} {033602} (\bibinfo {year} {2011})}\BibitemShut {NoStop}%
\bibitem [{\citenamefont {Pattinson}\ \emph {et~al.}(2013)\citenamefont
  {Pattinson}, \citenamefont {Billam}, \citenamefont {Gardiner}, \citenamefont
  {McCarron}, \citenamefont {Cho}, \citenamefont {Cornish}, \citenamefont
  {Parker},\ and\ \citenamefont {Proukakis}}]{pattinson_billam_2013}%
  \BibitemOpen
  \bibfield  {author} {\bibinfo {author} {\bibfnamefont {R.~W.}\ \bibnamefont
  {Pattinson}}, \bibinfo {author} {\bibfnamefont {T.~P.}\ \bibnamefont
  {Billam}}, \bibinfo {author} {\bibfnamefont {S.~A.}\ \bibnamefont
  {Gardiner}}, \bibinfo {author} {\bibfnamefont {D.~J.}\ \bibnamefont
  {McCarron}}, \bibinfo {author} {\bibfnamefont {H.~W.}\ \bibnamefont {Cho}},
  \bibinfo {author} {\bibfnamefont {S.~L.}\ \bibnamefont {Cornish}}, \bibinfo
  {author} {\bibfnamefont {N.~G.}\ \bibnamefont {Parker}}, \ and\ \bibinfo
  {author} {\bibfnamefont {N.~P.}\ \bibnamefont {Proukakis}},\ }\href {\doibase
  10.1103/PhysRevA.87.013625} {\bibfield  {journal} {\bibinfo  {journal} {Phys.
  Rev. A}\ }\textbf {\bibinfo {volume} {87}},\ \bibinfo {pages} {013625}
  (\bibinfo {year} {2013})}\BibitemShut {NoStop}%
\bibitem [{\citenamefont {Ronen}\ \emph {et~al.}(2008)\citenamefont {Ronen},
  \citenamefont {Bohn}, \citenamefont {Halmo},\ and\ \citenamefont
  {Edwards}}]{Ronen2008a}%
  \BibitemOpen
  \bibfield  {author} {\bibinfo {author} {\bibfnamefont {S.}~\bibnamefont
  {Ronen}}, \bibinfo {author} {\bibfnamefont {J.~L.}\ \bibnamefont {Bohn}},
  \bibinfo {author} {\bibfnamefont {L.~E.}\ \bibnamefont {Halmo}}, \ and\
  \bibinfo {author} {\bibfnamefont {M.}~\bibnamefont {Edwards}},\ }\href
  {\doibase 10.1103/PhysRevA.78.053613} {\bibfield  {journal} {\bibinfo
  {journal} {Phys. Rev. A}\ }\textbf {\bibinfo {volume} {78}},\ \bibinfo
  {pages} {053613} (\bibinfo {year} {2008})}\BibitemShut {NoStop}%
\bibitem [{\citenamefont {Achilleos}\ \emph {et~al.}(2012)\citenamefont
  {Achilleos}, \citenamefont {Yan}, \citenamefont {Kevrekidis},\ and\
  \citenamefont {Frantzeskakis}}]{Achilleos2012a}%
  \BibitemOpen
  \bibfield  {author} {\bibinfo {author} {\bibfnamefont {V.}~\bibnamefont
  {Achilleos}}, \bibinfo {author} {\bibfnamefont {D.}~\bibnamefont {Yan}},
  \bibinfo {author} {\bibfnamefont {P.~G.}\ \bibnamefont {Kevrekidis}}, \ and\
  \bibinfo {author} {\bibfnamefont {D.~J.}\ \bibnamefont {Frantzeskakis}},\
  }\href {http://stacks.iop.org/1367-2630/14/i=5/a=055006} {\bibfield
  {journal} {\bibinfo  {journal} {New J. Phys.}\ }\textbf {\bibinfo {volume}
  {14}},\ \bibinfo {pages} {055006} (\bibinfo {year} {2012})}\BibitemShut
  {NoStop}%
\bibitem [{\citenamefont {Pattinson}\ \emph
  {et~al.}(2014{\natexlab{a}})\citenamefont {Pattinson}, \citenamefont
  {Parker},\ and\ \citenamefont {Proukakis}}]{pattinson_2014}%
  \BibitemOpen
  \bibfield  {author} {\bibinfo {author} {\bibfnamefont {R.~W.}\ \bibnamefont
  {Pattinson}}, \bibinfo {author} {\bibfnamefont {N.~G.}\ \bibnamefont
  {Parker}}, \ and\ \bibinfo {author} {\bibfnamefont {N.~P.}\ \bibnamefont
  {Proukakis}},\ }\href {\doibase 10.1088/1742-6596/497/1/012029} {\bibfield
  {journal} {\bibinfo  {journal} {J. Phys. Conf. Ser.}\ }\textbf {\bibinfo
  {volume} {497}},\ \bibinfo {pages} {012029} (\bibinfo {year}
  {2014}{\natexlab{a}})}\BibitemShut {NoStop}%
\bibitem [{\citenamefont {Pattinson}\ \emph
  {et~al.}(2014{\natexlab{b}})\citenamefont {Pattinson}, \citenamefont
  {Proukakis},\ and\ \citenamefont {Parker}}]{pattinson_proukakis_14}%
  \BibitemOpen
  \bibfield  {author} {\bibinfo {author} {\bibfnamefont {R.~W.}\ \bibnamefont
  {Pattinson}}, \bibinfo {author} {\bibfnamefont {N.~P.}\ \bibnamefont
  {Proukakis}}, \ and\ \bibinfo {author} {\bibfnamefont {N.~G.}\ \bibnamefont
  {Parker}},\ }\href {\doibase 10.1103/PhysRevA.90.033625} {\bibfield
  {journal} {\bibinfo  {journal} {Phys. Rev. A}\ }\textbf {\bibinfo {volume}
  {90}},\ \bibinfo {pages} {033625} (\bibinfo {year}
  {2014}{\natexlab{b}})}\BibitemShut {NoStop}%
\bibitem [{\citenamefont {Sabbatini}\ \emph {et~al.}(2011)\citenamefont
  {Sabbatini}, \citenamefont {Zurek},\ and\ \citenamefont
  {Davis}}]{Sabbatini2011a}%
  \BibitemOpen
  \bibfield  {author} {\bibinfo {author} {\bibfnamefont {J.}~\bibnamefont
  {Sabbatini}}, \bibinfo {author} {\bibfnamefont {W.~H.}\ \bibnamefont
  {Zurek}}, \ and\ \bibinfo {author} {\bibfnamefont {M.~J.}\ \bibnamefont
  {Davis}},\ }\href {\doibase 10.1103/PhysRevLett.107.230402} {\bibfield
  {journal} {\bibinfo  {journal} {Phys. Rev. Lett.}\ }\textbf {\bibinfo
  {volume} {107}},\ \bibinfo {pages} {230402} (\bibinfo {year}
  {2011})}\BibitemShut {NoStop}%
\bibitem [{\citenamefont {Sabbatini}\ \emph {et~al.}(2012)\citenamefont
  {Sabbatini}, \citenamefont {Zurek},\ and\ \citenamefont
  {Davis}}]{Sabbatini2012a}%
  \BibitemOpen
  \bibfield  {author} {\bibinfo {author} {\bibfnamefont {J.}~\bibnamefont
  {Sabbatini}}, \bibinfo {author} {\bibfnamefont {W.~H.}\ \bibnamefont
  {Zurek}}, \ and\ \bibinfo {author} {\bibfnamefont {M.~J.}\ \bibnamefont
  {Davis}},\ }\href {\doibase 10.1088/1367-2630/14/9/095030} {\bibfield
  {journal} {\bibinfo  {journal} {New J. Phys.}\ }\textbf {\bibinfo {volume}
  {14}},\ \bibinfo {pages} {095030} (\bibinfo {year} {2012})}\BibitemShut
  {NoStop}%
\bibitem [{\citenamefont {\ifmmode~\acute{S}\else \'{S}\fi{}wis\l{}ocki}\ \emph
  {et~al.}(2013)\citenamefont {\ifmmode~\acute{S}\else \'{S}\fi{}wis\l{}ocki},
  \citenamefont {Witkowska}, \citenamefont {Dziarmaga},\ and\ \citenamefont
  {Matuszewski}}]{Swislocki2013a}%
  \BibitemOpen
  \bibfield  {author} {\bibinfo {author} {\bibfnamefont {T.}~\bibnamefont
  {\ifmmode~\acute{S}\else \'{S}\fi{}wis\l{}ocki}}, \bibinfo {author}
  {\bibfnamefont {E.}~\bibnamefont {Witkowska}}, \bibinfo {author}
  {\bibfnamefont {J.}~\bibnamefont {Dziarmaga}}, \ and\ \bibinfo {author}
  {\bibfnamefont {M.}~\bibnamefont {Matuszewski}},\ }\href {\doibase
  10.1103/PhysRevLett.110.045303} {\bibfield  {journal} {\bibinfo  {journal}
  {Phys. Rev. Lett.}\ }\textbf {\bibinfo {volume} {110}},\ \bibinfo {pages}
  {045303} (\bibinfo {year} {2013})}\BibitemShut {NoStop}%
\bibitem [{\citenamefont {Bradley}\ and\ \citenamefont
  {Blakie}(2014)}]{bradley_blakie_2014}%
  \BibitemOpen
  \bibfield  {author} {\bibinfo {author} {\bibfnamefont {A.~S.}\ \bibnamefont
  {Bradley}}\ and\ \bibinfo {author} {\bibfnamefont {P.~B.}\ \bibnamefont
  {Blakie}},\ }\href {\doibase 10.1103/PhysRevA.90.023631} {\bibfield
  {journal} {\bibinfo  {journal} {Phys. Rev. A}\ }\textbf {\bibinfo {volume}
  {90}},\ \bibinfo {pages} {023631} (\bibinfo {year} {2014})}\BibitemShut
  {NoStop}%
\bibitem [{\citenamefont {De}\ \emph {et~al.}(2014)\citenamefont {De},
  \citenamefont {Campbell}, \citenamefont {Price}, \citenamefont {Putra},
  \citenamefont {Anderson},\ and\ \citenamefont {Spielman}}]{De2014a}%
  \BibitemOpen
  \bibfield  {author} {\bibinfo {author} {\bibfnamefont {S.}~\bibnamefont
  {De}}, \bibinfo {author} {\bibfnamefont {D.~L.}\ \bibnamefont {Campbell}},
  \bibinfo {author} {\bibfnamefont {R.~M.}\ \bibnamefont {Price}}, \bibinfo
  {author} {\bibfnamefont {A.}~\bibnamefont {Putra}}, \bibinfo {author}
  {\bibfnamefont {B.~M.}\ \bibnamefont {Anderson}}, \ and\ \bibinfo {author}
  {\bibfnamefont {I.~B.}\ \bibnamefont {Spielman}},\ }\href {\doibase
  10.1103/PhysRevA.89.033631} {\bibfield  {journal} {\bibinfo  {journal} {Phys.
  Rev. A}\ }\textbf {\bibinfo {volume} {89}},\ \bibinfo {pages} {033631}
  (\bibinfo {year} {2014})}\BibitemShut {NoStop}%
\bibitem [{\citenamefont {Su}\ \emph {et~al.}(2013)\citenamefont {Su},
  \citenamefont {Gou}, \citenamefont {Bradley}, \citenamefont {Fialko},\ and\
  \citenamefont {Brand}}]{Su2013a}%
  \BibitemOpen
  \bibfield  {author} {\bibinfo {author} {\bibfnamefont {S.-W.}\ \bibnamefont
  {Su}}, \bibinfo {author} {\bibfnamefont {S.-C.}\ \bibnamefont {Gou}},
  \bibinfo {author} {\bibfnamefont {A.}~\bibnamefont {Bradley}}, \bibinfo
  {author} {\bibfnamefont {O.}~\bibnamefont {Fialko}}, \ and\ \bibinfo {author}
  {\bibfnamefont {J.}~\bibnamefont {Brand}},\ }\href {\doibase
  10.1103/PhysRevLett.110.215302} {\bibfield  {journal} {\bibinfo  {journal}
  {Phys. Rev. Lett.}\ }\textbf {\bibinfo {volume} {110}},\ \bibinfo {pages}
  {215302} (\bibinfo {year} {2013})}\BibitemShut {NoStop}%
\bibitem [{\citenamefont {Liu}\ \emph {et~al.}()\citenamefont {Liu},
  \citenamefont {Pattinson}, \citenamefont {Billam}, \citenamefont {Gardiner},
  \citenamefont {Cornish}, \citenamefont {Huang}, \citenamefont {Lin},
  \citenamefont {Gou}, \citenamefont {Parker},\ and\ \citenamefont
  {Proukakis}}]{liu_pattinson_2014}%
  \BibitemOpen
  \bibfield  {author} {\bibinfo {author} {\bibfnamefont {I.-K.}\ \bibnamefont
  {Liu}}, \bibinfo {author} {\bibfnamefont {R.~W.}\ \bibnamefont {Pattinson}},
  \bibinfo {author} {\bibfnamefont {T.~P.}\ \bibnamefont {Billam}}, \bibinfo
  {author} {\bibfnamefont {S.~A.}\ \bibnamefont {Gardiner}}, \bibinfo {author}
  {\bibfnamefont {S.~L.}\ \bibnamefont {Cornish}}, \bibinfo {author}
  {\bibfnamefont {T.-M.}\ \bibnamefont {Huang}}, \bibinfo {author}
  {\bibfnamefont {W.-W.}\ \bibnamefont {Lin}}, \bibinfo {author} {\bibfnamefont
  {S.-C.}\ \bibnamefont {Gou}}, \bibinfo {author} {\bibfnamefont {N.~G.}\
  \bibnamefont {Parker}}, \ and\ \bibinfo {author} {\bibfnamefont {N.~P.}\
  \bibnamefont {Proukakis}},\ }\href@noop {} {\bibinfo  {journal}
  {arXiv:1408.0891}\ }\BibitemShut {NoStop}%
\bibitem [{\citenamefont {Mason}\ and\ \citenamefont
  {Gardiner}(2014)}]{mason_gardiner_2014}%
  \BibitemOpen
\bibfield  {journal} {  }\bibfield  {author} {\bibinfo {author} {\bibfnamefont
  {P.}~\bibnamefont {Mason}}\ and\ \bibinfo {author} {\bibfnamefont {S.~A.}\
  \bibnamefont {Gardiner}},\ }\href {\doibase 10.1103/PhysRevA.89.043617}
  {\bibfield  {journal} {\bibinfo  {journal} {Phys. Rev. A}\ }\textbf {\bibinfo
  {volume} {89}},\ \bibinfo {pages} {043617} (\bibinfo {year}
  {2014})}\BibitemShut {NoStop}%
\bibitem [{\citenamefont {Rooney}\ \emph {et~al.}(2013)\citenamefont {Rooney},
  \citenamefont {Neely}, \citenamefont {Anderson},\ and\ \citenamefont
  {Bradley}}]{rooney_neely_13}%
  \BibitemOpen
  \bibfield  {author} {\bibinfo {author} {\bibfnamefont {S.~J.}\ \bibnamefont
  {Rooney}}, \bibinfo {author} {\bibfnamefont {T.~W.}\ \bibnamefont {Neely}},
  \bibinfo {author} {\bibfnamefont {B.~P.}\ \bibnamefont {Anderson}}, \ and\
  \bibinfo {author} {\bibfnamefont {A.~S.}\ \bibnamefont {Bradley}},\ }\href
  {\doibase 10.1103/PhysRevA.88.063620} {\bibfield  {journal} {\bibinfo
  {journal} {Phys. Rev. A}\ }\textbf {\bibinfo {volume} {88}},\ \bibinfo
  {pages} {063620} (\bibinfo {year} {2013})}\BibitemShut {NoStop}%
\bibitem [{\citenamefont {Bradley}\ \emph {et~al.}(2005)\citenamefont
  {Bradley}, \citenamefont {Blakie},\ and\ \citenamefont
  {Gardiner}}]{bradley_blakie_05}%
  \BibitemOpen
  \bibfield  {author} {\bibinfo {author} {\bibfnamefont {A.~S.}\ \bibnamefont
  {Bradley}}, \bibinfo {author} {\bibfnamefont {P.~B.}\ \bibnamefont {Blakie}},
  \ and\ \bibinfo {author} {\bibfnamefont {C.~W.}\ \bibnamefont {Gardiner}},\
  }\href {http://stacks.iop.org/0953-4075/38/i=23/a=008} {\bibfield  {journal}
  {\bibinfo  {journal} {Journal of Physics B: Atomic, Molecular and Optical
  Physics}\ }\textbf {\bibinfo {volume} {38}},\ \bibinfo {pages} {4259}
  (\bibinfo {year} {2005})}\BibitemShut {NoStop}%
\bibitem [{\citenamefont {Nikuni}\ and\ \citenamefont
  {Williams}(2003)}]{nikuni_williams_2003}%
  \BibitemOpen
  \bibfield  {author} {\bibinfo {author} {\bibfnamefont {T.}~\bibnamefont
  {Nikuni}}\ and\ \bibinfo {author} {\bibfnamefont {J.}~\bibnamefont
  {Williams}},\ }\href {\doibase 10.1023/A:1026206724886} {\bibfield  {journal}
  {\bibinfo  {journal} {J. Low Temp. Phys.}\ }\textbf {\bibinfo {volume}
  {133}},\ \bibinfo {pages} {323} (\bibinfo {year} {2003})}\BibitemShut
  {NoStop}%
\bibitem [{\citenamefont {Endo}\ and\ \citenamefont
  {Nikuni}(2011)}]{endo_nikuni_2011}%
  \BibitemOpen
  \bibfield  {author} {\bibinfo {author} {\bibfnamefont {Y.}~\bibnamefont
  {Endo}}\ and\ \bibinfo {author} {\bibfnamefont {T.}~\bibnamefont {Nikuni}},\
  }\href@noop {} {\bibfield  {journal} {\bibinfo  {journal} {J. Low Temp.
  Phys.}\ }\textbf {\bibinfo {volume} {163}},\ \bibinfo {pages} {92} (\bibinfo
  {year} {2011})}\BibitemShut {NoStop}%
\bibitem [{\citenamefont {Edmonds}\ \emph {et~al.}(2015)\citenamefont
  {Edmonds}, \citenamefont {Lee},\ and\ \citenamefont
  {Proukakis}}]{edmonds_lee_2015}%
  \BibitemOpen
  \bibfield  {author} {\bibinfo {author} {\bibfnamefont {M.~J.}\ \bibnamefont
  {Edmonds}}, \bibinfo {author} {\bibfnamefont {K.~L.}\ \bibnamefont {Lee}}, \
  and\ \bibinfo {author} {\bibfnamefont {N.~P.}\ \bibnamefont {Proukakis}},\
  }\href {\doibase 10.1103/PhysRevA.91.011602} {\bibfield  {journal} {\bibinfo
  {journal} {Phys. Rev. A}\ }\textbf {\bibinfo {volume} {91}},\ \bibinfo
  {pages} {011602} (\bibinfo {year} {2015})}\BibitemShut {NoStop}%
\bibitem [{Note1()}]{Note1}%
  \BibitemOpen
  \bibinfo {note} {Subtle issues associated with the validity of this statement
  are addressed in the related number-conserving approaches of Refs.~\cite
  {morgan_2000,morgan_2004,gardiner_morgan_2007,*gardiner_morgan_2007err}.}\BibitemShut
  {Stop}%
\bibitem [{\citenamefont {Davis}\ \emph
  {et~al.}(2001{\natexlab{b}})\citenamefont {Davis}, \citenamefont {Ballagh},\
  and\ \citenamefont {Burnett}}]{davis_ballagh_2001}%
  \BibitemOpen
  \bibfield  {author} {\bibinfo {author} {\bibfnamefont {M.~J.}\ \bibnamefont
  {Davis}}, \bibinfo {author} {\bibfnamefont {R.~J.}\ \bibnamefont {Ballagh}},
  \ and\ \bibinfo {author} {\bibfnamefont {K.}~\bibnamefont {Burnett}},\ }\href
  {http://stacks.iop.org/0953-4075/34/i=22/a=316} {\bibfield  {journal}
  {\bibinfo  {journal} {J. Phys. B: At. Mol. Opt. Phys.}\ }\textbf {\bibinfo
  {volume} {34}},\ \bibinfo {pages} {4487} (\bibinfo {year}
  {2001}{\natexlab{b}})}\BibitemShut {NoStop}%
\bibitem [{\citenamefont {Imamovi\ifmmode \acute{c}\else
  \'{c}\fi{}-Tomasovi\ifmmode~\acute{c}\else \'{c}\fi{}}\ and\ \citenamefont
  {Griffin}(1999)}]{imamovic_griffin_1999}%
  \BibitemOpen
  \bibfield  {author} {\bibinfo {author} {\bibfnamefont {M.}~\bibnamefont
  {Imamovi\ifmmode \acute{c}\else \'{c}\fi{}-Tomasovi\ifmmode~\acute{c}\else
  \'{c}\fi{}}}\ and\ \bibinfo {author} {\bibfnamefont {A.}~\bibnamefont
  {Griffin}},\ }\href {\doibase 10.1103/PhysRevA.60.494} {\bibfield  {journal}
  {\bibinfo  {journal} {Phys. Rev. A}\ }\textbf {\bibinfo {volume} {60}},\
  \bibinfo {pages} {494} (\bibinfo {year} {1999})}\BibitemShut {NoStop}%
\bibitem [{\citenamefont {Imamovi\ifmmode \acute{c}\else
  \'{c}\fi{}-Tomasovi\ifmmode~\acute{c}\else
  \'{c}\fi{}}(2001)}]{imamovic_2001}%
  \BibitemOpen
  \bibfield  {author} {\bibinfo {author} {\bibfnamefont {M.}~\bibnamefont
  {Imamovi\ifmmode \acute{c}\else \'{c}\fi{}-Tomasovi\ifmmode~\acute{c}\else
  \'{c}\fi{}}},\ }\emph {\bibinfo {title} {{K}adanoff-{B}aym kinetic theory for
  a trapped Bose-condensed gas}},\ \href@noop {} {Ph.D. thesis},\ \bibinfo
  {school} {University of Toronto} (\bibinfo {year} {2001})\BibitemShut
  {NoStop}%
\bibitem [{\citenamefont {Rusch}\ and\ \citenamefont
  {Burnett}(1999)}]{rusch_burnett_1999}%
  \BibitemOpen
  \bibfield  {author} {\bibinfo {author} {\bibfnamefont {M.}~\bibnamefont
  {Rusch}}\ and\ \bibinfo {author} {\bibfnamefont {K.}~\bibnamefont
  {Burnett}},\ }\href {\doibase 10.1103/PhysRevA.59.3851} {\bibfield  {journal}
  {\bibinfo  {journal} {Phys. Rev. A}\ }\textbf {\bibinfo {volume} {59}},\
  \bibinfo {pages} {3851} (\bibinfo {year} {1999})}\BibitemShut {NoStop}%
\bibitem [{\citenamefont {Morgan}(2000)}]{morgan_2000}%
  \BibitemOpen
  \bibfield  {author} {\bibinfo {author} {\bibfnamefont {S.~A.}\ \bibnamefont
  {Morgan}},\ }\href@noop {} {\bibfield  {journal} {\bibinfo  {journal} {J.
  Phys. B: At. Mol. Opt.}\ }\textbf {\bibinfo {volume} {33}},\ \bibinfo {pages}
  {3847} (\bibinfo {year} {2000})}\BibitemShut {NoStop}%
\bibitem [{\citenamefont {Bruus}\ and\ \citenamefont
  {Flensberg}(2006)}]{bruus_flensberg_2006}%
  \BibitemOpen
  \bibfield  {author} {\bibinfo {author} {\bibfnamefont {H.}~\bibnamefont
  {Bruus}}\ and\ \bibinfo {author} {\bibfnamefont {K.}~\bibnamefont
  {Flensberg}},\ }\href@noop {} {\emph {\bibinfo {title} {Many-Body Quantum
  Theory in Condensed Matter Physics}}}\ (\bibinfo  {publisher} {Oxford
  University Press},\ \bibinfo {year} {2006})\BibitemShut {NoStop}%
\bibitem [{\citenamefont {Nikuni}\ \emph {et~al.}(1999)\citenamefont {Nikuni},
  \citenamefont {Zaremba},\ and\ \citenamefont
  {Griffin}}]{nikuni_zaremba_1999}%
  \BibitemOpen
  \bibfield  {author} {\bibinfo {author} {\bibfnamefont {T.}~\bibnamefont
  {Nikuni}}, \bibinfo {author} {\bibfnamefont {E.}~\bibnamefont {Zaremba}}, \
  and\ \bibinfo {author} {\bibfnamefont {A.}~\bibnamefont {Griffin}},\ }\href
  {\doibase 10.1103/PhysRevLett.83.10} {\bibfield  {journal} {\bibinfo
  {journal} {Phys. Rev. Lett.}\ }\textbf {\bibinfo {volume} {83}},\ \bibinfo
  {pages} {10} (\bibinfo {year} {1999})}\BibitemShut {NoStop}%
\bibitem [{\citenamefont {Jackson}\ and\ \citenamefont
  {Zaremba}(2001{\natexlab{b}})}]{jackson_zaremba_2001}%
  \BibitemOpen
  \bibfield  {author} {\bibinfo {author} {\bibfnamefont {B.}~\bibnamefont
  {Jackson}}\ and\ \bibinfo {author} {\bibfnamefont {E.}~\bibnamefont
  {Zaremba}},\ }\href {\doibase 10.1103/PhysRevLett.87.100404} {\bibfield
  {journal} {\bibinfo  {journal} {Phys. Rev. Lett.}\ }\textbf {\bibinfo
  {volume} {87}},\ \bibinfo {pages} {100404} (\bibinfo {year}
  {2001}{\natexlab{b}})}\BibitemShut {NoStop}%
\bibitem [{\citenamefont {Jackson}\ and\ \citenamefont
  {Zaremba}(2002{\natexlab{b}})}]{jackson_zaremba_2002a}%
  \BibitemOpen
  \bibfield  {author} {\bibinfo {author} {\bibfnamefont {B.}~\bibnamefont
  {Jackson}}\ and\ \bibinfo {author} {\bibfnamefont {E.}~\bibnamefont
  {Zaremba}},\ }\href {\doibase 10.1103/PhysRevLett.89.150402} {\bibfield
  {journal} {\bibinfo  {journal} {Phys. Rev. Lett.}\ }\textbf {\bibinfo
  {volume} {89}},\ \bibinfo {pages} {150402} (\bibinfo {year}
  {2002}{\natexlab{b}})}\BibitemShut {NoStop}%
\bibitem [{\citenamefont {Jackson}\ and\ \citenamefont
  {Zaremba}(2002{\natexlab{c}})}]{jackson_zaremba_2002}%
  \BibitemOpen
  \bibfield  {author} {\bibinfo {author} {\bibfnamefont {B.}~\bibnamefont
  {Jackson}}\ and\ \bibinfo {author} {\bibfnamefont {E.}~\bibnamefont
  {Zaremba}},\ }\href {\doibase 10.1103/PhysRevA.66.033606} {\bibfield
  {journal} {\bibinfo  {journal} {Phys. Rev. A}\ }\textbf {\bibinfo {volume}
  {66}},\ \bibinfo {pages} {033606} (\bibinfo {year}
  {2002}{\natexlab{c}})}\BibitemShut {NoStop}%
\bibitem [{\citenamefont {Allen}\ \emph
  {et~al.}(2013{\natexlab{b}})\citenamefont {Allen}, \citenamefont {Zaremba},
  \citenamefont {Barenghi},\ and\ \citenamefont
  {Proukakis}}]{allen_zaremba_2013}%
  \BibitemOpen
  \bibfield  {author} {\bibinfo {author} {\bibfnamefont {A.~J.}\ \bibnamefont
  {Allen}}, \bibinfo {author} {\bibfnamefont {E.}~\bibnamefont {Zaremba}},
  \bibinfo {author} {\bibfnamefont {C.~F.}\ \bibnamefont {Barenghi}}, \ and\
  \bibinfo {author} {\bibfnamefont {N.~P.}\ \bibnamefont {Proukakis}},\ }\href
  {\doibase 10.1103/PhysRevA.87.013630} {\bibfield  {journal} {\bibinfo
  {journal} {Phys. Rev. A}\ }\textbf {\bibinfo {volume} {87}},\ \bibinfo
  {pages} {013630} (\bibinfo {year} {2013}{\natexlab{b}})}\BibitemShut
  {NoStop}%
\bibitem [{\citenamefont {Thalhammer}\ \emph {et~al.}(2008)\citenamefont
  {Thalhammer}, \citenamefont {Barontini}, \citenamefont {De~Sarlo},
  \citenamefont {Catani}, \citenamefont {Minardi},\ and\ \citenamefont
  {Inguscio}}]{thalhammer_barontini_2008}%
  \BibitemOpen
  \bibfield  {author} {\bibinfo {author} {\bibfnamefont {G.}~\bibnamefont
  {Thalhammer}}, \bibinfo {author} {\bibfnamefont {G.}~\bibnamefont
  {Barontini}}, \bibinfo {author} {\bibfnamefont {L.}~\bibnamefont {De~Sarlo}},
  \bibinfo {author} {\bibfnamefont {J.}~\bibnamefont {Catani}}, \bibinfo
  {author} {\bibfnamefont {F.}~\bibnamefont {Minardi}}, \ and\ \bibinfo
  {author} {\bibfnamefont {M.}~\bibnamefont {Inguscio}},\ }\href {\doibase
  10.1103/PhysRevLett.100.210402} {\bibfield  {journal} {\bibinfo  {journal}
  {Phys. Rev. Lett.}\ }\textbf {\bibinfo {volume} {100}},\ \bibinfo {pages}
  {210402} (\bibinfo {year} {2008})}\BibitemShut {NoStop}%
\bibitem [{\citenamefont {Holzmann}\ \emph {et~al.}(1999)\citenamefont
  {Holzmann}, \citenamefont {Krauth},\ and\ \citenamefont
  {Naraschewski}}]{holzmann_krauth_1999}%
  \BibitemOpen
  \bibfield  {author} {\bibinfo {author} {\bibfnamefont {M.}~\bibnamefont
  {Holzmann}}, \bibinfo {author} {\bibfnamefont {W.}~\bibnamefont {Krauth}}, \
  and\ \bibinfo {author} {\bibfnamefont {M.}~\bibnamefont {Naraschewski}},\
  }\href {\doibase 10.1103/PhysRevA.59.2956} {\bibfield  {journal} {\bibinfo
  {journal} {Phys. Rev. A}\ }\textbf {\bibinfo {volume} {59}},\ \bibinfo
  {pages} {2956} (\bibinfo {year} {1999})}\BibitemShut {NoStop}%
\bibitem [{\citenamefont {Dalfovo}\ \emph {et~al.}(1999)\citenamefont
  {Dalfovo}, \citenamefont {Giorgini}, \citenamefont {Pitaevskii},\ and\
  \citenamefont {Stringari}}]{dalfovo_giorgini_1999}%
  \BibitemOpen
  \bibfield  {author} {\bibinfo {author} {\bibfnamefont {F.}~\bibnamefont
  {Dalfovo}}, \bibinfo {author} {\bibfnamefont {S.}~\bibnamefont {Giorgini}},
  \bibinfo {author} {\bibfnamefont {L.~P.}\ \bibnamefont {Pitaevskii}}, \ and\
  \bibinfo {author} {\bibfnamefont {S.}~\bibnamefont {Stringari}},\ }\href
  {\doibase 10.1103/RevModPhys.71.463} {\bibfield  {journal} {\bibinfo
  {journal} {Rev. Mod. Phys.}\ }\textbf {\bibinfo {volume} {71}},\ \bibinfo
  {pages} {463} (\bibinfo {year} {1999})}\BibitemShut {NoStop}%
\bibitem [{\citenamefont {Hutchinson}\ \emph {et~al.}(1997)\citenamefont
  {Hutchinson}, \citenamefont {Zaremba},\ and\ \citenamefont
  {Griffin}}]{hutchinson_zaremba_1997}%
  \BibitemOpen
  \bibfield  {author} {\bibinfo {author} {\bibfnamefont {D.~A.~W.}\
  \bibnamefont {Hutchinson}}, \bibinfo {author} {\bibfnamefont
  {E.}~\bibnamefont {Zaremba}}, \ and\ \bibinfo {author} {\bibfnamefont
  {A.}~\bibnamefont {Griffin}},\ }\href {\doibase 10.1103/PhysRevLett.78.1842}
  {\bibfield  {journal} {\bibinfo  {journal} {Phys. Rev. Lett.}\ }\textbf
  {\bibinfo {volume} {78}},\ \bibinfo {pages} {1842} (\bibinfo {year}
  {1997})}\BibitemShut {NoStop}%
\bibitem [{\citenamefont {Grossmann}\ and\ \citenamefont
  {Holthaus}(1995)}]{grossmann_holthaus_1995}%
  \BibitemOpen
  \bibfield  {author} {\bibinfo {author} {\bibfnamefont {S.}~\bibnamefont
  {Grossmann}}\ and\ \bibinfo {author} {\bibfnamefont {M.}~\bibnamefont
  {Holthaus}},\ }\href {\doibase
  http://dx.doi.org/10.1016/0375-9601(95)00766-V} {\bibfield  {journal}
  {\bibinfo  {journal} {Phys. Lett. A}\ }\textbf {\bibinfo {volume} {208}},\
  \bibinfo {pages} {188} (\bibinfo {year} {1995})}\BibitemShut {NoStop}%
\bibitem [{\citenamefont {Ketterle}\ and\ \citenamefont {van
  Druten}(1996)}]{ketterle_vandruten_1996}%
  \BibitemOpen
  \bibfield  {author} {\bibinfo {author} {\bibfnamefont {W.}~\bibnamefont
  {Ketterle}}\ and\ \bibinfo {author} {\bibfnamefont {N.~J.}\ \bibnamefont {van
  Druten}},\ }\href {\doibase 10.1103/PhysRevA.54.656} {\bibfield  {journal}
  {\bibinfo  {journal} {Phys. Rev. A}\ }\textbf {\bibinfo {volume} {54}},\
  \bibinfo {pages} {656} (\bibinfo {year} {1996})}\BibitemShut {NoStop}%
\bibitem [{\citenamefont {Kirsten}\ and\ \citenamefont
  {Toms}(1996)}]{kirsten_toms_1996}%
  \BibitemOpen
  \bibfield  {author} {\bibinfo {author} {\bibfnamefont {K.}~\bibnamefont
  {Kirsten}}\ and\ \bibinfo {author} {\bibfnamefont {D.~J.}\ \bibnamefont
  {Toms}},\ }\href {\doibase 10.1103/PhysRevA.54.4188} {\bibfield  {journal}
  {\bibinfo  {journal} {Phys. Rev. A}\ }\textbf {\bibinfo {volume} {54}},\
  \bibinfo {pages} {4188} (\bibinfo {year} {1996})}\BibitemShut {NoStop}%
\bibitem [{\citenamefont {Giorgini}\ \emph {et~al.}(1996)\citenamefont
  {Giorgini}, \citenamefont {Pitaevskii},\ and\ \citenamefont
  {Stringari}}]{giorgini_pitaevskii_1996}%
  \BibitemOpen
  \bibfield  {author} {\bibinfo {author} {\bibfnamefont {S.}~\bibnamefont
  {Giorgini}}, \bibinfo {author} {\bibfnamefont {L.~P.}\ \bibnamefont
  {Pitaevskii}}, \ and\ \bibinfo {author} {\bibfnamefont {S.}~\bibnamefont
  {Stringari}},\ }\href {\doibase 10.1103/PhysRevA.54.R4633} {\bibfield
  {journal} {\bibinfo  {journal} {Phys. Rev. A}\ }\textbf {\bibinfo {volume}
  {54}},\ \bibinfo {pages} {R4633} (\bibinfo {year} {1996})}\BibitemShut
  {NoStop}%
\bibitem [{\citenamefont {Gr\"uter}\ \emph {et~al.}(1997)\citenamefont
  {Gr\"uter}, \citenamefont {Ceperley},\ and\ \citenamefont
  {Lalo\"e}}]{gruter_ceperley_1997}%
  \BibitemOpen
  \bibfield  {author} {\bibinfo {author} {\bibfnamefont {P.}~\bibnamefont
  {Gr\"uter}}, \bibinfo {author} {\bibfnamefont {D.}~\bibnamefont {Ceperley}},
  \ and\ \bibinfo {author} {\bibfnamefont {F.}~\bibnamefont {Lalo\"e}},\ }\href
  {\doibase 10.1103/PhysRevLett.79.3549} {\bibfield  {journal} {\bibinfo
  {journal} {Phys. Rev. Lett.}\ }\textbf {\bibinfo {volume} {79}},\ \bibinfo
  {pages} {3549} (\bibinfo {year} {1997})}\BibitemShut {NoStop}%
\bibitem [{\citenamefont {Baym}\ \emph {et~al.}(1999)\citenamefont {Baym},
  \citenamefont {Blaizot}, \citenamefont {Holzmann}, \citenamefont {Lalo\"e},\
  and\ \citenamefont {Vautherin}}]{baym_blaizot_1999}%
  \BibitemOpen
  \bibfield  {author} {\bibinfo {author} {\bibfnamefont {G.}~\bibnamefont
  {Baym}}, \bibinfo {author} {\bibfnamefont {J.-P.}\ \bibnamefont {Blaizot}},
  \bibinfo {author} {\bibfnamefont {M.}~\bibnamefont {Holzmann}}, \bibinfo
  {author} {\bibfnamefont {F.}~\bibnamefont {Lalo\"e}}, \ and\ \bibinfo
  {author} {\bibfnamefont {D.}~\bibnamefont {Vautherin}},\ }\href {\doibase
  10.1103/PhysRevLett.83.1703} {\bibfield  {journal} {\bibinfo  {journal}
  {Phys. Rev. Lett.}\ }\textbf {\bibinfo {volume} {83}},\ \bibinfo {pages}
  {1703} (\bibinfo {year} {1999})}\BibitemShut {NoStop}%
\bibitem [{\citenamefont {Wu}\ and\ \citenamefont {Foot}(1996)}]{wu_foot_1996}%
  \BibitemOpen
  \bibfield  {author} {\bibinfo {author} {\bibfnamefont {H.}~\bibnamefont
  {Wu}}\ and\ \bibinfo {author} {\bibfnamefont {C.~J.}\ \bibnamefont {Foot}},\
  }\href {http://stacks.iop.org/0953-4075/29/i=8/a=003} {\bibfield  {journal}
  {\bibinfo  {journal} {J. Phys. B: At. Mol. Opt.}\ }\textbf {\bibinfo {volume}
  {29}},\ \bibinfo {pages} {L321} (\bibinfo {year} {1996})}\BibitemShut
  {NoStop}%
\bibitem [{\citenamefont {Delannoy}\ \emph {et~al.}(2001)\citenamefont
  {Delannoy}, \citenamefont {Murdoch}, \citenamefont {Boyer}, \citenamefont
  {Josse}, \citenamefont {Bouyer},\ and\ \citenamefont
  {Aspect}}]{delannoy_murdoch_2001}%
  \BibitemOpen
  \bibfield  {author} {\bibinfo {author} {\bibfnamefont {G.}~\bibnamefont
  {Delannoy}}, \bibinfo {author} {\bibfnamefont {S.~G.}\ \bibnamefont
  {Murdoch}}, \bibinfo {author} {\bibfnamefont {V.}~\bibnamefont {Boyer}},
  \bibinfo {author} {\bibfnamefont {V.}~\bibnamefont {Josse}}, \bibinfo
  {author} {\bibfnamefont {P.}~\bibnamefont {Bouyer}}, \ and\ \bibinfo {author}
  {\bibfnamefont {A.}~\bibnamefont {Aspect}},\ }\href {\doibase
  10.1103/PhysRevA.63.051602} {\bibfield  {journal} {\bibinfo  {journal} {Phys.
  Rev. A}\ }\textbf {\bibinfo {volume} {63}},\ \bibinfo {pages} {051602}
  (\bibinfo {year} {2001})}\BibitemShut {NoStop}%
\bibitem [{\citenamefont {Mosk}\ \emph {et~al.}(2001)\citenamefont {Mosk},
  \citenamefont {Kraft}, \citenamefont {Mudrich}, \citenamefont {Singer},
  \citenamefont {Wohlleben}, \citenamefont {Grimm},\ and\ \citenamefont
  {Weidem\"uller}}]{mosk_kraft_2001}%
  \BibitemOpen
  \bibfield  {author} {\bibinfo {author} {\bibfnamefont {A.}~\bibnamefont
  {Mosk}}, \bibinfo {author} {\bibfnamefont {S.}~\bibnamefont {Kraft}},
  \bibinfo {author} {\bibfnamefont {M.}~\bibnamefont {Mudrich}}, \bibinfo
  {author} {\bibfnamefont {K.}~\bibnamefont {Singer}}, \bibinfo {author}
  {\bibfnamefont {W.}~\bibnamefont {Wohlleben}}, \bibinfo {author}
  {\bibfnamefont {R.}~\bibnamefont {Grimm}}, \ and\ \bibinfo {author}
  {\bibfnamefont {M.}~\bibnamefont {Weidem\"uller}},\ }\href {\doibase
  10.1007/s003400100743} {\bibfield  {journal} {\bibinfo  {journal} {Appl.
  Phys. B}\ }\textbf {\bibinfo {volume} {73}},\ \bibinfo {pages} {791}
  (\bibinfo {year} {2001})}\BibitemShut {NoStop}%
\bibitem [{\citenamefont {Anderlini}\ \emph {et~al.}(2005)\citenamefont
  {Anderlini}, \citenamefont {Ciampini}, \citenamefont {Cossart}, \citenamefont
  {Courtade}, \citenamefont {Cristiani}, \citenamefont {Sias}, \citenamefont
  {Morsch},\ and\ \citenamefont {Arimondo}}]{anderlini_ciampini_2005}%
  \BibitemOpen
  \bibfield  {author} {\bibinfo {author} {\bibfnamefont {M.}~\bibnamefont
  {Anderlini}}, \bibinfo {author} {\bibfnamefont {D.}~\bibnamefont {Ciampini}},
  \bibinfo {author} {\bibfnamefont {D.}~\bibnamefont {Cossart}}, \bibinfo
  {author} {\bibfnamefont {E.}~\bibnamefont {Courtade}}, \bibinfo {author}
  {\bibfnamefont {M.}~\bibnamefont {Cristiani}}, \bibinfo {author}
  {\bibfnamefont {C.}~\bibnamefont {Sias}}, \bibinfo {author} {\bibfnamefont
  {O.}~\bibnamefont {Morsch}}, \ and\ \bibinfo {author} {\bibfnamefont
  {E.}~\bibnamefont {Arimondo}},\ }\href {\doibase 10.1103/PhysRevA.72.033408}
  {\bibfield  {journal} {\bibinfo  {journal} {Phys. Rev. A}\ }\textbf {\bibinfo
  {volume} {72}},\ \bibinfo {pages} {033408} (\bibinfo {year}
  {2005})}\BibitemShut {NoStop}%
\bibitem [{\citenamefont {Burger}\ \emph {et~al.}(1999)\citenamefont {Burger},
  \citenamefont {Bongs}, \citenamefont {Dettmer}, \citenamefont {Ertmer},
  \citenamefont {Sengstock}, \citenamefont {Sanpera}, \citenamefont
  {Shlyapnikov},\ and\ \citenamefont {Lewenstein}}]{burger_bongs_1999}%
  \BibitemOpen
  \bibfield  {author} {\bibinfo {author} {\bibfnamefont {S.}~\bibnamefont
  {Burger}}, \bibinfo {author} {\bibfnamefont {K.}~\bibnamefont {Bongs}},
  \bibinfo {author} {\bibfnamefont {S.}~\bibnamefont {Dettmer}}, \bibinfo
  {author} {\bibfnamefont {W.}~\bibnamefont {Ertmer}}, \bibinfo {author}
  {\bibfnamefont {K.}~\bibnamefont {Sengstock}}, \bibinfo {author}
  {\bibfnamefont {A.}~\bibnamefont {Sanpera}}, \bibinfo {author} {\bibfnamefont
  {G.~V.}\ \bibnamefont {Shlyapnikov}}, \ and\ \bibinfo {author} {\bibfnamefont
  {M.}~\bibnamefont {Lewenstein}},\ }\href {\doibase
  10.1103/PhysRevLett.83.5198} {\bibfield  {journal} {\bibinfo  {journal}
  {Phys. Rev. Lett.}\ }\textbf {\bibinfo {volume} {83}},\ \bibinfo {pages}
  {5198} (\bibinfo {year} {1999})}\BibitemShut {NoStop}%
\bibitem [{\citenamefont {Becker}\ \emph {et~al.}(2008)\citenamefont {Becker},
  \citenamefont {Stellmer}, \citenamefont {Soltan-Panahi}, \citenamefont
  {D\"orscher}, \citenamefont {Baumert}, \citenamefont {Richter}, \citenamefont
  {Kronj\"ager}, \citenamefont {Bongs},\ and\ \citenamefont
  {Sengstock}}]{becker_stellmer_2008}%
  \BibitemOpen
  \bibfield  {author} {\bibinfo {author} {\bibfnamefont {C.}~\bibnamefont
  {Becker}}, \bibinfo {author} {\bibfnamefont {S.}~\bibnamefont {Stellmer}},
  \bibinfo {author} {\bibfnamefont {P.}~\bibnamefont {Soltan-Panahi}}, \bibinfo
  {author} {\bibfnamefont {S.}~\bibnamefont {D\"orscher}}, \bibinfo {author}
  {\bibfnamefont {M.}~\bibnamefont {Baumert}}, \bibinfo {author} {\bibfnamefont
  {E.-M.}\ \bibnamefont {Richter}}, \bibinfo {author} {\bibfnamefont
  {J.}~\bibnamefont {Kronj\"ager}}, \bibinfo {author} {\bibfnamefont
  {K.}~\bibnamefont {Bongs}}, \ and\ \bibinfo {author} {\bibfnamefont
  {K.}~\bibnamefont {Sengstock}},\ }\href {\doibase 10.1038/nphys962}
  {\bibfield  {journal} {\bibinfo  {journal} {Nat. Phys.}\ }\textbf {\bibinfo
  {volume} {4}},\ \bibinfo {pages} {496} (\bibinfo {year} {2008})}\BibitemShut
  {NoStop}%
\bibitem [{\citenamefont {Donadello}\ \emph {et~al.}(2014)\citenamefont
  {Donadello}, \citenamefont {Serafini}, \citenamefont {Tylutki}, \citenamefont
  {Pitaevskii}, \citenamefont {Dalfovo}, \citenamefont {Lamporesi},\ and\
  \citenamefont {Ferrari}}]{donadello_serafini_2014}%
  \BibitemOpen
  \bibfield  {author} {\bibinfo {author} {\bibfnamefont {S.}~\bibnamefont
  {Donadello}}, \bibinfo {author} {\bibfnamefont {S.}~\bibnamefont {Serafini}},
  \bibinfo {author} {\bibfnamefont {M.}~\bibnamefont {Tylutki}}, \bibinfo
  {author} {\bibfnamefont {L.~P.}\ \bibnamefont {Pitaevskii}}, \bibinfo
  {author} {\bibfnamefont {F.}~\bibnamefont {Dalfovo}}, \bibinfo {author}
  {\bibfnamefont {G.}~\bibnamefont {Lamporesi}}, \ and\ \bibinfo {author}
  {\bibfnamefont {G.}~\bibnamefont {Ferrari}},\ }\href {\doibase
  10.1103/PhysRevLett.113.065302} {\bibfield  {journal} {\bibinfo  {journal}
  {Phys. Rev. Lett.}\ }\textbf {\bibinfo {volume} {113}},\ \bibinfo {pages}
  {065302} (\bibinfo {year} {2014})}\BibitemShut {NoStop}%
\bibitem [{\citenamefont {Madison}\ \emph {et~al.}(2001)\citenamefont
  {Madison}, \citenamefont {Chevy}, \citenamefont {Bretin},\ and\ \citenamefont
  {Dalibard}}]{madison_cheby_2001}%
  \BibitemOpen
  \bibfield  {author} {\bibinfo {author} {\bibfnamefont {K.~W.}\ \bibnamefont
  {Madison}}, \bibinfo {author} {\bibfnamefont {F.}~\bibnamefont {Chevy}},
  \bibinfo {author} {\bibfnamefont {V.}~\bibnamefont {Bretin}}, \ and\ \bibinfo
  {author} {\bibfnamefont {J.}~\bibnamefont {Dalibard}},\ }\href {\doibase
  10.1103/PhysRevLett.86.4443} {\bibfield  {journal} {\bibinfo  {journal}
  {Phys. Rev. Lett.}\ }\textbf {\bibinfo {volume} {86}},\ \bibinfo {pages}
  {4443} (\bibinfo {year} {2001})}\BibitemShut {NoStop}%
\bibitem [{\citenamefont {Abo-Shaeer}\ \emph {et~al.}(2001)\citenamefont
  {Abo-Shaeer}, \citenamefont {Raman}, \citenamefont {Vogels},\ and\
  \citenamefont {Ketterle}}]{abo_shaeer_raman_2001}%
  \BibitemOpen
  \bibfield  {author} {\bibinfo {author} {\bibfnamefont {J.~R.}\ \bibnamefont
  {Abo-Shaeer}}, \bibinfo {author} {\bibfnamefont {C.}~\bibnamefont {Raman}},
  \bibinfo {author} {\bibfnamefont {J.~M.}\ \bibnamefont {Vogels}}, \ and\
  \bibinfo {author} {\bibfnamefont {W.}~\bibnamefont {Ketterle}},\ }\href
  {\doibase 10.1126/science.1060182} {\bibfield  {journal} {\bibinfo  {journal}
  {Science}\ }\textbf {\bibinfo {volume} {292}},\ \bibinfo {pages} {476}
  (\bibinfo {year} {2001})}\BibitemShut {NoStop}%
\bibitem [{\citenamefont {Nikuni}\ and\ \citenamefont
  {Griffin}(2001)}]{nikuni_griffin_2001}%
  \BibitemOpen
  \bibfield  {author} {\bibinfo {author} {\bibfnamefont {T.}~\bibnamefont
  {Nikuni}}\ and\ \bibinfo {author} {\bibfnamefont {A.}~\bibnamefont
  {Griffin}},\ }\href {\doibase 10.1103/PhysRevA.63.033608} {\bibfield
  {journal} {\bibinfo  {journal} {Phys. Rev. A}\ }\textbf {\bibinfo {volume}
  {63}},\ \bibinfo {pages} {033608} (\bibinfo {year} {2001})}\BibitemShut
  {NoStop}%
\bibitem [{\citenamefont {Armaitis}\ \emph {et~al.}(2015)\citenamefont
  {Armaitis}, \citenamefont {Stoof},\ and\ \citenamefont
  {Duine}}]{armaitis_stoof_2015}%
  \BibitemOpen
  \bibfield  {author} {\bibinfo {author} {\bibfnamefont {J.}~\bibnamefont
  {Armaitis}}, \bibinfo {author} {\bibfnamefont {H.~T.~C.}\ \bibnamefont
  {Stoof}}, \ and\ \bibinfo {author} {\bibfnamefont {R.~A.}\ \bibnamefont
  {Duine}},\ }\href {\doibase 10.1103/PhysRevA.91.043641} {\bibfield  {journal}
  {\bibinfo  {journal} {Phys. Rev. A}\ }\textbf {\bibinfo {volume} {91}},\
  \bibinfo {pages} {043641} (\bibinfo {year} {2015})}\BibitemShut {NoStop}%
\bibitem [{\citenamefont {Hou}\ and\ \citenamefont {Yu}()}]{hou_yu_2015}%
  \BibitemOpen
  \bibfield  {author} {\bibinfo {author} {\bibfnamefont {Y.}~\bibnamefont
  {Hou}}\ and\ \bibinfo {author} {\bibfnamefont {Z.}~\bibnamefont {Yu}},\
  }\href@noop {} {\bibinfo  {journal} {arXiv:1504.04786}\ }\BibitemShut
  {NoStop}%
\bibitem [{\citenamefont {Liu}\ \emph {et~al.}(2014)\citenamefont {Liu},
  \citenamefont {Pattinson}, \citenamefont {Billam}, \citenamefont {Gardiner},
  \citenamefont {Cornish}, \citenamefont {Huang}, \citenamefont {Lin},
  \citenamefont {Gou}, \citenamefont {Parker},\ and\ \citenamefont
  {Proukakis}}]{liu_pattinson_15}%
  \BibitemOpen
\bibfield  {journal} {  }\bibfield  {author} {\bibinfo {author} {\bibfnamefont
  {I.-K.}\ \bibnamefont {Liu}}, \bibinfo {author} {\bibfnamefont {R.~W.}\
  \bibnamefont {Pattinson}}, \bibinfo {author} {\bibfnamefont {T.~P.}\
  \bibnamefont {Billam}}, \bibinfo {author} {\bibfnamefont {S.~A.}\
  \bibnamefont {Gardiner}}, \bibinfo {author} {\bibfnamefont {S.~L.}\
  \bibnamefont {Cornish}}, \bibinfo {author} {\bibfnamefont {T.-M.}\
  \bibnamefont {Huang}}, \bibinfo {author} {\bibfnamefont {W.-W.}\ \bibnamefont
  {Lin}}, \bibinfo {author} {\bibfnamefont {S.-C.}\ \bibnamefont {Gou}},
  \bibinfo {author} {\bibfnamefont {N.~G.}\ \bibnamefont {Parker}}, \ and\
  \bibinfo {author} {\bibfnamefont {N.~P.}\ \bibnamefont {Proukakis}},\
  }\href@noop {} {} (\bibinfo {year} {2014}),\ \Eprint
  {http://arxiv.org/abs/1408.0891} {arXiv:1408.0891} \BibitemShut {NoStop}%
\bibitem [{\citenamefont {Billam}\ and\ \citenamefont
  {Gardiner}(2012)}]{billam_gardiner_2012}%
  \BibitemOpen
  \bibfield  {author} {\bibinfo {author} {\bibfnamefont {T.~P.}\ \bibnamefont
  {Billam}}\ and\ \bibinfo {author} {\bibfnamefont {S.~A.}\ \bibnamefont
  {Gardiner}},\ }\href {http://stacks.iop.org/1367-2630/14/i=1/a=013038}
  {\bibfield  {journal} {\bibinfo  {journal} {New J. Phys.}\ }\textbf {\bibinfo
  {volume} {14}},\ \bibinfo {pages} {013038} (\bibinfo {year}
  {2012})}\BibitemShut {NoStop}%
\end{thebibliography}%

\end{document}